\documentclass[twocolumn,floatfix]{aastex631}
\usepackage{multirow, makecell}
\usepackage{microtype}
\usepackage{graphicx,subfigure}

\usepackage{textcomp}
\usepackage{rotating}
\usepackage{csquotes}

\begin{document}

\title{Investigating the Young Stellar Populations and Hierarchies in Nearby Galaxies with the UVIT. III. Evidence for a Largest Scale of Correlated Stellar Structures and a Non-universal Fractal Dimension}

\author[0009-0008-1250-6128]{Gairola Shashank}
\affiliation{Indian Institute of Astrophysics, Koramangala II Block, Bangalore-560034, India}
\affiliation{Pondicherry University, R.V. Nagar, Kalapet, 605014, Puducherry, India}

\author[0000-0002-5331-6098]{Smitha Subramanian}
\affiliation{Indian Institute of Astrophysics, Koramangala II Block, Bangalore-560034, India}
\affiliation{Pondicherry University, R.V. Nagar, Kalapet, 605014, Puducherry, India}
\affiliation{Leibniz-Institut für Astrophysik Potsdam, An der Sternwarte 16, D-14482 Potsdam, Germany}

\author[0000-0001-5944-291X]{Shyam H. Menon}
\affiliation{Center for Computational Astrophysics, Flatiron Institute, 162 5th Avenue, New York, NY 10010, USA}
\affiliation{Department of Physics and Astronomy, Rutgers University, 136 Frelinghuysen Road, Piscataway, NJ 08854, USA}

\author[0000-0003-4531-0945]{Chayan Mondal}
\affiliation{S. N. Bose National Centre for Basic Sciences Block-JD, Sector-III, Salt Lake, Kolkata-700106, India}

\author[0000-0002-3247-5321]{Kathryn Grasha}
\affiliation{Research School of Astronomy and Astrophysics, Australian National University, Canberra, ACT 2611, Australia}

\author[0000-0002-7203-5996]{Richard de Grijs}
\affiliation{School of Mathematical and Physical Sciences, Macquarie University, Balaclava Road, Sydney, NSW 2109, Australia}
\affiliation{Astrophysics and Space Technologies Research Centre, Macquarie University, Balaclava Road, Sydney, NSW 2109, Australia}

\author[0000-0002-1122-6270]{Prasun Dutta}
\affiliation{Department of Physics, IIT (BHU), Varanasi 221005, India}

\author[0000-0003-4531-0945]{Annapurni Subramaniam}
\affiliation{Indian Institute of Astrophysics, Koramangala II Block, Bangalore-560034, India}

\begin{abstract}

Scale-free turbulent motions, gravitational collapse and galactic dynamics govern galactic-scale, hierarchical organization of star formation (SF) within galaxies. Past studies suggest that properties of SF hierarchies depend upon host galaxy properties and interstellar medium (ISM) conditions. We explored stellar hierarchies in a sample of 17 morphologically diverse galaxies, including 8 grand design, 6 flocculent spirals and 3 dwarf irregulars. We performed two-point correlation function analysis on $\sim$25000 UV-selected star-forming clumps (SFCs) across 17 galaxies to characterize SF hierarchies. We found that SF hierarchies in galaxies exhibit a maximum spatial scale — the correlation length ($l_{\rm corr}$) — representing the largest scale up to which SF is spatially correlated - presumably owing to the ISM turbulence. The $l_{\rm corr}$ values range from $\sim$100 pc to 3.4 kpc and exhibit strong dependence on the galaxy’s stellar mass, morphology and nature of spiral arms. This suggests that a galaxy’s gravitational potential and spiral structure place an upper limit on the sizes of the largest, hierarchically structured SF complexes. Connecting $l_{\rm corr}$ values with turbulence injection sources and scales suggests that stellar feedback in dwarf irregulars, whereas disk instabilities and spiral structure in flocculent/classic spirals dominate towards sustaining their SF hierarchies up to the $l_{\rm corr}$ scale. These hierarchies disperse to near-random distributions on timescales ($T_{\rm dis}$) ranging from 20$-$160 Myr. The broad range of derived $l_{\rm corr}$, projected fractal dimension ($D_2$ $\in$ 0.71$-$1.73), and $T_{\rm dis}$ indicates a non-universal, galaxy-specific nature of SF hierarchies. The distinctive feature of our study is the full coverage of each galaxy’s star-forming extent provided by the AstroSat$-$UltraViolet Imaging Telescope, which allowed us to connect global parameters of SF hierarchies with large-scale galaxy properties.

\end{abstract}


\section{Introduction}
\label{sec:intro}

\begin{figure*}
    \centering
    \includegraphics[width=0.80\textwidth]{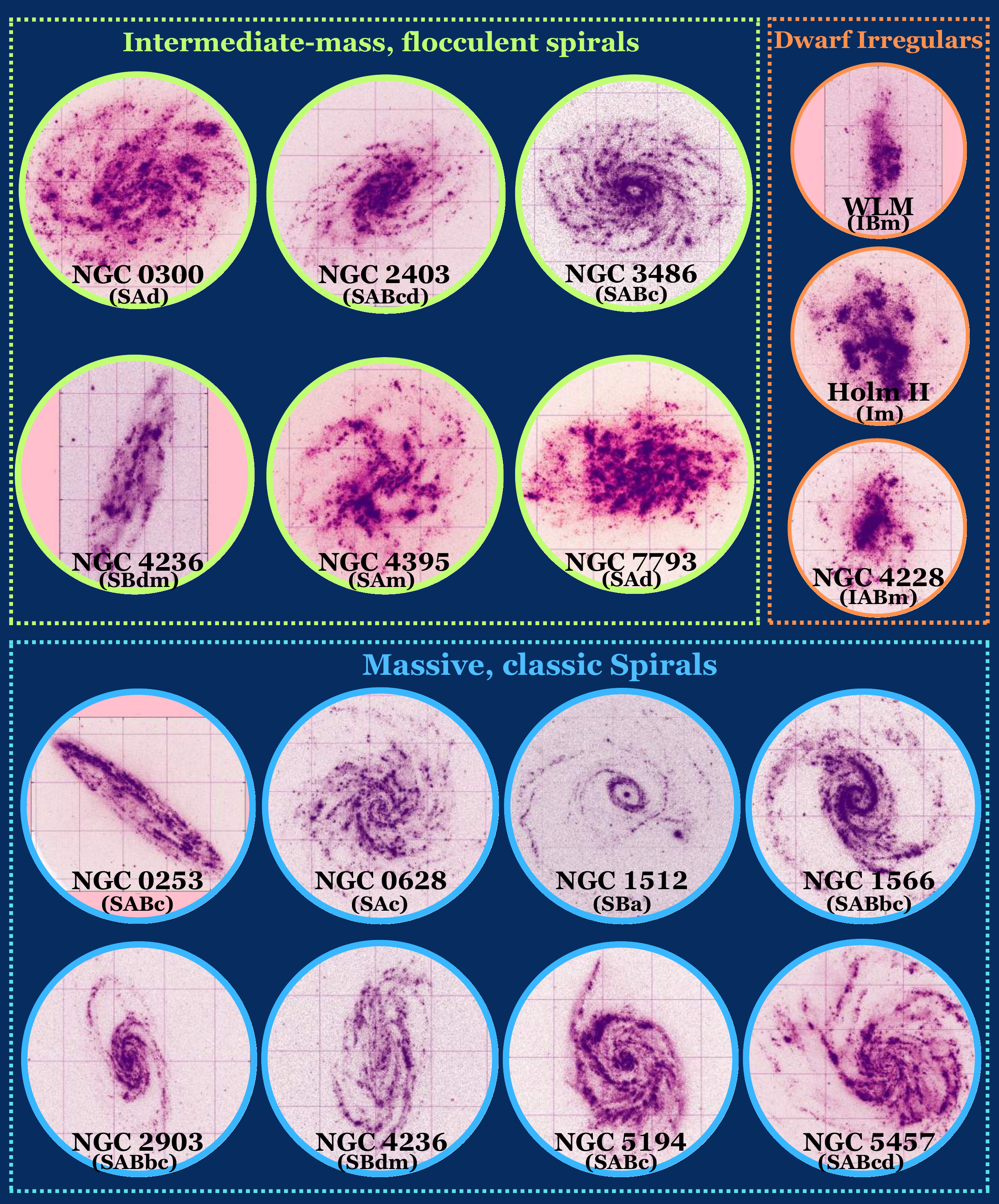}
    \caption{UVIT FUV images of our 17 sample galaxies, separated by morphological subclass. See Figure 1 in Paper II for a higher quality, UVIT FUV + NUV color-composite images of these galaxies.}
    \label{fig:UVIT_images}
\end{figure*}

Star formation in galaxies is a multi-scale, hierarchical process in which single stars, stellar multiplets, star clusters, stellar associations, and kilo-parsec (kpc) sized stellar complexes can form within relatively short timescales, ranging from tens of thousands to tens of millions of years (Myrs) \citep{1998MNRAS.299..588E, 2003ARA&A..41...57L, 2000ApJ...530..277E, 2003MNRAS.343..413B, elmegreen2006hierarchical, elmegreen2014hierarchical, 2017ApJ...840..113G, 2017ApJ...842...25G, 2018PASP..130g2001G, 2019ARA&A..57..227K}. These stellar structures share hierarchical relationships in which denser structures are progressively embedded in larger, less dense structures. This hierarchy of stellar structures arises from the hierarchical organization of the star-forming interstellar medium (ISM) components such as neutral hydrogen (HI) gas \citep{2006ApJ...652.1339M, 2009MNRAS.398..887D, 2013NewA...19...89D, 2020MNRAS.496.1803N} and molecular gas \citep{1987ApJ...312L..45B, 1981MNRAS.194..809L, 1996ApJ...471..816E, Shadmehri_2011}. The scale-free processes of supersonic turbulence (induced by feedback, galactic dynamics, spiral structure, shear etc.) and gravitational interaction (in the form of gravitational collapse, disk instabilities, accretion, galaxy interactions etc.) are believed to be the sculpting mechanisms for the hierarchical structure of the ISM \citep{1981MNRAS.194..809L, 2004RvMP...76..125M, 2007ARA&A..45..565M, 2009ApJ...692..364F, 2013MNRAS.436.1245F, Federrath_2018, 2023A&A...672A.193F}. 

Our current understanding of the star formation hierarchy comes from two different types of studies. Some studies combined multi-tracer observations of individual galaxies to connect the hierarchy of gaseous ISM with that of the stellar matter, e.g. \cite{Zhang_2001} in Antennae Galaxies, \cite{2010ApJ...720..541S} in M33,  \cite{Grasha_2018, Grasha_2019} in NGC 7793 and NGC 5194 respectively, \cite{2022MNRAS.516.4612T} in 11 nearby galaxies. Other studies focused separately on large galaxy samples to investigate the hierarchical distribution of either the gaseous ISM \citep{1996ApJ...471..816E, Sanchez_2008, Shadmehri_2011, 2026arXiv260407450H} or stellar matter \citep{elmegreen2006hierarchical, elmegreen2014hierarchical, 2017ApJ...840..113G, Rodriguez_2020, 2021MNRAS.507.5542M, Meena_2025, lapeer2026feast}. Moreover, hierarchical star formation was also investigated extensively in the Magellanic Clouds, which serve as our nearest extra-galactic laboratories of star formation \citep{2008MNRAS.391L..93G, 2009MNRAS.392..868B, Sun_2017, 2017ApJ...849..149S, 2018ApJ...858...31S, 2022MNRAS.512.1196M, 2025ApJ...989..216H}. These studies showed that young stellar structures retain the hierarchical characteristics of their natal gas clouds for a few tens of Myrs (a timescale termed the hierarchy dispersal timescale), before galactic dynamics and turbulent motions in the ISM sweep these features away. Therefore, examining the distribution of young star-forming regions by itself can provide valuable insights into the physical mechanisms governing the hierarchical star formation process.

On small scales, the adiabatic limit of gravitational collapse limits the star formation hierarchy at $\sim$0.1 pc. However, it is still unclear whether these hierarchical structures span the full extent of the galaxy or they have a limiting upper scale. Recent studies suggest that star formation hierarchies may not be entirely scale-free, rather they may exhibit a characteristic maximum scale called the correlation length ($l_{\rm corr}$) \citep{2017ApJ...840..113G, 2017ApJ...849..149S, 2021MNRAS.507.5542M, shashank2025tracing, lapeer2026feast}, which is usually much smaller than the full galaxy size. These studies propose that the correlation length may represent the largest scale up to which ISM turbulence can dictate the locations of spatially correlated star formation events within a galaxy. Beyond the correlation length scale, star-forming regions may be uncorrelated, and their positions would be dictated by large-scale galactic structure and dynamics. In this paradigm, the upper limit of stellar hierarchies can be equal to or smaller than the largest scales up to which the turbulence is injected in the ISM \citep{2008MNRAS.384L..34D, 2009MNRAS.398..887D, 2013NewA...19...89D, 2020MNRAS.496.1803N}. The existence of a correlation length also implies that certain physical mechanisms within galaxies may be limiting the star-formation hierarchies from permeating indefinitely, up to the full size of the galaxy. Disk instabilities and galactic shear have been proposed as suitable mechanisms for this phenomenon in the literature \citep{elmegreen2014hierarchical, 2016MNRAS.460.2360R, 2017MNRAS.469..286R, 2017ApJ...840..113G, 2018PASP..130g2001G, 2021MNRAS.507.5542M}. 

\begin{table*}
\caption{Key physical properties of the 17 galaxies studied in this paper.}
\centering
\hspace*{-\dimexpr\oddsidemargin+1in\relax}%
\makebox[\paperwidth][c]{%
\resizebox{\textwidth}{!}{%
\begin{tabular}{ccccccccccc}
\hline
Galaxy   & Morphological & P.A.           & Incl.         & Distance     & $M_\star$                        & $R_{25}$         & SFR              & $\Sigma$$_{\rm{SFR}}$                     & $N_{\rm{SFC}}$ & Mag. error \\
         & subclass      & (deg.)         & (deg.)        & (Mpc)        & (M$_{\odot}$)                  & (kpc)            & (M$_{\odot}$/yr) & (M$_{\odot}$/yr/kpc$^{2}$) &           & (mag)      \\
(1)      & (2)           & (3)            & (4)           & (5)          & (6)                            & (7)              & (8)           & (9)   & (10)  & (11)       \\\hline
Holmberg II  & dIrr          & 177.0$^{(d)}$  & 41.0$^{(d)}$  & 3.4$^{(d)}$  & 2.0 $\times$10$^{9}$ $^{(d)}$  & 3.7$^{(d)}$      & 0.05$^{(d)}$  & 1.54 $\times$10$^{-3}$  & 1181  & 0.10\\
NGC 4228 & dIrr          & 65.0$^{(d)}$   & 44.0$^{(d)}$  & 2.9$^{(d)}$  & 6.3 $\times$10$^{8}$ $^{(d)}$  & 2.9$^{(d)}$      & 0.11$^{(d)}$  & 5.79 $\times$10$^{-3}$  & 648   & 0.25\\
WLM      & dIrr          & 181.0$^{(m)}$  & 69.0$^{(m)}$  & 1.0$^{(m)}$  & 4.3 $\times$10$^{7}$ $^{(n)}$  & 1.8$^{(f)}$$^{*}$  & 0.01$^{(m)}$  & 2.74 $\times$10$^{-3}$  & 1357  & 0.20\\
NGC 0300 & IMFS          & 114.3$^{(a)}$  & 39.8$^{(a)}$  & 2.1$^{(a)}$  & 1.9 $\times$10$^{9}$ $^{(a)}$  & 5.9$^{(c)}$      & 0.15$^{(a)}$  & 1.79 $\times$10$^{-3}$  & 4947  & 0.10\\
NGC 2403 & IMFS          & 124.0$^{(d)}$  & 63.0$^{(d)}$  & 3.2$^{(d)}$  & 5.0 $\times$10$^{9}$ $^{(d)}$  & 7.3$^{(d)}$      & 1.30$^{(d)}$  & 17.10 $\times$10$^{-3}$  & 2144  & 0.10\\
NGC 3486 & IMFS          & 80.0$^{(e)}$   & 50.0$^{(e)}$  & 11.4$^{(f)}$ & 6.3 $\times$10$^{9}$ $^{(g)}$  & 11.7$^{(e)}$     & 1.10$^{(e)}$  & 3.98 $\times$10$^{-3}$  & 864   & 0.25\\
NGC 4236 & IMFS          & 162.0$^{(h)}$  & 75.0$^{(h)}$  & 4.5$^{(h)}$  & 9.2 $\times$10$^{8}$ $^{(i)}$  & 15.2$^{(h)}$     & 0.21$^{(h)}$  & 1.11 $\times$10$^{-3}$  & 1194  & 0.20\\
NGC 4395 & IMFS          & 147.0$^{(j)}$  & 38.0$^{(j)}$  & 4.3$^{(j)}$  & 2.5 $\times$10$^{9}$ $^{(j)}$  & 8.2$^{(j)}$      & 0.47$^{(j)}$  & 2.82 $\times$10$^{-3}$  & 947   & 0.20\\
NGC 7793 & IMFS          & 98.0$^{(b)}$   & 55.0$^{(b)}$  & 3.6$^{(b)}$  & 3.2 $\times$10$^{9}$ $^{(b)}$  & 4.9$^{(b)}$      & 0.52$^{(b)}$  & 12.02 $\times$10$^{-3}$  & 1775  & 0.10\\
NGC 0253 & MCS           & 52.5$^{(a)}$   & 75.0$^{(a)}$  & 3.7$^{(a)}$  & 4.4 $\times$10$^{10}$ $^{(a)}$ & 14.4$^{(c)}$     & 5.00$^{(a)}$  & 29.65 $\times$10$^{-3}$  & 1348  & 0.10\\
NGC 0628 & MCS           & 20.7$^{(b)}$   & 8.9$^{(b)}$   & 9.8$^{(b)}$  & 1.1 $\times$10$^{10}$ $^{(b)}$ & 15.0$^{(b)}$     & 3.67$^{(b)}$  & 5.26 $\times$10$^{-3}$  & 883   & 0.25\\
NGC 1512 & MCS           & 261.9$^{(c)}$  & 42.5$^{(c)}$  & 18.8$^{(c)}$ & 5.2 $\times$10$^{10}$ $^{(c)}$ & 23.1$^{(c)}$     & 5.16$^{(a)}$  & 4.17 $\times$10$^{-3}$  & 385   & 0.20\\
NGC 1566 & MCS           & 214.7$^{(b)}$  & 29.6$^{(b)}$  & 17.7$^{(b)}$ & 2.7 $\times$10$^{10}$ $^{(b)}$ & 21.4$^{(b)}$     & 5.67$^{(b)}$  & 4.53 $\times$10$^{-3}$  & 1194  & 0.20\\
NGC 2903 & MCS           & 203.7$^{(c)}$  & 66.8$^{(c)}$  & 10.0$^{(c)}$ & 4.4 $\times$10$^{10}$ $^{(c)}$ & 17.4$^{(c)}$     & 4.30$^{(a)}$  & 11.48 $\times$10$^{-3}$  & 1180  & 0.20\\
NGC 5033 & MCS           & 352.0$^{(k)}$  & 68.0$^{(k)}$  & 16.5$^{(k)}$ & 3.7 $\times$10$^{10}$ $^{(l)}$ & 29.7$^{(l)}$     & 1.50$^{(l)}$  & 1.44 $\times$10$^{-3}$  & 647   & 0.25\\
NGC 5194 & MCS           & 173.0$^{(b)}$  & 22.0$^{(b)}$  & 8.6$^{(b)}$  & 2.4 $\times$10$^{10}$ $^{(b)}$ & 13.9$^{(b)}$     & 6.88$^{(b)}$  & 12.22 $\times$10$^{-3}$ & 1694  & 0.20\\
NGC 5457 & MCS           & 39.0$^{(b)}$   & 18.0$^{(b)}$  & 6.7$^{(b)}$  & 1.9 $\times$10$^{10}$ $^{(b)}$ & 27.9$^{(b)}$     & 6.72$^{(b)}$  & 2.89 $\times$10$^{-3}$  & 1507  & 0.10\\
\hline
\end{tabular}
}
}
\tablecomments{(1) Galaxy name, (2) morphological subclass (dIrr = dwarf irregular, IMFS = intermediate-mass flocculent spiral, MCS: massive classic spiral), (3) position angle measured anti-clockwise from the celestial north, (4) inclination, (5) distance, (6) stellar mass, (7) $R_{25}$, (8) SFR, (9) SFR surface density (= SFR/($\pi$$\times$$R_{25}$$^{2}$), (10) number of SFCs belonging to the galaxy and (11) FUV and NUV magnitude error cuts associated with the SFCs, respectively. The galaxies are sorted by morphological subclasses. References indicated as bracketed superscripts are as follows (a) - \cite{Hassani_2024}, (b) - \cite{2021MNRAS.507.5542M}, (c) - \cite{2021ApJS..257...43L}, (d) - \cite{2008AJ....136.2782L}, (e) - \cite{2015AJ....149....1Z}, (f) - NED, (g) - \cite{2022MNRAS.515.3270S}, (h) - \cite{2007A&A...462..933C}, (i) - \cite{2019A&A...621A..51H}, (j) - \cite{2023ApJ...950...81N}, (k) - \cite{1997MNRAS.290...15T}, (l) - \cite{2019MNRAS.488.3826B}, (m) - \cite{Mondal_2018}, (n) - \cite{2019MNRAS.490..467Z}. $^{*}$ $-$ For WLM, the galaxy radius is taken as $R_{25}$.}
\label{table1}
\end{table*}

The properties of star formation hierarchies also vary with galaxy morphology and the environment within the galaxy. Here, the word environment indicates the physical conditions of the ISM, such as ambient pressure, gas density, galactic shear, and metallicity. Results from different studies point towards a remarkable diversity in the \enquote{hierarchy parameters}\footnote{In this paper, we collectively refer to the correlation length, hierarchy dispersal timescales and fractal dimension as hierarchy parameters.} derived for different galaxies, such as the correlation length, hierarchy dispersal timescale, and fractal dimension ($D_2$ : it quantifies the fractal complexity and space-filling nature of the distribution). The wide diversity in hierarchy parameters can arise because physical conditions of the star-forming ISM can vary significantly across different galaxy morphologies (dwarfs, irregulars, flocculent spirals, grand design spirals, interacting galaxies) and different spatial locations within individual galaxies (galactic center, spiral arms, bars, galaxy outskirts) \citep{Sanchez_2008, 2017ApJ...840..113G, 2017ApJ...849..149S, 2021MNRAS.507.5542M, 2022MNRAS.512.1196M, shashank2025tracing}. Therefore, comparing hierarchy parameters for a statistically large sample of galaxies, spanning a wide range in morphology and environments, will be essential to understand the different physical mechanisms that influence hierarchical star formation.

High resolution ($\lesssim$0.1") studies using Hubble Space Telescope (HST) data have provided many of the aforementioned insights about the hierarchical star formation process. Although these studies covered the inner disks of galaxies with high angular resolution, in \cite{shashank2025tracing} (Paper I from here on), we showed that global hierarchy parameters of galaxies, derived using full galaxy coverage data, can differ significantly from those derived for the HST-covered regions of the same galaxies. This implied that, for statistical studies of hierarchical star formation where each galaxy serves as a single data point, full galaxy coverage is essential, allowing us to connect global hierarchy parameters with host galaxy properties\footnote{This assumes that star formation within a galaxy is relatively homogeneous and enhanced localised star formation within the galaxy is absent.}. In Paper I, we had focused on four nearby spiral galaxies and used archival far-UV (FUV) and near-UV (NUV) observations from the Ultra-Violet Imaging Telescope (UVIT) onboard AstroSat satellite \citep{2012SPIE.8443E..1NK}. UVIT's 28\arcmin~field of view (FoV) enabled us to probe the complete star-forming extent of these galaxies at $\le$1.5\arcsec~ angular resolution. We showed that all the spiral galaxies are expected to have a correlation length. Additionally, we observed significant galaxy-to-galaxy variation in the hierarchy parameters of our galaxies.

Anchored by the UVIT FUV and NUV observations, in this paper, we extend our analysis of hierarchical star formation to a larger and morphologically diverse sample of 17 nearby galaxies, spanning three orders of magnitude in mass range ($\sim$10$^{8}$M$_{\odot}$ to 10$^{11}$M$_{\odot}$). We utilize the catalog of $\sim$25000 star-forming clumps (SFCs) characterized in these 17 galaxies, presented in \cite{2026arXiv260612254S} (Paper II from here on), as our probes of stellar hierarchies within galaxies. We aim to address the following points in this paper - 1) Do all galaxies, irrespective of their morphology, have a correlation length and which physical mechanisms determine it?, 2) To probe the galaxy-to-galaxy variation of fractal dimension and hierarchy dispersal timescale, 3) To investigate the nature of the dependence of hierarchy parameters on host galaxy properties.

The paper is structured as follows. Section \ref{sec:data} describes our sample of 17 galaxies and the SFC catalog used in this paper. Section \ref{sec:TPCF} and \ref{sec:TPCF_maths_models} provide details of the two-point correlation (TPCF) method, which enabled us to parametrize the stellar hierarchy of our sample galaxies. In Section \ref{sec:obs_TPCF}, we present our observed galaxy-by-galaxy TPCF results for different SFC age groups. In Section \ref{sec:discussion}, we present an interpretation of our observed results and discuss them in the context of existing literature on hierarchical star formation. Finally, we summarize our key findings and outline future plans in Section \ref{sec:summary}.\

\begin{figure*}
    \centering
    \includegraphics[width=0.95\textwidth]{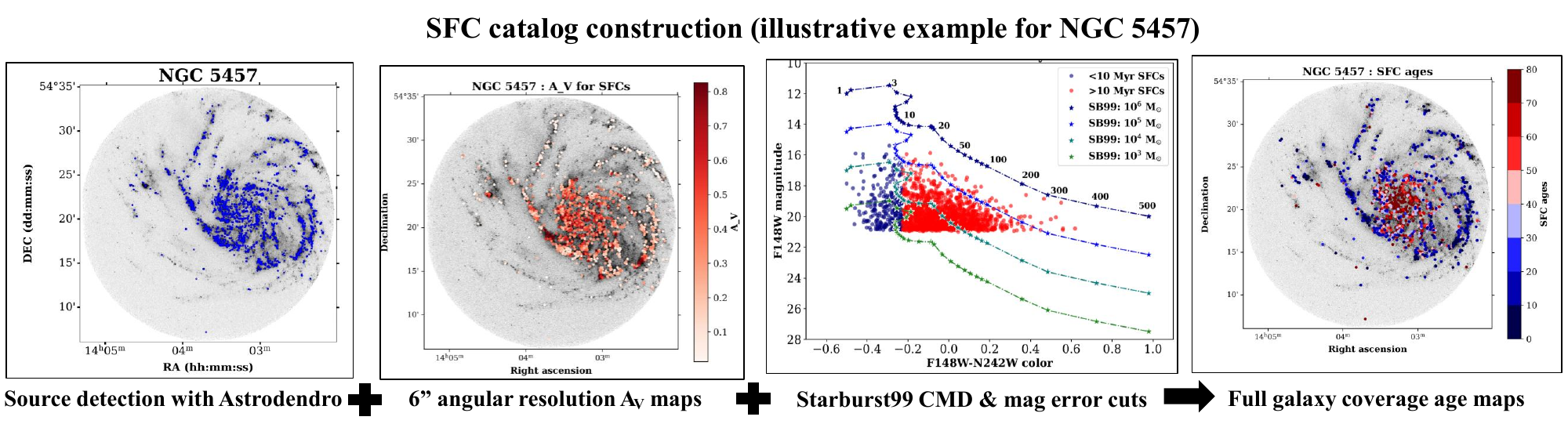}
    \caption{A summary of the important steps involved in the construction of our SFC catalog in Paper II. The SFCs are identified using Astrodendro software. Their FUV and NUV magnitudes are corrected using our own dust attenuation maps, and the SFC ages are determined by comparing their reddening-corrected FUV$-$NUV colors with synthetic colors from Starburst99 SSP models. These steps helped us create full galaxy coverage SFC age maps for our galaxies (refer to Section \ref{sec:data} for details).}
    \label{fig:catalog_construction}
\end{figure*}

\section{Galaxy Sample and a UV-selected catalog of $\sim$25000 SFCs}
\label{sec:data}
In this study, we investigated the hierarchical distribution of SFCs in 17 nearby galaxies, all located within 20 Mpc. These galaxies were selected in Paper II primarily based on the availability of UVIT FUV and NUV data. In paper II, we used a homogeneous SFC detection and characterization methodology to construct a catalog of $\sim$25000 SFCs in these 17 galaxies and studied their stellar population demographic. In this paper, we directly use the SFC catalog and ages presented in paper II for our investigation of hierarchical star formation. We refer the interested readers to Paper II for a detailed description of our galaxy sample selection and SFC catalog construction process. We briefly present the relevant details in this section.

We classified our 17 galaxies into three morphological subclasses using the stellar mass of the galaxy and the visual inspection of their spiral arms. Our galaxy sample includes 8 massive, classic spirals ($M_\star$$>$10$^{10}$M$_{\odot}$ with well-defined spiral structure), 6 intermediate mass, flocculent spirals (10$^{10}$M$_{\odot}$$>$$M_\star$$\gtrsim$10$^{9}$M$_{\odot}$ with fragmented spiral structure) and 3 dwarf irregulars ($M_\star$$\lesssim$10$^{9}$M$_{\odot}$ with no prominent spiral or disk structure) (see Figure \ref{fig:UVIT_images}). A summary of our morphological subclasses, along with the important physical properties of the galaxies used in our analysis, is provided in Table \ref{table1}. Our sample of 17 galaxies is the largest studied to date in the context of the hierarchical organization of stellar matter in galaxies. We note that previously, \cite{Sanchez_2008} had explored the hierarchical organization of ionized gas in 93 galaxies using HII regions $-$ which are indirect, gaseous tracers of star formation. However, the present study focuses on UV-emitting SFCs that provide a more direct tracer of the star formation hierarchy.\

To identify the SFCs, we applied the Astrodendro \citep{rosolowsky2008structural} software and a homogeneous source detection criterion on the UVIT FUV images of our galaxies. Astrodendro identifies hierarchical connections between different regions of peak intensity in two-dimensional (2D) flux maps, such as UVIT fits images. It creates a hierarchical dendrogram tree by scanning the images from the peak flux point down to a specified minimum flux threshold. A typical dendrogram consists of trunk, branch, and leaf structures, where leaves represent the most compact and indivisible structures identifiable in a flux map. We have used Astrodendro leaves as SFCs in our study. Leaves in Astrodendro are identified using three input parameters - min\_value, min\_delta and min\_npix. The min\_value parameter acts as the detection threshold and we set it equal to three times the standard deviation ($\sigma$) of flux in the sky-background subtracted image. The min\_delta parameter acts as a de-blending criterion for separating adjacent leaves and it is set to 1$\sigma$. Finally, min\_npix helps us identify leaves above the spatial resolution of the UVIT, and we set it equal to 11 pixels which is equivalent to a circular PSF of 1.5\arcsec~diameter FWHM. 

For the detected sources (SFC candidates), we performed aperture photometry on the sky background corrected FUV and NUV images and determined the source magnitudes. We employed FUV and NUV magnitude error cuts in order to restrict our source catalog to only those SFCs for which ages can be reliably constrained. This allows us to keep the age errors to a minimum (see Section 3.2 in Paper II for a discussion of age errors). The final number of SFCs characterized in our galaxies and the associated magnitude error cuts are provided in Table 1. The SFC ages were determined by comparing the observed UV color-magnitude diagrams (FUV$-$NUV color v/s FUV magnitude) against the Starburst99 simple stellar population (SSP) synthesis models \citep{leitherer1999starburst99, leitherer2014effects}. Several studies in the past have demonstrated that the GALEX/UVIT-detected star-forming complexes and LEGUS based OB associations can be effectively modeled as SSPs \citep{Iglesias-Paramo:2004mnq, 2005ApJ...619L..79T, bianchi2005recent, 2007ApJ...658.1006M, Pasquali_2008, 2017ApJ...841..131A, 2019MNRAS.484.4897C}. This makes the use of SSPs in deriving SFC ages a reasonable choice. We refer interested readers to Section 3.3 of Paper II for a detailed discussion about how modeling SFCs with continuous star formation history instead of SSPs affects the derived ages. 

The attenuation corrected FUV$-$NUV color was used as the direct indicator of the age of SFCs. The dust attenuation for each SFC was corrected using our own spatially-resolved, full galaxy coverage A$_V$ maps, created at 6\arcsec~angular resolution. We used the UVIT FUV, Two Micron All-Sky Survey (2MASS) $J$-band \citep{2003AJ....125..525J} and Multi-band Imaging Photometer for Spitzer (MIPS) 24$\mu$$m$ observations \citep{2009ApJ...693.1821D} and the method outlined by \cite{2016A&A...591A...6B} to create the A$_V$ maps. These steps ultimately led to a catalog of $\sim$25000 SFCs in the 17 galaxies, within the full spatial extent (i.e. up to the radial position of the outermost UV-selected SFC in the full-coverage UVIT FUV images) of our 17 galaxies. Figure \ref{fig:catalog_construction} presents an illustration showing the key steps involved in the construction of this catalog. 

\section{Two-point correlation function: A statistical method to probe the hierarchical distribution of SFCs}
\label{sec:TPCF}

\begin{table*}
\caption{The four TPCF models used to fit all the observed TPCF plots presented in this paper.}
\resizebox{1.0\textwidth}{!}{%
\begin{tabular}{cccc}
\hline
S.N. & TPCF model                          & Mathematical expression                                                                     & Interpretation and notes                                               \\
 & & \\\hline\hline

1 & Single power-law                    & A$_1$$x^{-\alpha_1}$                                                                         & Single power-law valid at all scales;                                  \\
& Model S; F$_{S}(x)$                 &                                                                                             & Fractal dimension ($D_2$) = 2 + $\alpha_1$                              \\\hline

2 & Broken power-law                    & A$_1$$x^{-\alpha_1}$ :  $x$ $<$ $l_{\rm corr}$                                              & Steep power-law slope $\alpha_1$ on scales $<$ $l_{\rm corr}$,         \\
& Model PW; F$_{PW}(x)$               & A$_2$$x^{-\alpha_2}$ :  $x$ $>$ $l_{\rm corr}$                                              & Shallow power-law slope $\alpha_2$$\sim$0 on scales $>$ $l_{\rm corr}$; \\
                                    & where, A$_2$ = A$_1$l$_{\rm{corr}}^{(\alpha_1-\alpha_2)}$                                   & $D_2$ = 1 + $\alpha_1$                                                   \\\hline

3 & Single power-law + exponential decline & A$_1$$x^{-\alpha_1}$ exp($\frac{-x}{r_c}$)                                                   & Shallow power-law slope $\alpha_1$,          \\
& Model PF; F$_{PF}(x)$               &                                                                                             & Exponential decline on larger scales (r$_c$ = the fall-off scale) \\\hline

4 & Broken power-law + exponential decline & A$_1$$x^{-\alpha_1}$ : $x$ $<$ $l_{\rm corr}$                                               &  Steep power law slope $\alpha_1$ on scales $<$ $l_{\rm corr}$,                   \\
& Model PWF; F$_{PWF}(x)$             & A$_2$$x^{-\alpha_2}$ exp($\frac{-x}{r_c}$) : $x$ $>$ $l_{\rm corr}$                         & Shallow power law slope $\alpha_1$$\sim$0 on scales $>$ $l_{\rm corr}$,           \\
& (hybrid of PW and PF model)         & where, A$_2$ = A$_1$l$_{\rm{corr}}^{(\alpha_1-\alpha_2)}$exp($\frac{+l_{\rm{corr}}}{r_c}$) & followed by an exponential decline on largest scales                             \\
                                    &                                                     &                                        & (r$_c$ being the fall-off scale); $D_2$ = 2 + $\alpha_1$                      \\\hline
\end{tabular}
}
\label{table2}
\end{table*}

We used the two-point correlation function (TPCF) to study the hierarchical distribution of SFCs and their evolution as a function of age. The shape of the TPCF can enable distinguishing between different types of distributions, such as hierarchical, Poissonian, or exponential. TPCF quantifies the amount of clustering in a given distribution as a function of spatial scale. Mathematically, TPCF measures the excess probability of finding any two data points separated by a given distance $x$ as compared to a random or Poissonian distribution \citep{1980lssu.book.....P}. In this work, we have used the Landy-Szalay TPCF estimator \citep{Landy_1993}, which works on a pair-counting principle as follows, 

\begin{equation}
\begin{array}{l}
\mathrm{TPCF}(x) = 1 + \frac{\mathrm{DD}(x) - 2 \mathrm{DR}(x)+\mathrm{RR}(x)}{\mathrm{RR}(x)}
\end{array}
\end{equation}

\begin{equation}
\begin{array}{l}
\mathrm{DD}(x)=\frac{N_{\mathrm{DD}}(x)}{N_D(N_D-1)};\ \mathrm{DR}(x)=\frac{N_{\mathrm{DR}}(x)}{N_D N_{R}}; \ \mathrm{RR}(x)=\frac{N_{\mathrm{RR}}(x)}{N_{R}\left(N_{R}-1\right)}\\
\end{array}
\end{equation} 

where, $N_{\mathrm{DD}}$(\textit{x}) $N_{\mathrm{DR}}$(\textit{x}), and $N_{\mathrm{RR}}$(\textit{x}) represent the number of data-data, data-random and random-random pairs of points found at separation $x$. $N_D$, $N_R$ are the total number of data and random points taken into consideration for the TPCF calculation. The Landy-Szalay TPCF estimator is defined such that a truly Poissonian distribution will produce a TPCF value of 1 at all the spatial scales. Any distribution exhibiting more clustering than a Poissonian distribution at a given physical scale will produce a TPCF value greater than 1. Fractal-like distributions exhibit a power-law form of the TPCF, indicating a scale-free, hierarchical distribution. Moreover, if at a given scale, the distribution of data points is less clustered than a Poissonian distribution, then the TPCF value will fall below 1. 

Practical implementation of the TPCF formula requires a distribution of data points for which TPCF is being estimated, and an approximately equal number of randomly distributed points, occupying the same spatial footprint. In order to ensure that data and random points span the same spatial extent, we used our customized random distribution generation scheme developed in Paper I. Before performing the TPCF analysis, we de-projected the SFC positions to correct for the inclination of the galaxy in the sky plane. Using the position angle of the galaxies, we rotated the SFC positions so that the major axis of the galaxy and the celestial North-South axis are aligned. Next, the SFC positions along the minor axis were divided by the cosine of the inclination angle. This ensured that the major and minor axis of the galaxy are nearly equal, as expected for a face-on disc galaxy.

We computed the TPCF in 20 logarithmically spaced bins between the physical scales corresponding to the UVIT's angular resolution (i.e.$\sim$1.5\arcsec) and the scale corresponding to the largest de-projected separation between any two SFCs within a galaxy. To robustly calculate TPCF and minimize noise, we iterate the TPCF estimation procedure 100 times. In each iteration, the data distribution remains the same, but a new distribution of random points is generated. The TPCF value at a given $x$ is taken as the mean of the TPCF($x$) distribution over the 100 runs, whereas the error bar on TPCF($x$) is taken as the standard deviation of the distribution. 

Similar to the observation made in Paper I, we note that the number of data and random pairs separated by scales close to the UVIT resolution is usually very small. The TPCF values corresponding to these first few bins are quite noisy, possessing error bars of the same order as the actual TPCF value. Therefore, such bins are not shown in our TPCF plots (Figures \ref{fig:tpcf_plots}, \ref{fig:tpcf_plots_contd}, and \ref{fig:tpcf_revised}) and are excluded during the process of fitting the observed TPCF with the mathematical models. The first spatial separation bin considered in the fitting of the observed TPCF was taken to be the one after which the TPCF starts to fall approximately monotonically as a function of the spatial scale. This first bin typically corresponds to $\sim$2$-$4 times the UVIT resolution.

\section{TPCF mathematical models}
\label{sec:TPCF_maths_models}

\begin{figure*}[t]
      \centering
		\includegraphics[width=0.70\linewidth]{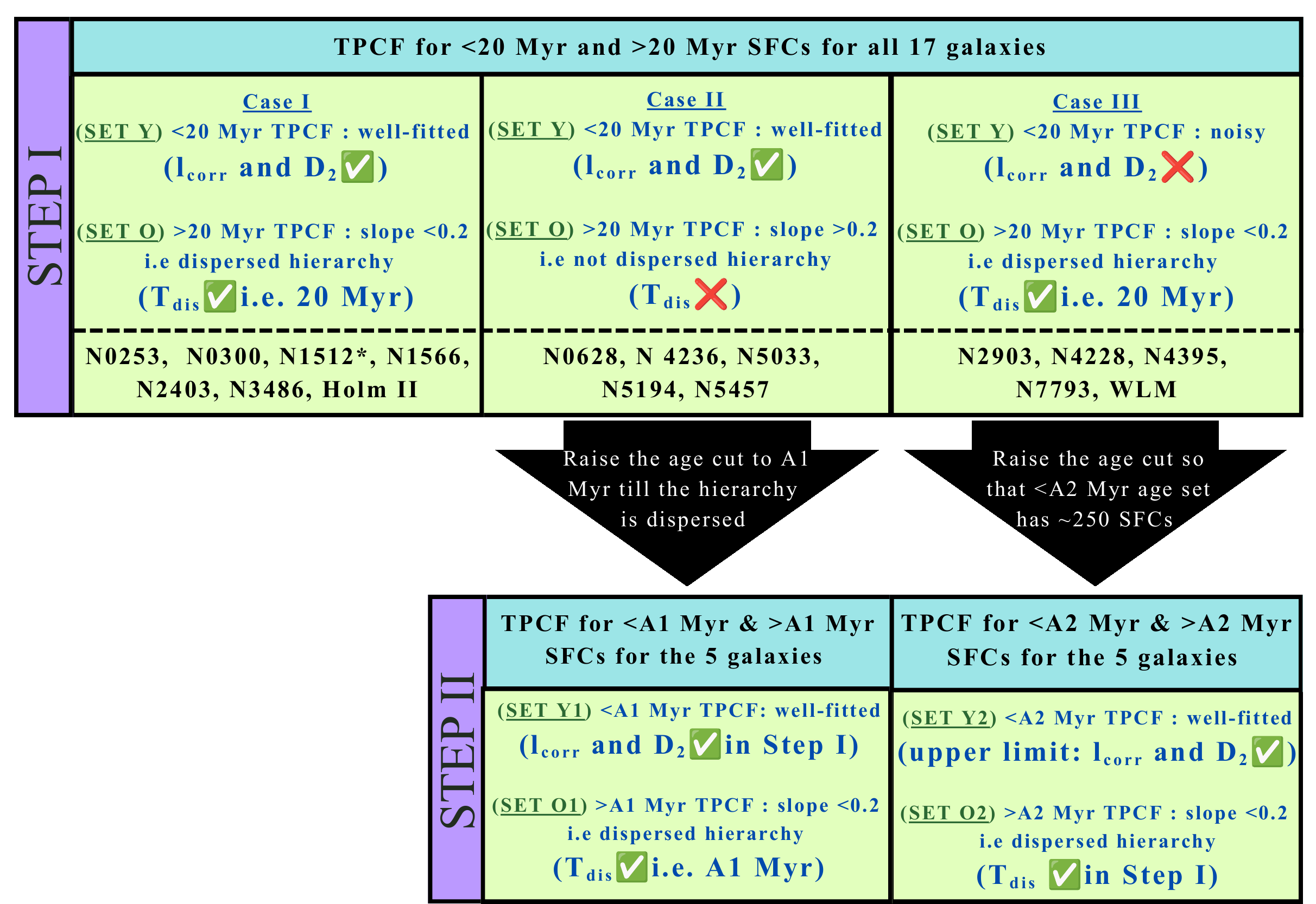} 
    \caption{An illustrative overview of our complete TPCF measurement methodology which is also elaborated in Section \ref{sec:obs_TPCF}. Here, \enquote{Y} and \enquote{O} denote young and old SFC sets, respectively, with the classification defined in relative terms.}
  \label{fig:schematic}
\end{figure*}

Before presenting the observed TPCF in our galaxies, we first outline the expected behavior of the TPCF. Past studies suggest that the shape of the TPCF of star-forming regions sensitively depends on the youthfulness of the SFCs \citep{Zhang_2001, Sanchez_2005, 2017ApJ...840..113G, 2021MNRAS.507.5542M, shashank2025tracing}. If the SFCs can be broadly classified into young and old based on an age cut, say 20 Myrs (where young: $<$20 Myr SFCs and old: $>$20 Myr SFCs), then the observed TPCF of different SFC sets follow well-behaved mathematical models. These models correspond to characteristic shapes in the TPCF plots. Each shape corresponds to a unique physical interpretation about the SFC distribution \citep{2021MNRAS.507.5542M}. 

For instance, the young SFCs may follow a single power law model (Model S) with a characteristic steep, negative slope (absolute slope $>$ 0.2), across all spatial scales considered. It represents a purely hierarchical distribution presumed to be governed by turbulence-driven density structures and dynamics in the ISM. We note that star-forming regions tend to exhibit more clustering on small scales and less clustering on large scales. Therefore, the observed TPCF slopes are usually negative. The young SFCs can also follow a broken, piece-wise power-law model (Model PW) with steep, negative power-law behavior on small scales which transitions sharply to a flatter, near-zero slope on larger scales, approaching a TPCF value of 1. Model PW represents a distribution that is consistent with a turbulence-driven hierarchy on small scales but transitions to a nearly random distribution on larger scales, where the SFC distribution is governed by the large-scale structure of the galaxy. The break in the power-law represents the transition scale from a hierarchical distribution to a random distribution. This break scale was termed correlation length ($l_{\rm corr}$) in the recent literature and it signifies the largest scale up to which stellar structures can be spatially correlated with each other \citep{2017ApJ...840..113G, 2021MNRAS.507.5542M, shashank2025tracing, lapeer2026feast}. We note that in TPCF based studies, the slope of the TPCF as derived from Model S or Model PW fit can be used to constrain the projected fractal dimension ($D_2$) of the SFC distribution. $D_2$ (= 1 + TPCF slope) quantifies the fractal-like complexity and space-filling nature of the distribution in a two-dimensional plane. Hierarchical distributions often have $D_2$$<$2, whereas Poissonian distributions has $D_2$$\sim$2 \citep{Grasha_2018, 2020MNRAS.493.4643M}. 

The old SFCs predominantly exhibit a nearly-flat TPCF model, characterized by a small negative slope (absolute slope $\lesssim$0.2) on small scales, transitioning into an exponential decline at the larger scales (Model PF). Model PF best describes any distribution in which the number of SFCs falls exponentially as a function of scale, similar to the exponential disk of a galaxy. Unlike the Model S and Model PW, the Model PF describes a distribution for which the hierarchical structuring has been almost entirely dispersed towards randomness. Still, owing to the exponential disk-like behavior, some clustering of SFCs remains, which leads to an absolute TPCF value slightly greater than 1. Even in the dwarf irregular galaxies, old SFC sets may obey the PF model TPCF despite lacking an underlying disc structure. This can simply be caused by the number of SFCs declining exponentially at large separations, which can mimic an exponential disc in the TPCF analysis. Finally, in some rare cases, young as well as old SFCs can follow a hybrid model (model PWF) which combines the PW model with an exponential decline behavior of model PF. This model is best suited to describe a distribution of SFCs that exhibits signatures of a hierarchical distribution on small scales (characterized by a steep negative power-law slope till the break scale), which is ultimately part of an exponential disk (as evidenced by the Model PF like exponential decline on larger scales). Notably, in Paper I, when all the SFCs in NGC 5194 and NGC 5457 were considered for the TPCF calculation, the PWF model was found to provide the best-fit. 

In Table \ref{table2}, we present the important details of our four TPCF models. As we shall see in the upcoming sections, our four models can efficiently describe the observed TPCF plots for the SFCs in our galaxies. The choice of which of the above four models best fits the observed TPCF is made based on the visual inspection of the TPCF plots. Additionally, the best fitting parameters corresponding to the fitted model are derived using a standard chi-square minimization scheme. In the rare cases of confusion between any two models, the model with the smaller reduced chi square value is chosen. In the case of young SFC TPCF, the fitting is performed with an aim to detect the TPCF break corresponding to the correlation length of the galaxy. In Appendix \ref{binning_effect}, we demonstrate that the measured correlation values are robust against variation in the number of bins and the fitting range of TPCF values. In Section \ref{subsec:best_fit}, we will describe how our TPCF method and the mathematical models enable us to constrain the hierarchy parameters of our sample galaxies.

\section{Galaxy-wise TPCF measurement methodology}
\label{sec:obs_TPCF}

The hierarchical distribution of star forming regions is understood to persist for a few tens of Myrs after which the effect of feedback, turbulent motions in the ISM and internal galactic dynamics slowly start to disperse this distribution towards randomness \citep{Elmegreen_2006, Sun_2017, Grasha_2018, Grasha_2019, 2021MNRAS.507.5542M, 2022MNRAS.512.1196M}. Therefore, young SFCs are ideal probes to characterize the hierarchical distribution of stellar matter in a galaxy. Young SFCs best resemble the hierarchical properties of the ISM close to the time when they were formed. On the other hand, old SFCs can inform us about the overall structure of the galaxy, e.g. a spiral galaxy's exponential disk, or the structure of the dwarf irregular galaxies. Moreover, old SFCs can trace the timescales over which the stellar hierarchies disperse. In Paper I, we had used the old SFC to estimate the hierarchy dispersal timescale ($T_{\rm dis}$) of a galaxy by setting the reasonable condition that the slope of the TPCF for SFCs with ages greater than $T_{\rm dis}$ must be shallower than an absolute value of 0.2. Since our SFC ages range from 1 to $\sim$400 Myrs, dividing our SFCs into different age sets can enable us to investigate the initial conditions, evolution and dispersal of the stellar hierarchies. Additionally, after compiling the hierarchy parameters of our galaxies using these different age sets, we can compare the hierarchical distribution of SFCs across our diverse galaxy sample. 

Our TPCF analysis method follows a two-step approach, where in each step we compute the TPCF for \enquote{young} and \enquote{old} SFCs. The \enquote{young} SFCs are denoted as either Y or Y1 or Y2 in different cases and \enquote{old} SFCs are denoted as either O or O1 or O2 in different cases. In step I, the age cut distinguishing between young and old SFCs is 20 Myr. As we shall see in the coming sections and Figures \ref{fig:tpcf_plots} and \ref{fig:tpcf_plots_contd}, the 20 Myr age cut is not sufficient to constrain the hierarchy parameters for all of our galaxies. Therefore, in step II, we had to raise the age cut to a higher value than 20 Myr. An illustrative overview of our complete TPCF measurement methodology, which we next describe in this section, is presented in Figure \ref{fig:schematic}.

\subsection{Step I : 20 Myr age cut} 
\label{subsec:step_I}
In \cite{2021MNRAS.507.5542M}, an age cut of 10 Myr was used characterize the TPCF of young star-forming regions. This timescale roughly corresponds to the time over which star-forming regions dissociate from their parent, hierarchically structured molecular clouds \citep{Chevance_2020, 2022MNRAS.516.3006K}. Using the $<$10 Myr star clusters enabled \cite{2021MNRAS.507.5542M} to constrain the correlation length ($l_{\rm corr}$) and the two-dimensional fractal dimension ($D_2$) of their galaxies. We too aimed to derive these hierarchy parameters for our galaxies using the 10 Myr age cut and the corresponding young ($<$10 Myr) SFC TPCF. However, in Paper I, we observed that using small numbers of SFCs to measure the TPCF leads to an extremely noisy TPCF shape which cannot be described by the mathematical models. Therefore, at least $\sim$250 SFCs need to be considered in order to robustly estimate the TPCF. For 16 out of 17 of our galaxies, we did not have $\sim$250 SFCs with ages less than 10 Myrs, so as to allow for a robust calculation of the young SFC TPCF. Therefore, in step I, we raised the age cut from 10 Myr to 20 Myr with an assumption that $<$20 Myr SFCs can probe the hierarchical distribution of star formation within our galaxies. We applied this homogeneous age cut of 20 Myr across our galaxy sample with an aim to obtain an unbiased view of the stellar hierarchies in galaxies. 

In paper I, we showed that increasing the age cut used to distinguish young and old SFCs from 10 Myr to 20 Myr does not significantly affect the derived hierarchy parameters (see the TPCF analysis for NGC 5457 in Figure 3 and 5 in Paper I). Moreover, \cite{2017ApJ...840..113G, Grasha_2019, 2021MNRAS.507.5542M} and \cite{lapeer2026feast} have demonstrated that the hierarchical distribution of star formation can be effectively probed with SFCs younger than 10 to 50 Myr, which makes our choice of 20 Myr age cut well-justified. These studies set the expectation that young, $<$20 Myr SFCs should exhibit a strongly hierarchical distribution, characterized by a well-fitted power-law form of the TPCF (model S or PW). The break scale in PW model-fitted TPCF of $<$20 Myr SFCs would correspond to the $l_{\rm corr}$ and the TPCF slope can be used to determine the $D_2$ of the galaxy. Additionally, $>$20 Myr SFCs are expected to have dispersed hierarchy, thereby showing flat TPCF with slopes less than an absolute value of 0.2, within the error-bars. Therefore, the $T_{\rm dis}$ can be considered as 20 Myr. We present the $<$20 Myr and $>$20 Myr TPCF of our galaxies in Figure \ref{fig:tpcf_plots} and \ref{fig:tpcf_plots_contd}, along with the best-fit model provided in the legend.

In step I, we observed three different cases among our galaxies. Case I was according to the aforementioned expectations and 7 galaxies belonged to it - Holmberg II, NGC 0253, NGC 0300, NGC 1512\footnote{In NGC 1512, we had a total of only 385 SFCs. Due to this shortage of SFCs, we only had 162 SFCs in the $<$20 Myr set $-$ well below the recommended minimum of 250. Nevertheless, the resulting TPCF is reasonably well-fit with the PW model.}, NGC 1566, NGC 2403, and NGC 3486. For these galaxies, $l_{\rm corr}$ and $D_2$ can be measured with the break scale and the slope of $<$20 Myr TPCF, respectively. The shallower than 0.2 slope of the $>$20 Myr SFC TPCF would suggest a $T_{\rm dis}$ value of $\sim$20 Myr. 

In case II, $<$20 Myr SFCs exhibited well-behaved, well-fitting TPCF and the $l_{\rm corr}$ and $D_2$ for these galaxies could be measured, similar to case I. However, the $>$20 Myr TPCF slope was observed to be greater than 0.2, indicating that the SFC distribution still exhibits significant hierarchical signatures. This meant that for these galaxies, the hierarchy is not dispersed by 20 Myr. $T_{\rm dis}$ for such galaxies is expected to be greater than 20 Myrs, and we will constrain it in step II. NGC 0628, NGC 4236, NGC 5033, NGC 5194 and NGC 5457 belonged to this case. 

Finally, in case III, the  $<$20 Myr age set had an insufficient ($<$250) number of SFCs, which led to an extremely noisy TPCF. We were unable to reliably fit these observed TPCF plots with our mathematical models. This meant that their $l_{\rm corr}$ and $D_2$ could not be constrained with the $<$20 Myr age set. For these galaxies, upper limits of $l_{\rm corr}$ and $D_2$ can still be derived in step II, with a sufficient ($\sim$250) number of SFCs if the age cut is increased beyond 20 Myr. The $>$20 Myr TPCF in this case exhibits shallower slopes than a value of 0.2, indicating dispersed hierarchies and that the $T_{\rm dis}$ for these galaxies is 20 Myr. NGC 2903, NGC 4228, NGC 4395, NGC 7793 and WLM belonged to this case. 

To summarize step I, the $l_{\rm corr}$, $D_2$ and $T_{\rm dis}$ (=20 Myr) were estimated for 7 galaxies (case I). For 5 more galaxies, $l_{\rm corr}$ and $D_2$ was measured, but their $T_{\rm dis}$ could not be constrained (case II). Finally, for the remaining 5 galaxies, $l_{\rm corr}$ and $D_2$ could not be constrained, but $T_{\rm dis}$ was estimated to be 20 Myr (case III). To measure the $T_{\rm dis}$ for the 5 case II galaxies and $l_{\rm corr}$ and $D_2$ for case III galaxies, we move on to step II. 

\begin{figure*}[b]
      \centering
		\includegraphics[width=0.31\linewidth]{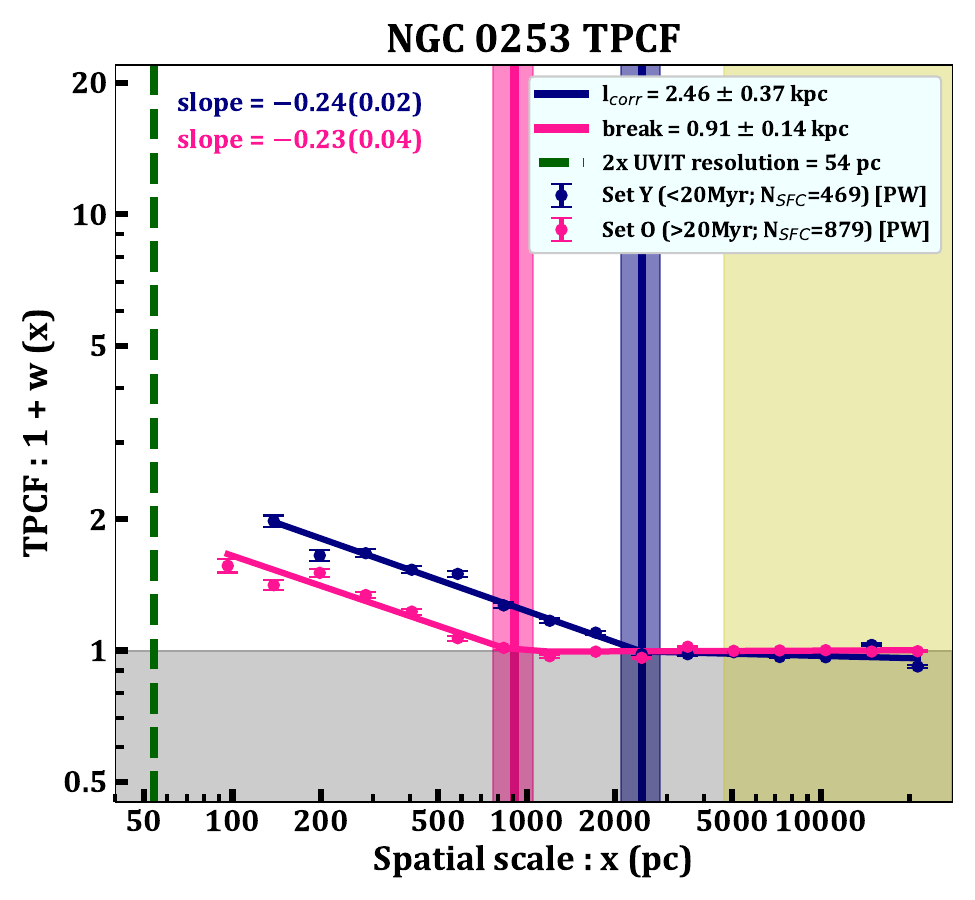}
        \hfill
		\includegraphics[width=0.31\linewidth]{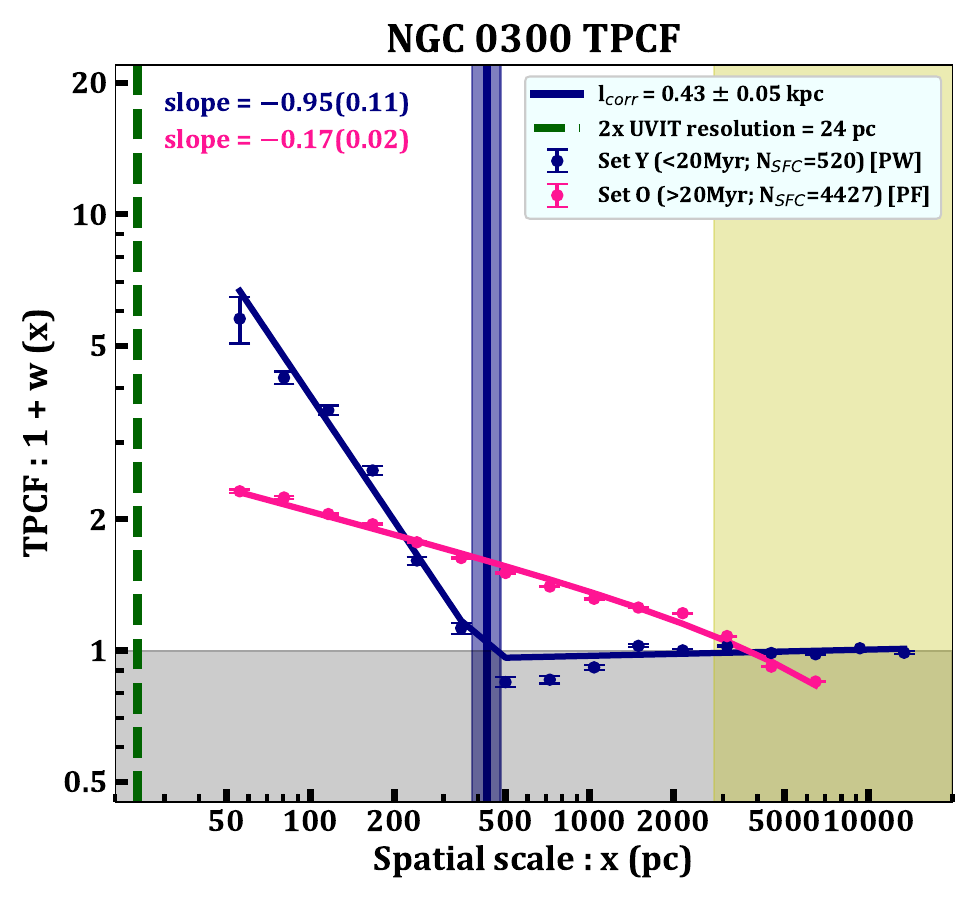}
        \hfill
		\includegraphics[width=0.31\linewidth]{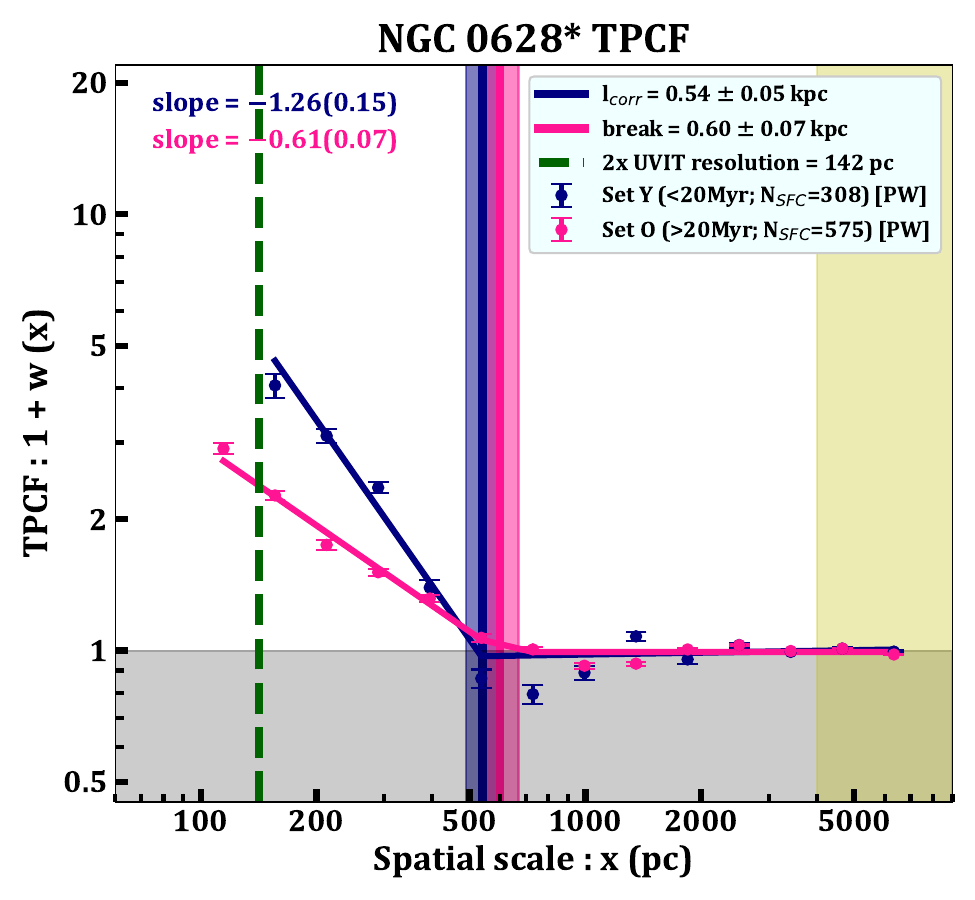}
    	\vfill
		\includegraphics[width=0.31\linewidth]{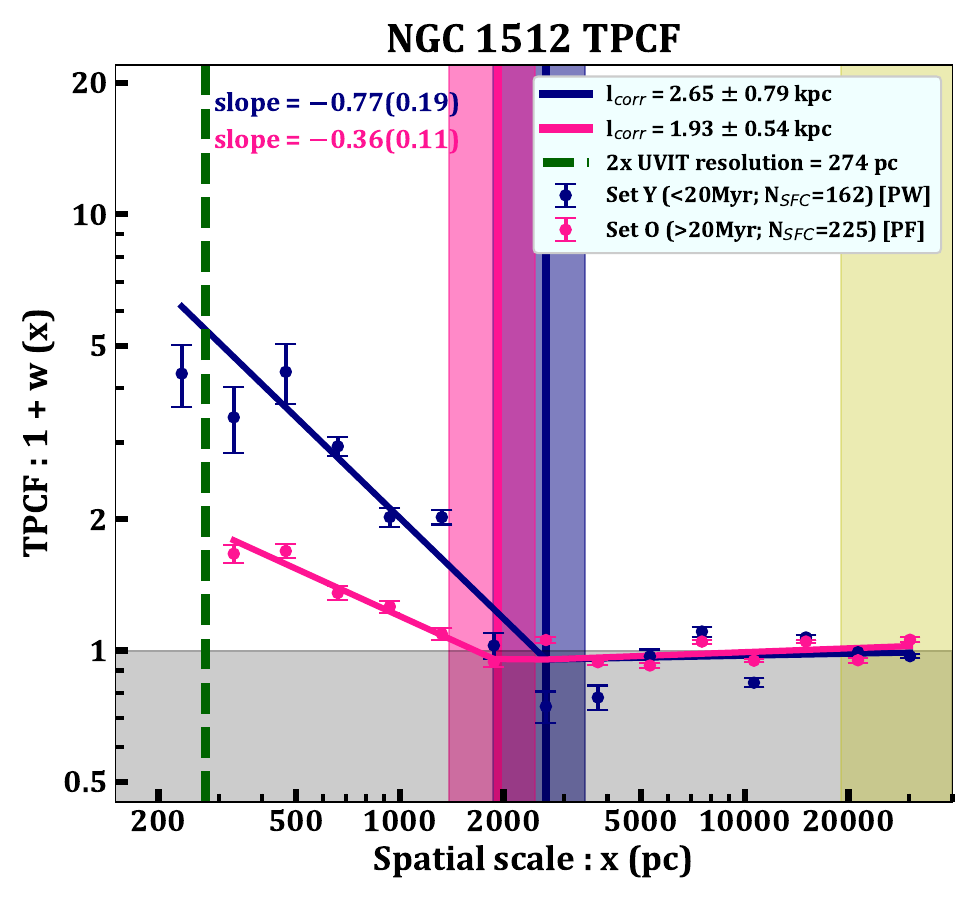}
	    \hfill
		\includegraphics[width=0.31\linewidth]{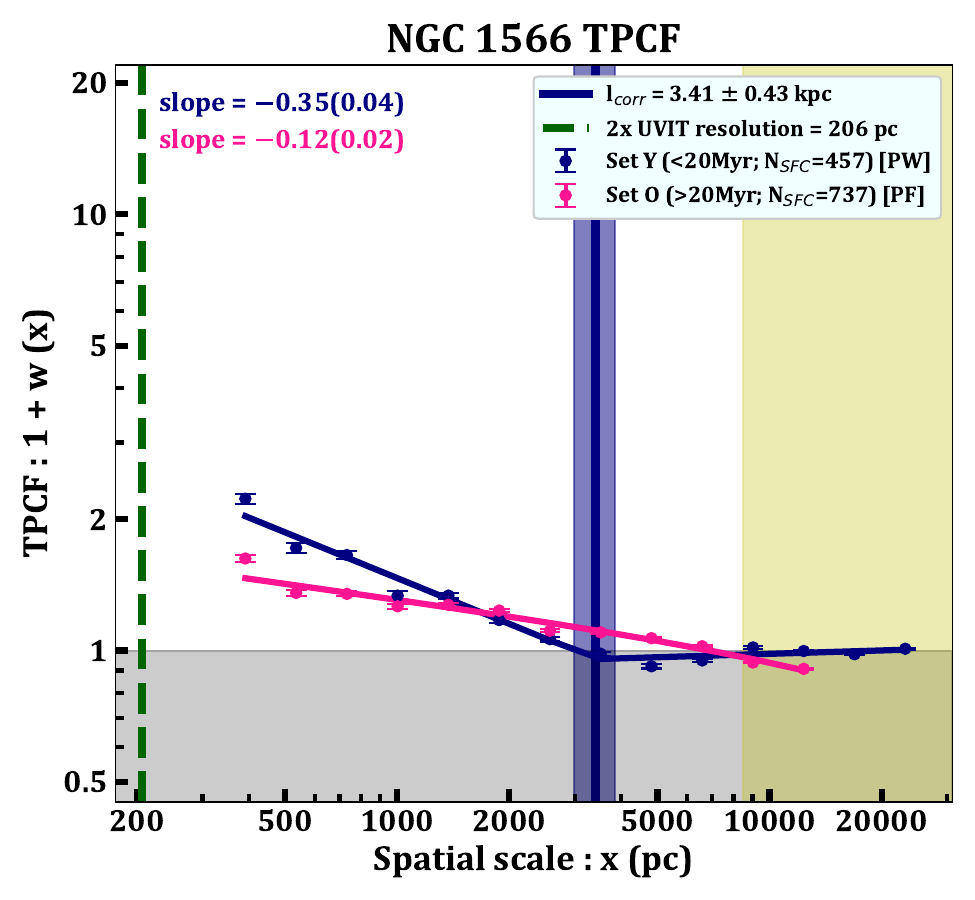}
        \hfill
		\includegraphics[width=0.31\linewidth]{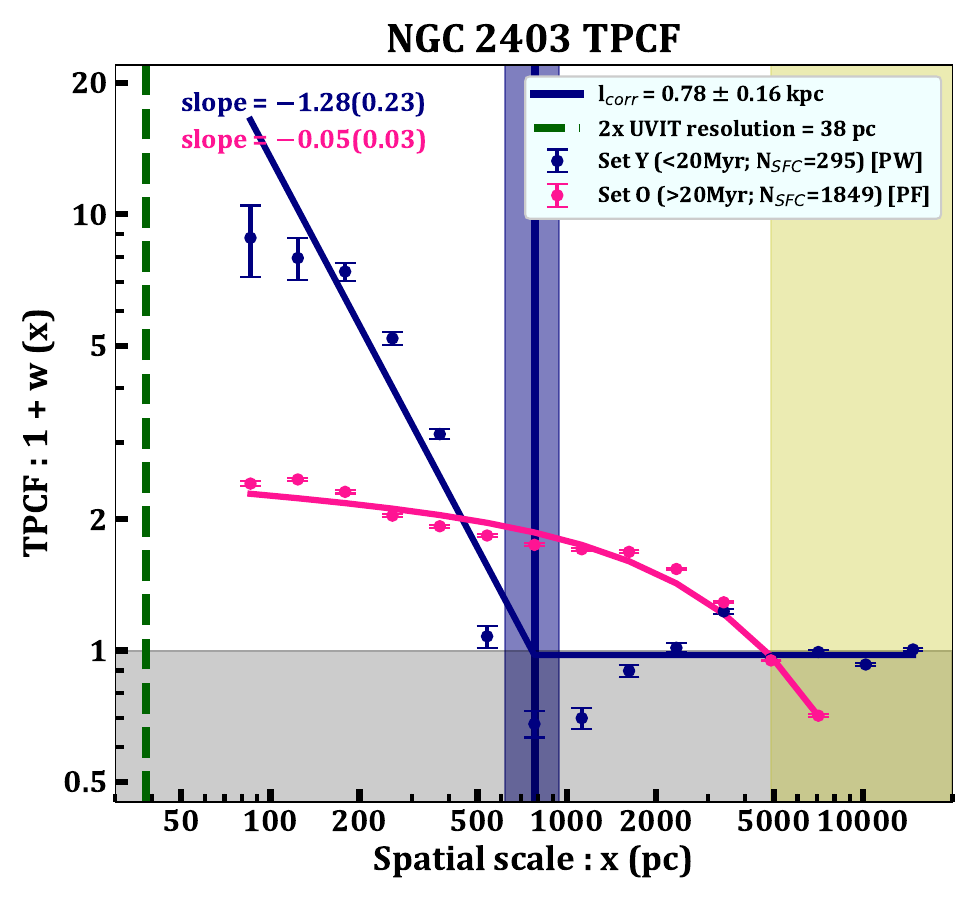}
        \vfill
		\includegraphics[width=0.31\linewidth]{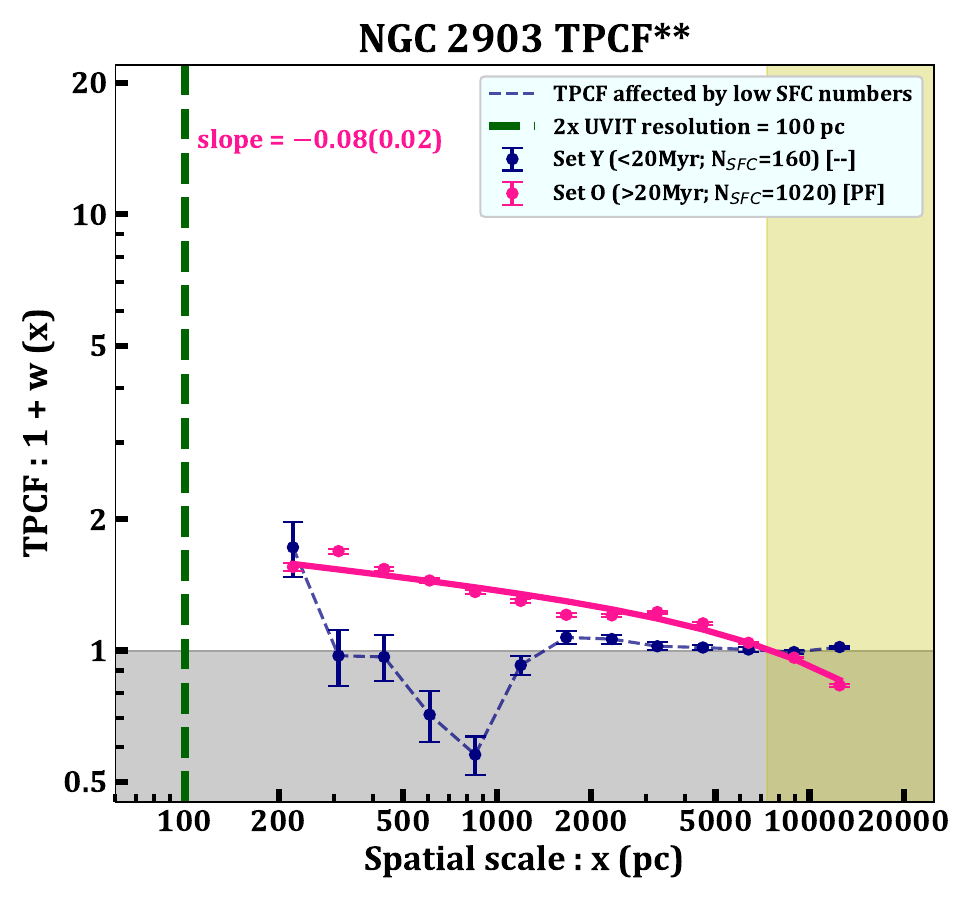}
    	\hfill
		\includegraphics[width=0.31\linewidth]{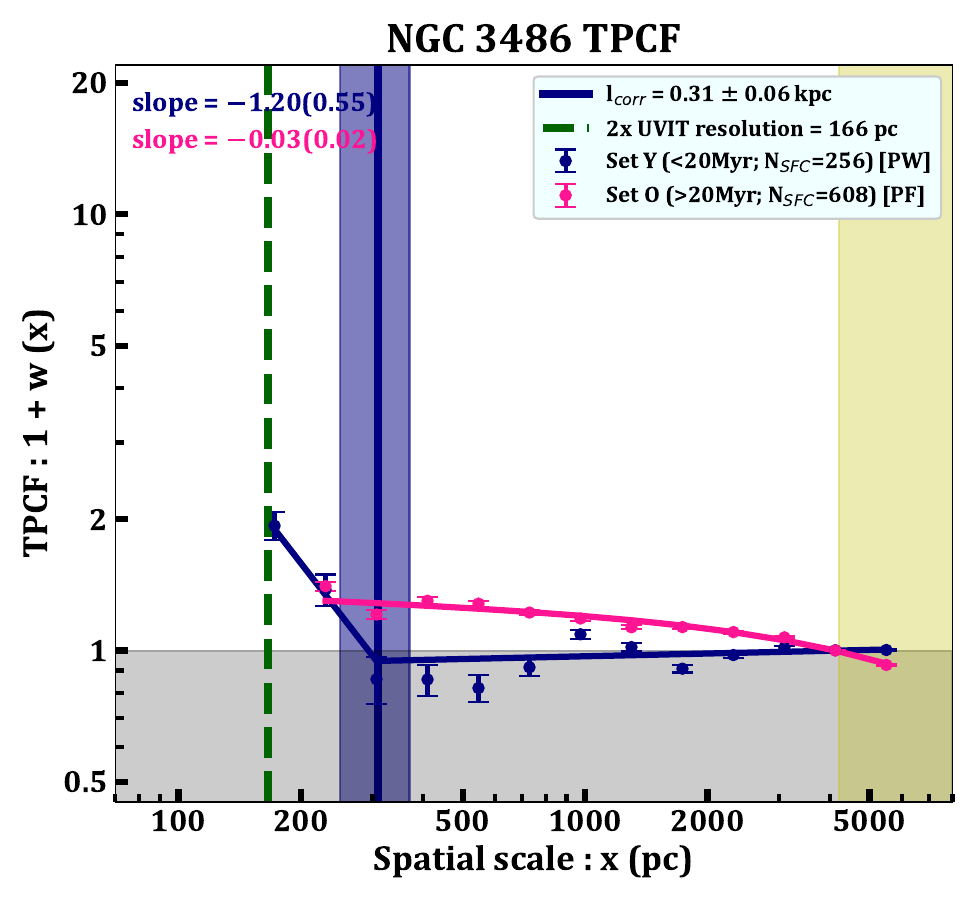}
        \hfill
		\includegraphics[width=0.31\linewidth]{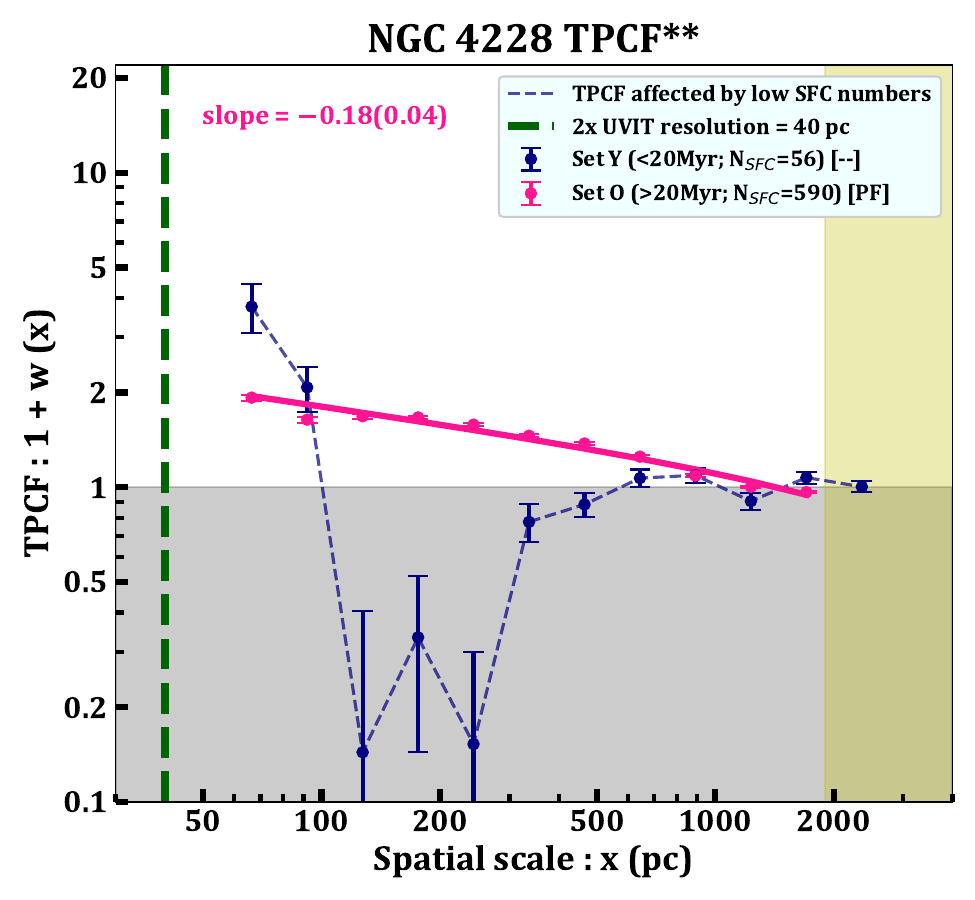}
        \vfill
        \includegraphics[width=0.31\linewidth]{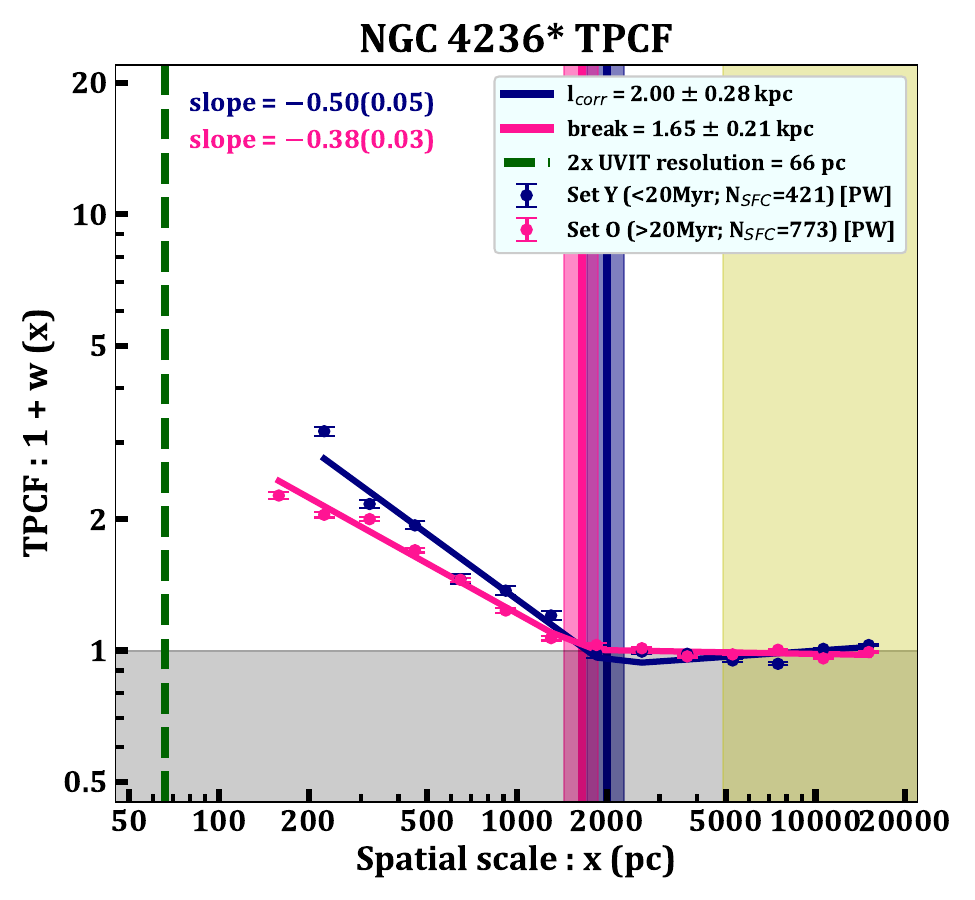}
        \hfill
        \includegraphics[width=0.31\linewidth]{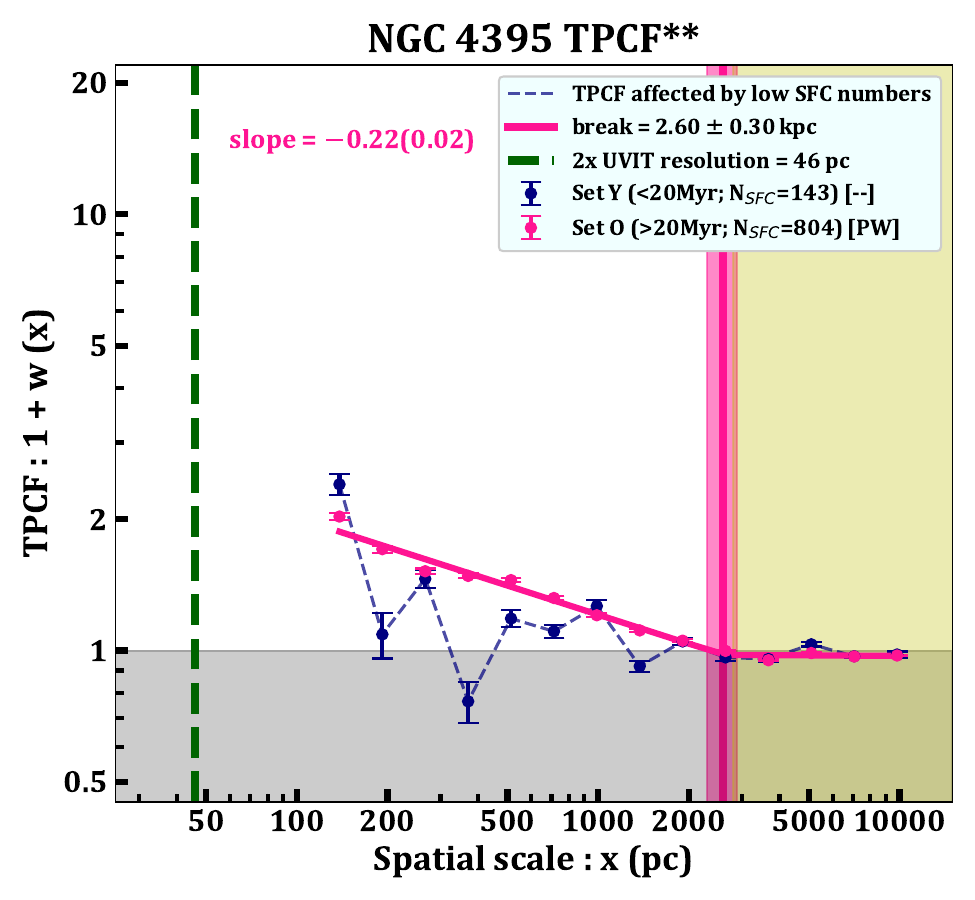}
        \hfill
		\includegraphics[width=0.31\linewidth]{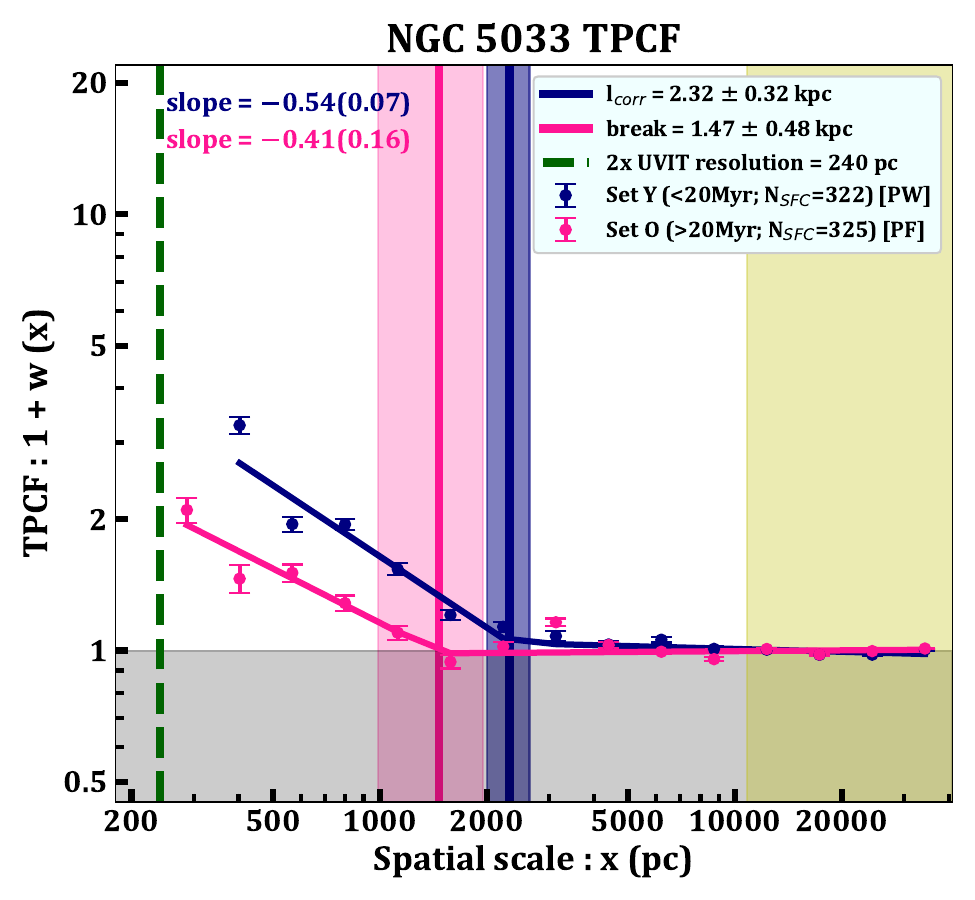}
    \caption{Step I TPCF for 12 out of 17 galaxies: Observed TPCF, corresponding best model fits and slopes for $<$20 Myr (set Y) and $>$20 Myr (set O) are given in navy and pink, respectively. Number of SFCs used in the TPCF measurement and the best fit models are mentioned in the legend. The green dotted line marks two times the UVIT's spatial resolution at the galaxy's distance. The vertical navy line and band marks the correlation length measured using $<$20 Myr TPCF. The pink vertical line and band indicates the break scale for $>$20 Myr TPCF. * or ** in the figure title indicate case II or III galaxies, respectively. The yellow vertical band marks the edge-effect limit where TPCF values may be unreliable due to the finite extent of the SFC distribution. This limit is 1/5 times the largest separation between any two SFCs \citep{2021MNRAS.507.5542M}. The black horizontal shaded region marks the TPCF expected from a clustering-free, Poisson-like distribution of SFCs.}
  \label{fig:tpcf_plots}
\end{figure*}

\begin{figure*}[t]
      \centering
		\includegraphics[width=0.31\linewidth]{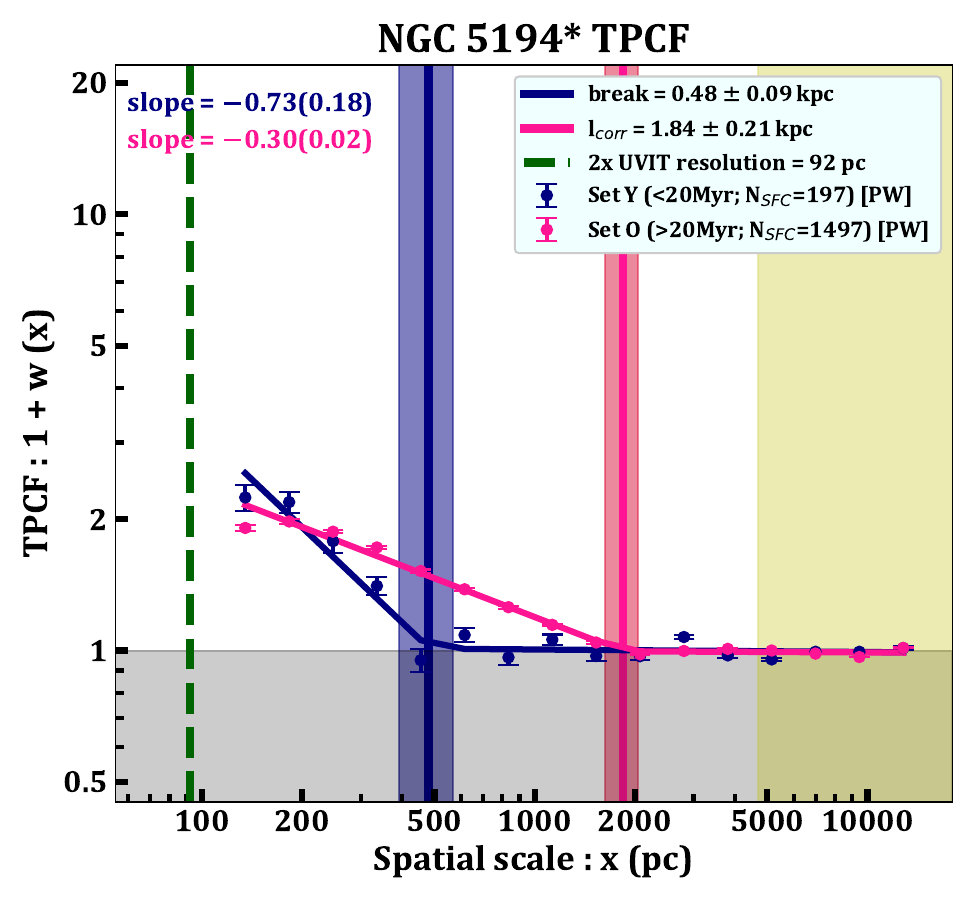}
    	\hfill
		\includegraphics[width=0.31\linewidth]{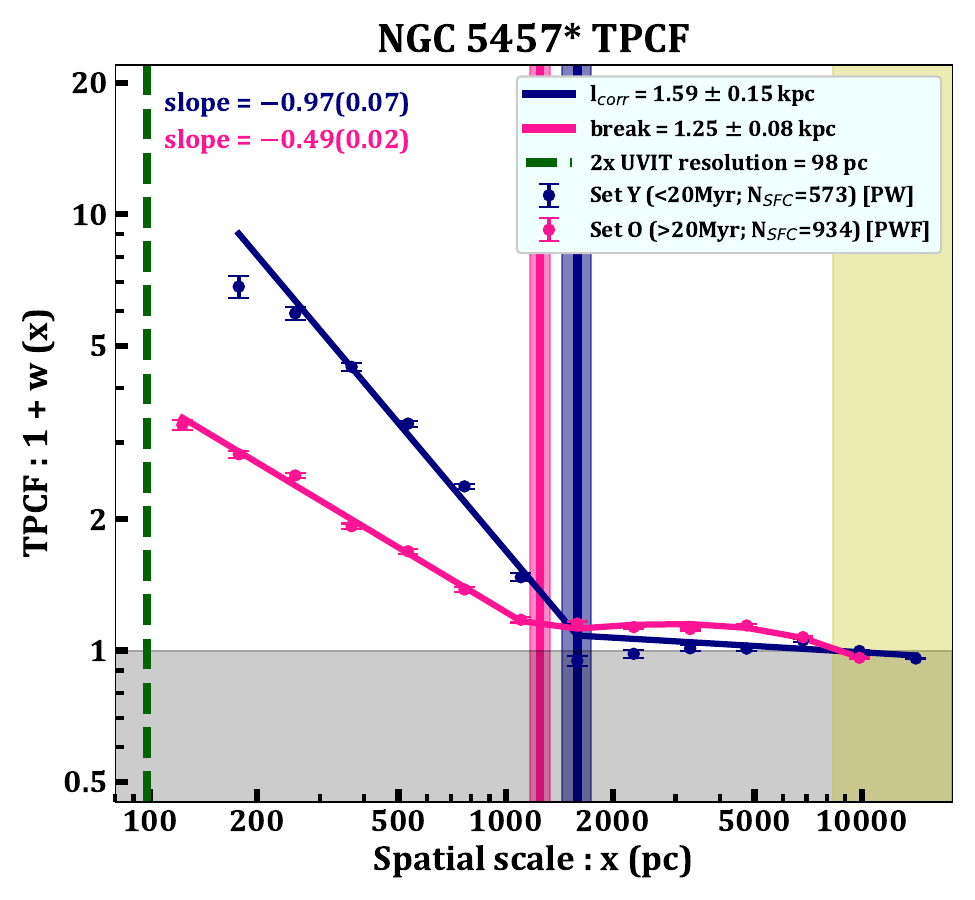}
        \hfill
		\includegraphics[width=0.31\linewidth]{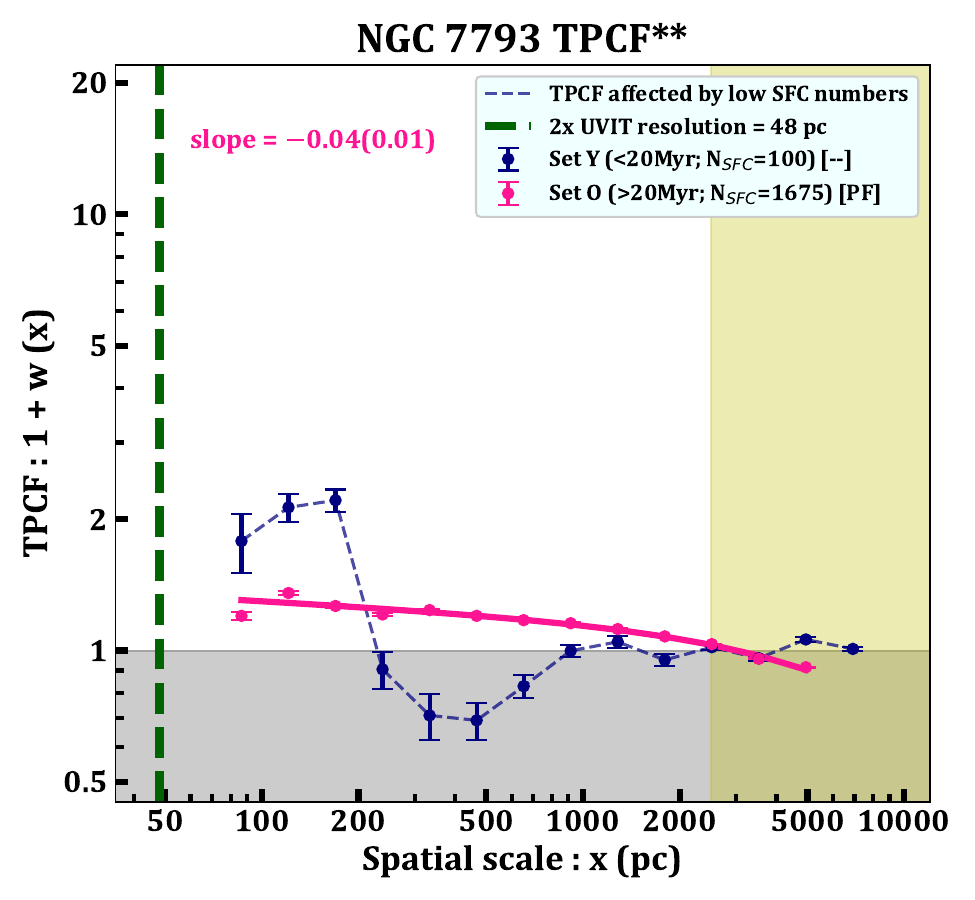}
        \vfill
		\includegraphics[width=0.31\linewidth]{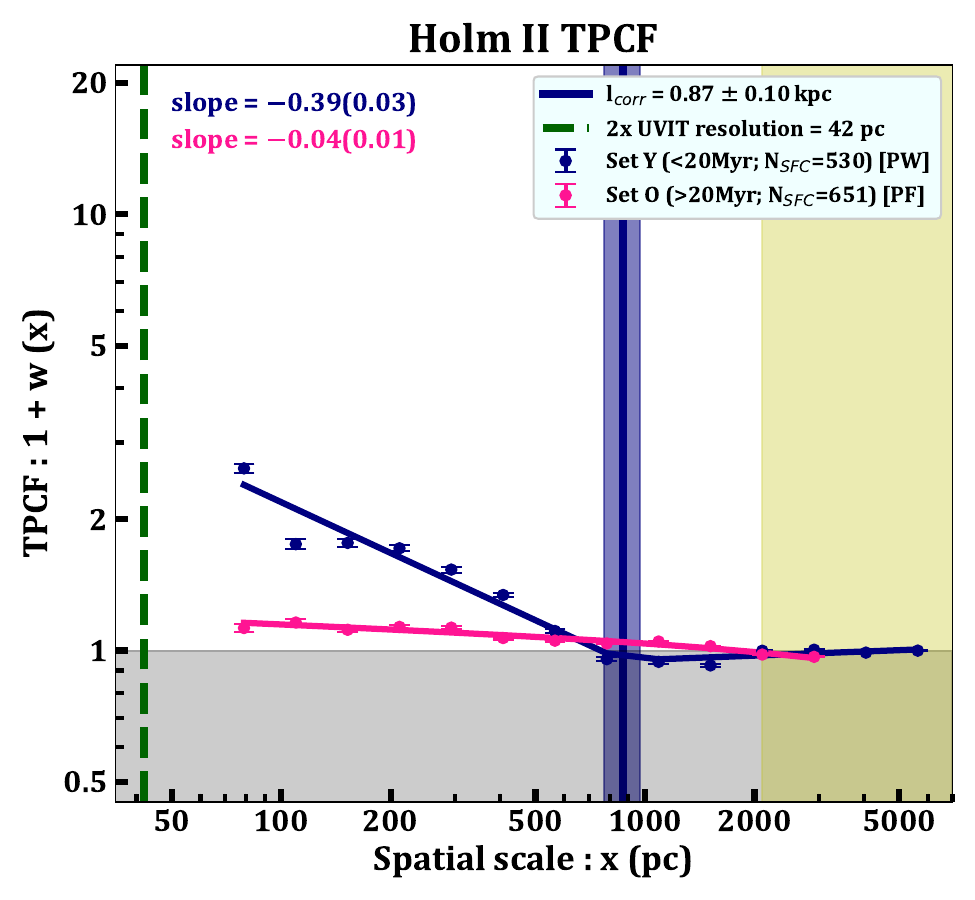}
    	\hfill
		\includegraphics[width=0.31\linewidth]{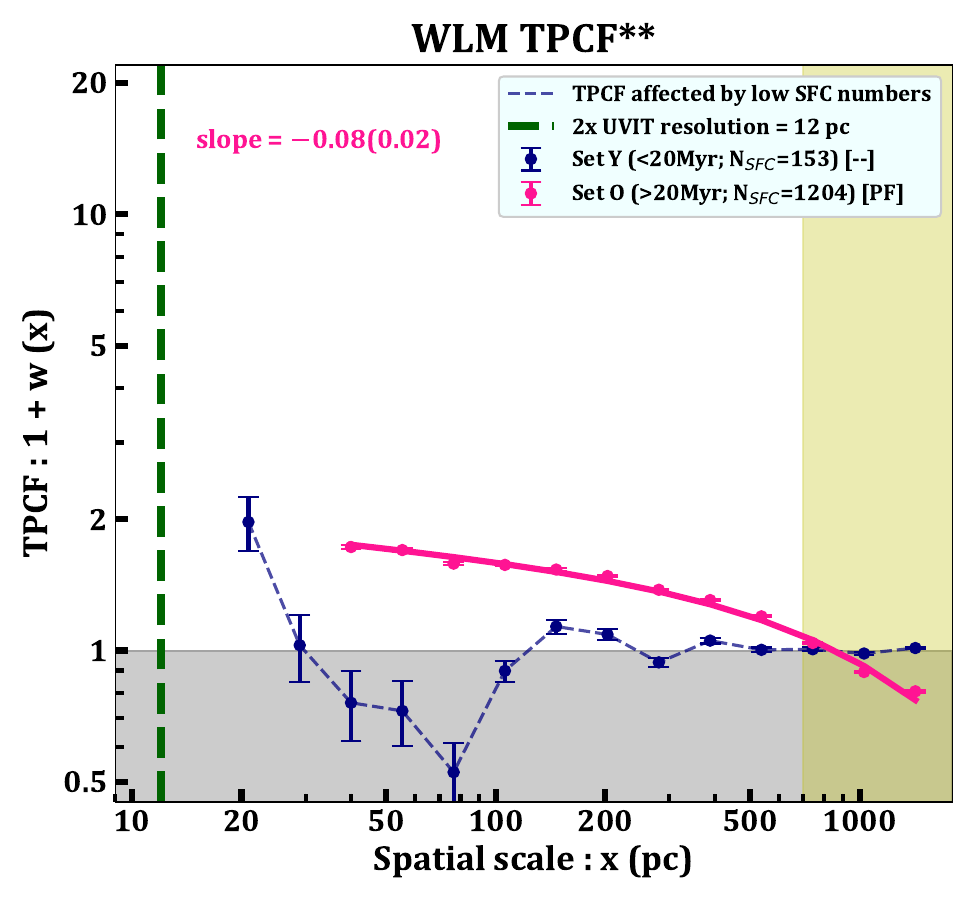}
    \caption{Continued from Figure \ref{fig:tpcf_plots} - TPCF plots for the remaining 5 out of 17 galaxies from step I.}
  \label{fig:tpcf_plots_contd}
\end{figure*}

\subsection{Step II : A1 or A2 Myr age cut} 
\label{subsec:step_II}

\begin{figure*}[t]
      \centering
		\includegraphics[width=0.31\linewidth]{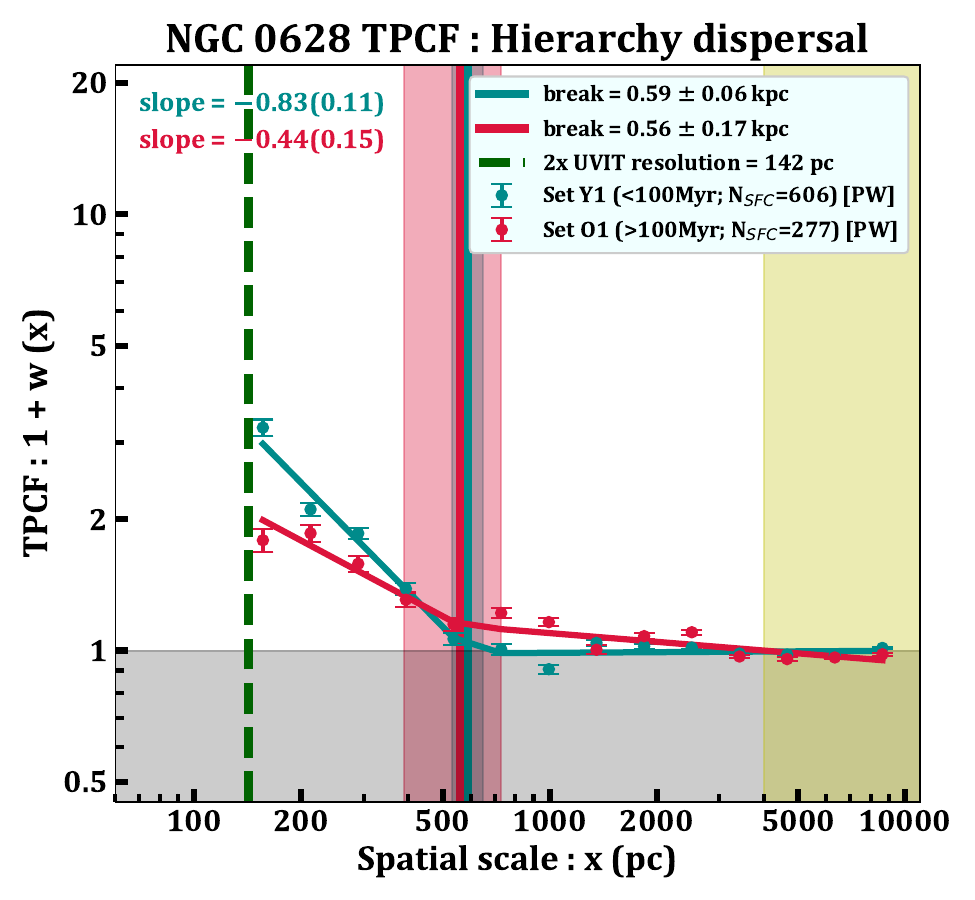}
	    \hfill
		\includegraphics[width=0.31\linewidth]{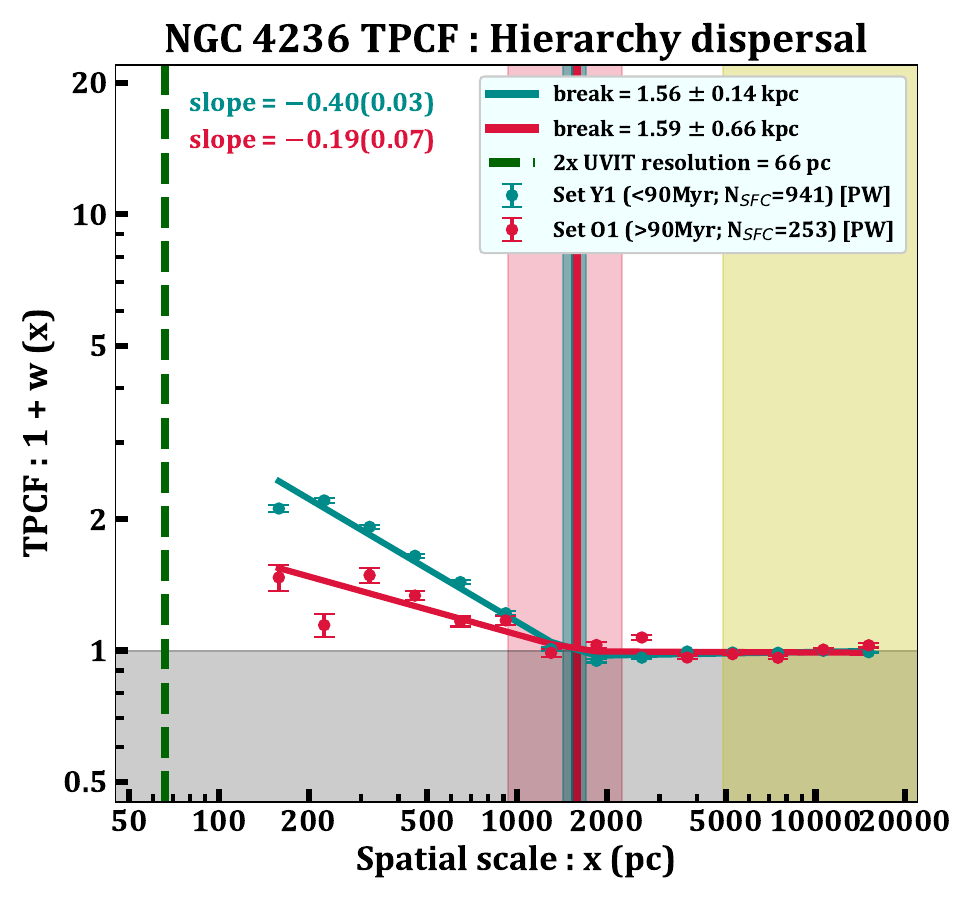}
        \hfill
		\includegraphics[width=0.31\linewidth]{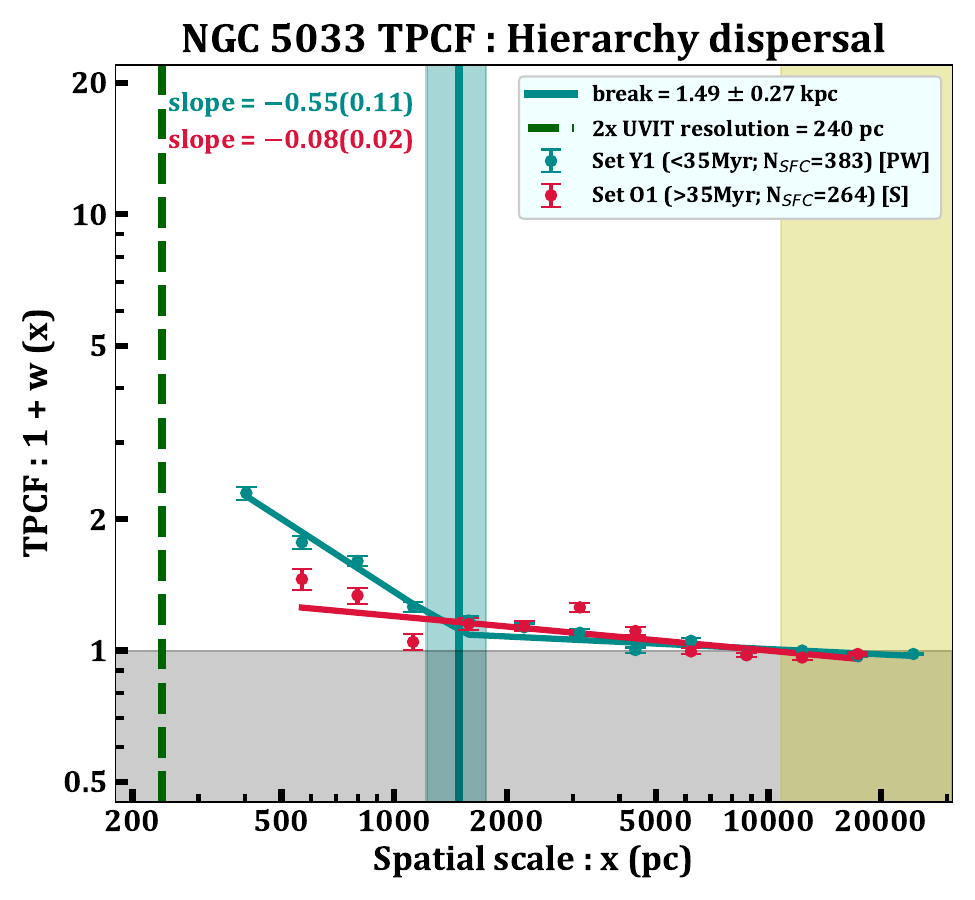}
        \hfill
		\includegraphics[width=0.31\linewidth]{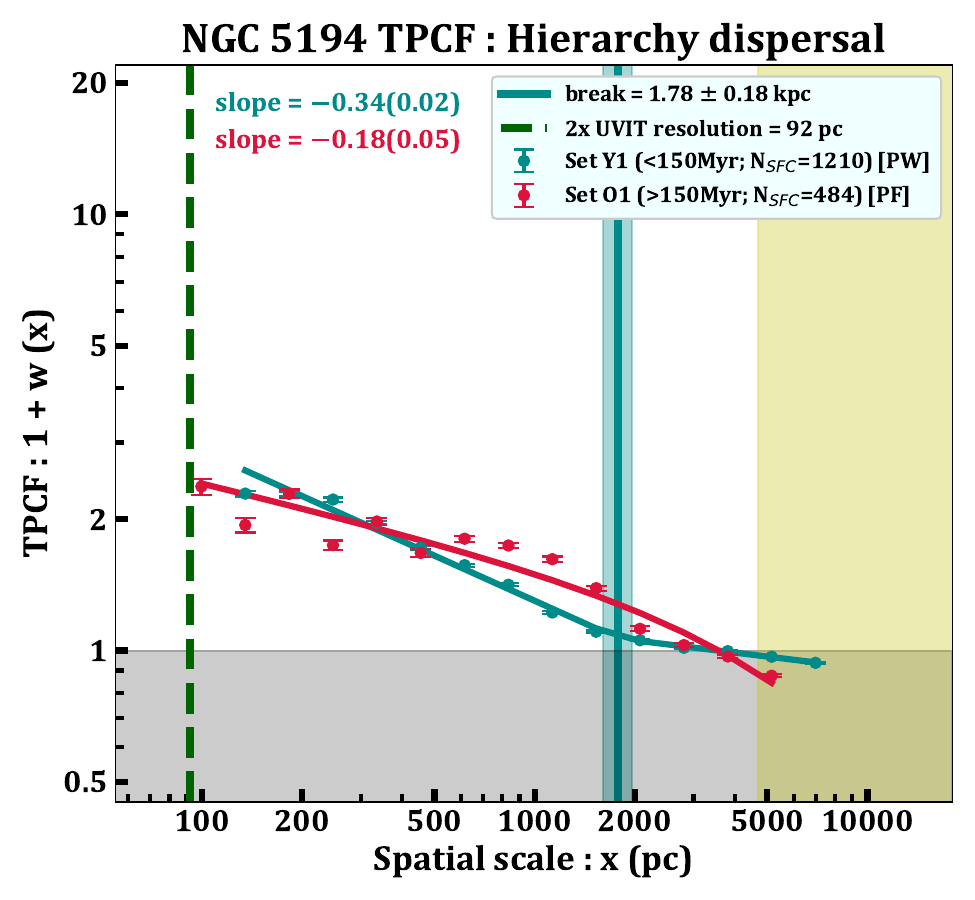}
        \hfill
		\includegraphics[width=0.31\linewidth]{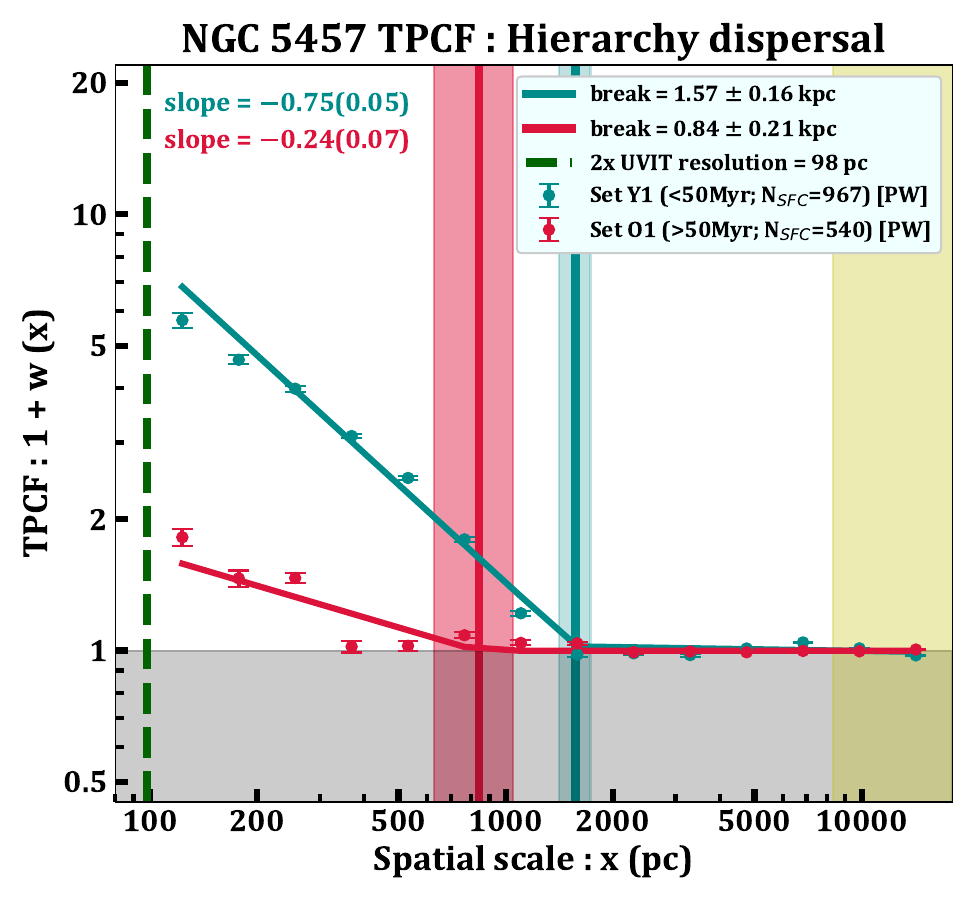}   
        \vfill
        \includegraphics[width=0.31\linewidth]{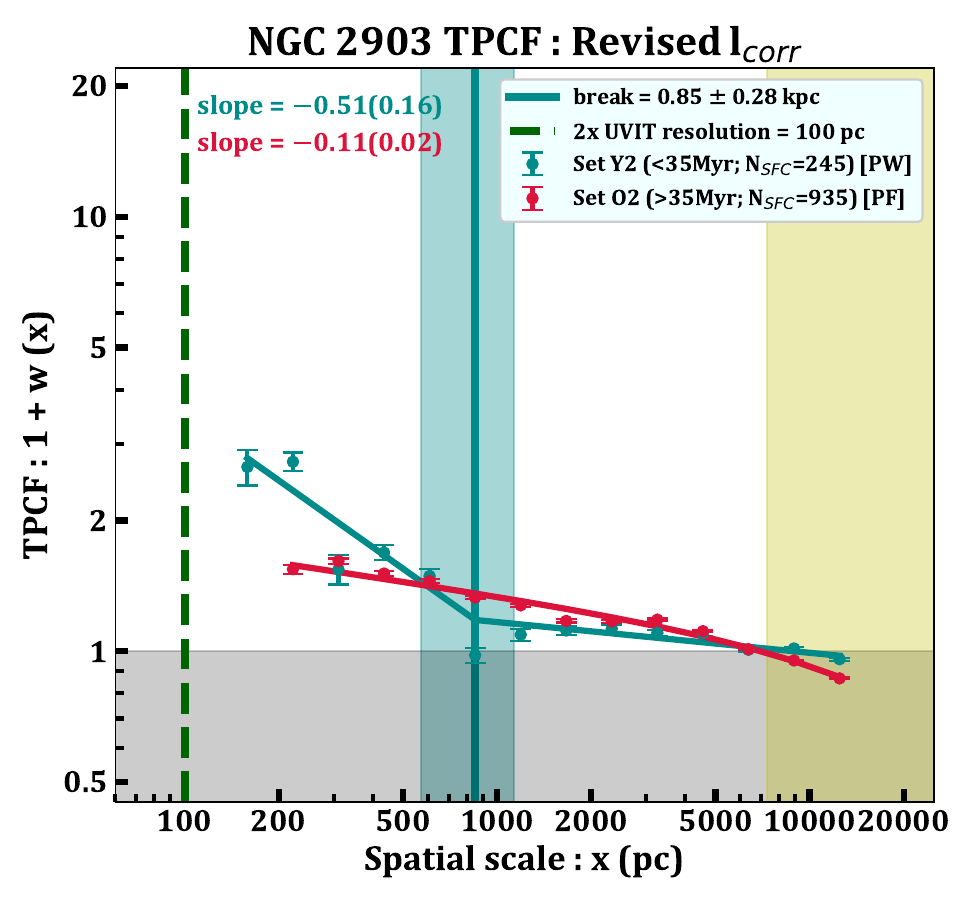}
	    \hfill
		\includegraphics[width=0.31\linewidth]{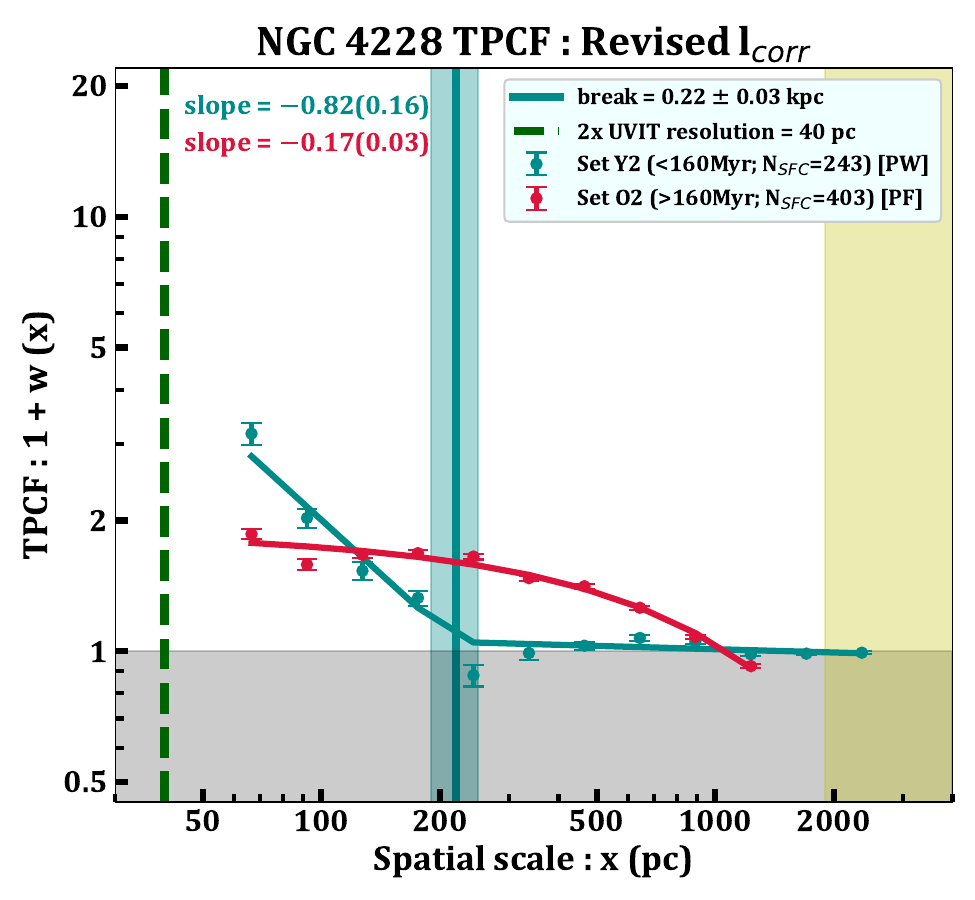}
        \hfill
		\includegraphics[width=0.31\linewidth]{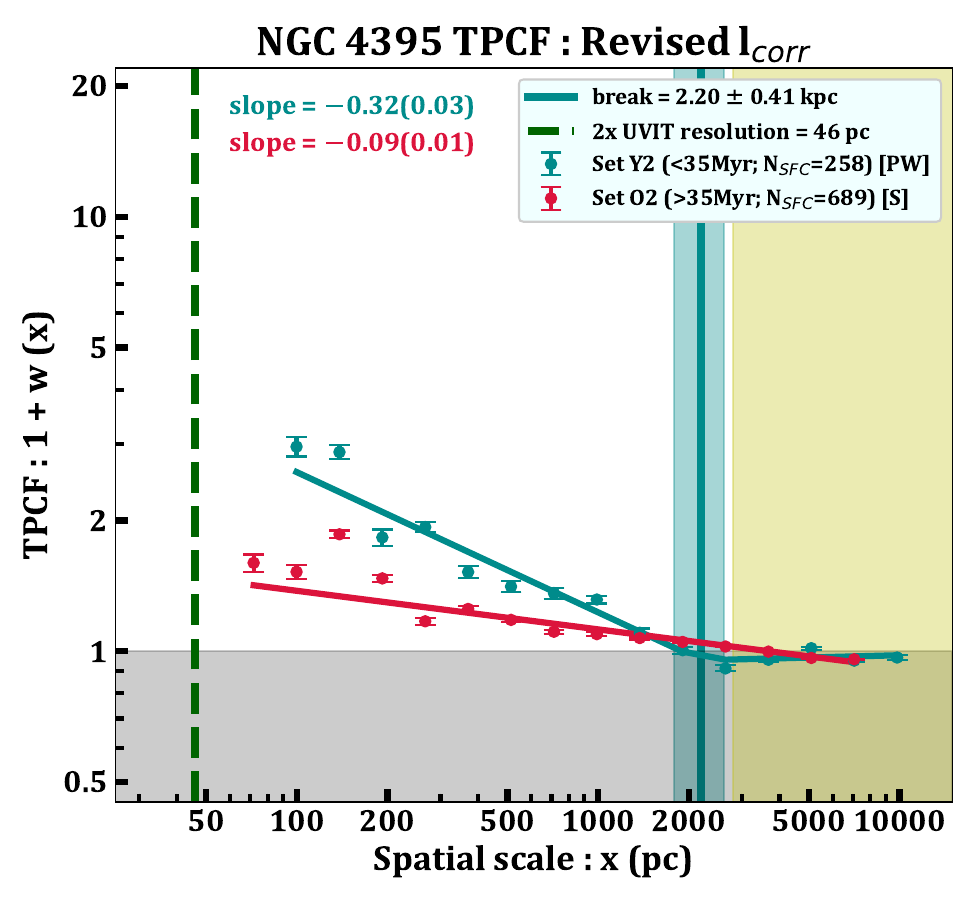}
        \hfill
		\includegraphics[width=0.31\linewidth]{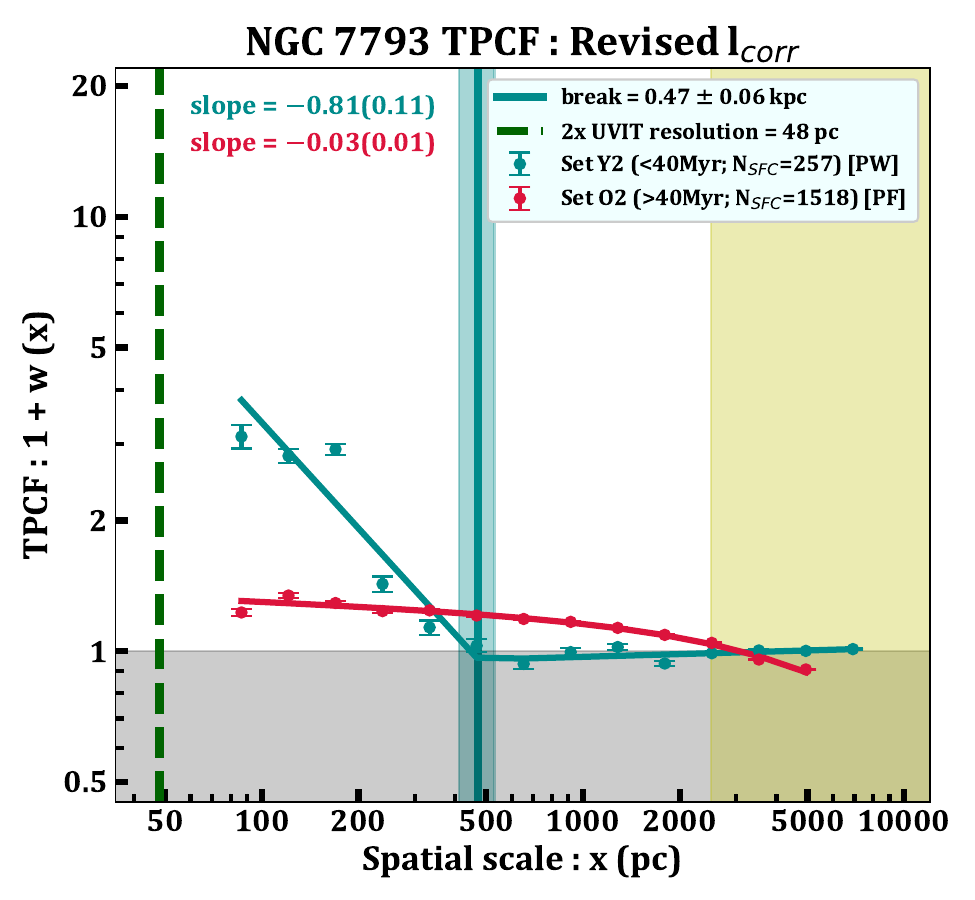}
        \hfill
		\includegraphics[width=0.31\linewidth]{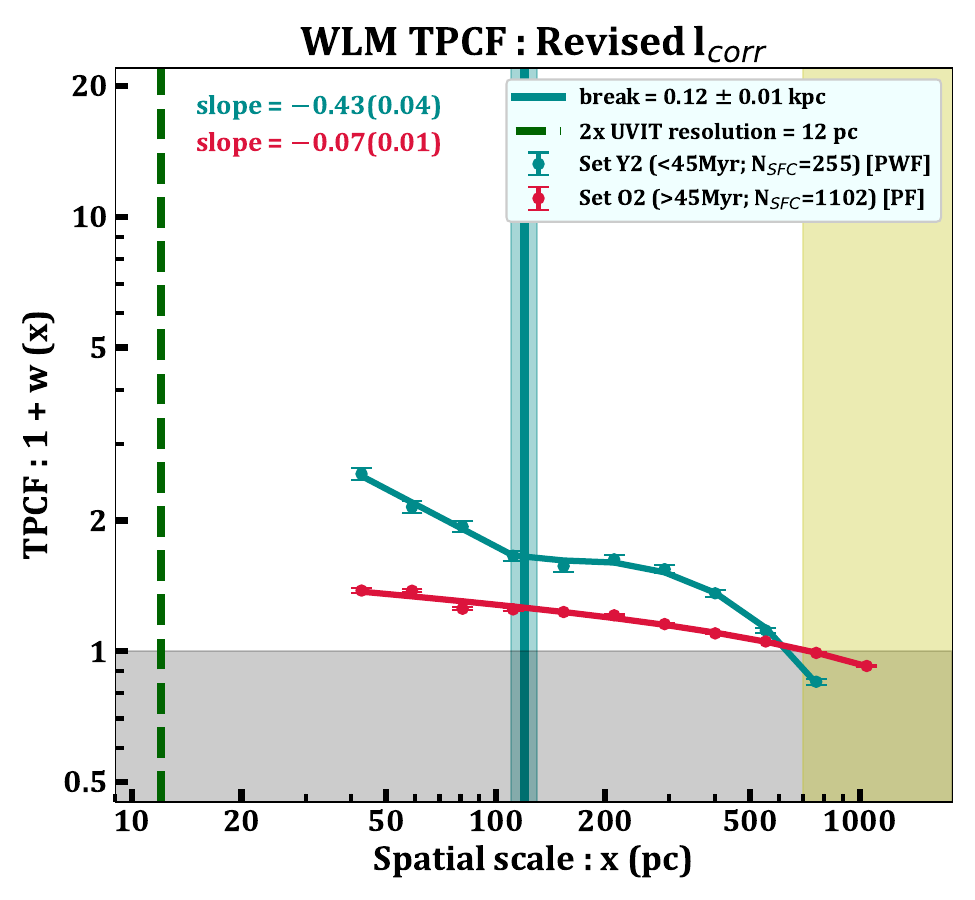}
    \caption{Top two rows: Observed TPCF and best-fit lines from Step II for the 5 case II galaxies, for which $T_{\rm dis}$ is being constrained in this step. Bottom two rows: Observed TPCF and best-fit lines from Step II for the 5 case III galaxies for which $l_{\rm corr}$ and $D_2$ is being constrained in this step. $<$A1 (or $<$A2) Myr TPCF, best-fit models and slopes in the top two (or bottom two) rows are shown in teal whereas, $>$A1 (or $>$A2) Myr TPCF, best-fit models and slopes in the top two (or bottom two) rows are shown in red. In the legend, Y1 and O1 indicate SFCs younger and older than A1 Myr, respectively, whereas Y2 and O2 indicate SFCs younger and older than A2 Myr, respectively. The description for the yellow and grey shaded areas as well as the green dashed line is the same as Figure \ref{fig:tpcf_plots}. The number of SFCs used in the TPCF measurement and the best-fit models are mentioned in the legend. The teal and red vertical line and band marks the TPCF brake scale, which in the case III galaxies (bottom two rows) represents the upper limit of correlation length values.}
  \label{fig:tpcf_revised}
\end{figure*}

For the 5 case II galaxies for which $T_{\rm dis}$ could not be constrained with a $>$20 Myr TPCF, we followed the following approach. We progressively increased the age cut from 20 Myr in steps of 5 Myrs and tested when the best-fit TPCF slope becomes shallower than 0.2. The corresponding age cut (termed A1 Myr) would indicate the $T_{\rm dis}$ of the galaxy. We present the TPCF plots with A1 Myr age cut being used in Figure \ref{fig:tpcf_revised} (top two rows). For the sake of completeness, we present both the $>$A1 Myr and $<$A1 Myr TPCF plots. The $>$A1 Myr plots exhibit slopes shallower than 0.2, indicating dispersed hierarchies. The $<$A1 Myr plots are similar to $<$20 Myr plots, characterized by a steep, negative power law TPCF behavior (model PW). A1 ranges from 35 Myr to 160 Myr for our 5 galaxies. In the case of NGC 0628, $>$A1 Myr TPCF measured with 277 SFCs exhibits a slope value of $\sim$0.44, which is steeper than 0.2, where A1 = 100 Myr. However, we do not have sufficient older SFCs to raise the age cut beyond 100 Myr. Therefore, the $T_{\rm dis}$ of 100 Myr for NGC 0628 may be considered as a lower limit.

For the 5 case III galaxies, the $l_{\rm corr}$ could not be constrained within step I, owing to the insufficient number of SFCs in the $<$20 Myr age set. So, we raised the age cut from 20 Myr to a high enough age value (termed A2 Myr), so that $\sim$250 SFCs are part of the $<$A2 Myr age set. Subsequently, the TPCF for $<$A2 Myr SFCs was computed, which we were able to fit with the PW model (see Figure \ref{fig:tpcf_revised} (bottom two rows)). Since the age cut has been increased from 20 Myr to A2 Myr and inclusion of older SFCs tends to raise the $l_{\rm corr}$ value \citep{shashank2025tracing, 2026ApJ..1002..220A}, the break scale in the $<$A2 Myr TPCF represents an upper limit of $l_{\rm corr}$ for the galaxy. Moreover, the $D_2$ measured with the $<$A2 Myr TPCF represents an upper limit, as the inclusion of older SFCs tends to flatten the TPCF slope, thereby raising the $D_2$ value.  For the sake of completeness, we present the $>$A2 Myr TPCF, which exhibits slopes shallower than 0.2. This is expected since their $T_{\rm dis}$ was already measured to be 20 Myr in Step I.

In summary, step II allowed us to constrain the $T_{\rm dis}$ for the 5 case II galaxies and the $l_{\rm corr}$ and $D_2$ for the 5 case III galaxies.

\subsection{Summarizing best-fit models corresponding to the observed TPCF}
\label{subsec:best_fit}
We observed in Figures \ref{fig:tpcf_plots}, \ref{fig:tpcf_plots_contd}, and \ref{fig:tpcf_revised} that young SFCs (age $<$20 or $<$A1 or $<$A2 Myr) always follow model PW TPCF (with the exception of WLM, for which young SFCs follow model PWF). We observed that for old SFCs ($>$20 Myr) in step I, 11 out of the 17 galaxies exhibit model PF TPCF, consistent with an underlying exponential disk-like distribution. Their absolute correlation is close to 1, and the power law slope is shallower than or comparable to $\sim$0.2, within error bars. For NGC 1512, $>$20 Myr SFCs exhibit a slope of -0.36 $\pm$ 0.11, which implies a significantly hierarchical distribution. However, we were unable to raise the age cut beyond 20 Myr to derive the $T_{\rm dis}$ of NGC 1512 due to a shortage of SFCs i.e. only 225 SFCs with ages greater than 20 Myr. Therefore, we suggest a lower limit of 20 Myr to be the $T_{\rm dis}$ of NGC 1512. For the remaining 5 out of 17 galaxies in step I i.e. case II, the $>$20 Myr SFCs exhibited PW model TPCF with slopes greater than 0.2. This indicated a reasonably hierarchical distribution for these \enquote{old} SFCs. However, in step II, the $>$A1 Myr TPCF for most of these galaxies, except NGC 0628, exhibited a slope shallower than or comparable to 0.2, within error-bars. This marked the dispersal of hierarchy for these galaxies at A1 Myr.   

We note that for 3 out of our 5 case II galaxies (NGC 0628, NGC 4236 and NGC 5457), the TPCF breaks observed in step II are comparable to the correlation length measured with $<$20 Myr TPCF in step I. However, in the remaining two galaxies - NGC 5194 and NGC 5033, the TPCF breaks observed in step II and the measured correlation length have no overlap. In NGC 5033, this difference may arise due to our TPCF fitting process and the accuracy with which the break scale can be measured in the model PW TPCF. In NGC 5194, this difference can be attributed to the low number of SFCs used in the $<$20 Myr TPCF i.e. 197 SFCs. This is also supported by our previous measurement of NGC 5194's $l_{\rm corr}$ to be 2.0 $\pm$ 0.2 kpc in paper I, which is closer to the value measured using the $<$A1 Myr SFCs than that obtained using the $<$20 Myr SFCs.


Overall, this galaxy-wise TPCF measurement methodology helped us derive the three key hierarchy parameters for our galaxies. First parameter is the $l_{\rm corr}$ and a detailed discussion about its physical interpretation and dependence on galaxy properties is provided in Section \ref{subsec:lcorr_interpret}. We emphasize that the measured $l_{\rm corr}$ values are significantly larger than the UVIT spatial resolution at the corresponding galaxy distances. To illustrate this, we have marked twice the UVIT spatial resolution for each galaxy with vertical green dashed lines in the TPCF plots (Figures \ref{fig:tpcf_plots}, \ref{fig:tpcf_plots_contd}, and \ref{fig:tpcf_revised}). 
Moreover, in Appendix \ref{appdx:dist_resolution_effect}, we demonstrate that the measured $l_{\rm corr}$ values are robust against a 2.5 to 4 times variation in spatial resolution and galaxy distances. The second parameter, $D_2$ is a measure of fractal-like complexity and space-filling nature of the SFC distribution within a two-dimensional plane. We provide a separate discussion of $D_2$ in Section \ref{subsec:D2}. The third parameter is $T_{\rm dis}$ of the galaxy, for which we provide a separate discussion in Section \ref{subsec:Tdis}. These three hierarchy parameters, which are tabulated in Table \ref{table3}, (along with the best-fit parameters for all the TPCF measurements performed with various age cuts) provide a nearly complete description of the stellar hierarchy in a galaxy.

\begin{table*}[t]
\caption{Best fit parameters for the observed TPCF in our sample galaxies.}
\hspace*{-\dimexpr\oddsidemargin+1in\relax}%
\makebox[\paperwidth][c]{%
\resizebox{\textwidth}{!}{%
\begin{tabular}{cccccccccc}
\hline
Galaxy    & Young set & Young TPCF      & Young TPCF         & Old set   & Old TPCF          & Old TPCF           & $l_{\rm corr}$          & $D_2$ = 2 + $\alpha_1$   & $T_{\rm dis}$    \\
          & (Myr)     & break(kpc)     & slope              & (Myr)     & break(kpc)       & slope              & (kpc)                    &                          & (Myr)             \\
(1)       & (2)       & (3)             & (4)                & (5)       & (6)               & (7)                & (8)                      & (9)                      & (10)              \\\hline
N0253  & $<$20     & 2.46 $\pm$ 0.37 & $-$0.24 $\pm$ 0.02 & $>$20     & 0.91 $\pm$ 0.14   & $-$0.23 $\pm$ 0.04 & \textbf{2.46 $\pm$ 0.37} & \textbf{1.76 $\pm$ 0.02} & \textbf{20}       \\\hline

N0300  & $<$20     & 0.43 $\pm$ 0.05 & $-$0.95 $\pm$ 0.11 & $>$20     & $-$               & $-$0.17 $\pm$ 0.02 & \textbf{0.43 $\pm$ 0.05} & \textbf{1.05 $\pm$ 0.11} & \textbf{20}       \\\hline

N0628*  & $<$20     & 0.54 $\pm$ 0.05 & $-$1.26 $\pm$ 0.15 & $>$20     & 0.60 $\pm$ 0.07  & $-$0.61 $\pm$ 0.07 & \textbf{0.54 $\pm$ 0.05} & \textbf{0.74 $\pm$ 0.15} & $-$               \\
revised $T_{\rm dis}$ & $<$100 & 0.59 $\pm$ 0.06 & $-$0.83 $\pm$ 0.11 & $>$100              & 0.56 $\pm$ 0.17   & $-$0.44 $\pm$ 0.15 &  $-$                     &  $-$                     & \textbf{$>$100}   \\\hline   

N1512  & $<$20     & 2.65 $\pm$ 0.79 & $-$0.77 $\pm$ 0.19 & $>$20     & 1.93 $\pm$ 0.54   & $-$0.36 $\pm$ 0.11 & \textbf{2.65 $\pm$ 0.79} & \textbf{1.23 $\pm$ 0.19} & \textbf{$>$20}    \\\hline

N1566  & $<$20     & 3.41 $\pm$ 0.43 & $-$0.35 $\pm$ 0.04 & $>$20     & $-$               & $-$0.12 $\pm$ 0.02 & \textbf{3.41 $\pm$ 0.43} & \textbf{1.65 $\pm$ 0.04} & \textbf{20}       \\\hline   

N2403  & $<$20     & 0.78 $\pm$ 0.16 & $-$1.28 $\pm$ 0.23 & $>$20     & $-$               & $-$0.05 $\pm$ 0.03 & \textbf{0.78 $\pm$ 0.16} & \textbf{0.72 $\pm$ 0.23} & \textbf{20}       \\\hline

N2903**  & $<$20     & noisy TPCF      & noisy TPCF         & $>$20     & $-$               & $-$0.08 $\pm$ 0.02 & $-$                      & $-$                      & \textbf{20}       \\
revised $l_{\rm corr}$        & $<$35     & $<$0.85 $\pm$ 0.28 & $-$0.51 $\pm$ 0.16 & $>$35     & $-$               & $-$0.11 $\pm$ 0.02 & \textbf{$<$0.85 $\pm$ 0.28} & \textbf{$<$1.49 $\pm$ 0.16} &  $-$              \\\hline    

N3486  & $<$20     & 0.31 $\pm$ 0.06 & $-$1.20 $\pm$ 0.55 & $>$20     & $-$               & $-$0.03 $\pm$ 0.02 & \textbf{0.31 $\pm$ 0.06} & \textbf{0.80 $\pm$ 0.55} & \textbf{20}       \\\hline

N4228**  & $<$20     & noisy TPCF      & noisy TPCF         & $>$20     & $-$               & $-$0.18 $\pm$ 0.04 & $-$                      & $-$                      & \textbf{20}       \\
revised $l_{\rm corr}$        & $<$160    & $<$0.22 $\pm$ 0.03 & $-$0.82 $\pm$ 0.16 & $>$160    & $-$               & $-$0.17 $\pm$ 0.03 & \textbf{$<$0.22 $\pm$ 0.03} & \textbf{$<$1.18 $\pm$ 0.16} &  $-$              \\\hline   

N4236*  & $<$20     & 2.00 $\pm$ 0.28 & $-$0.50 $\pm$ 0.05 & $>$20     & 1.65 $\pm$ 0.21   & $-$0.38 $\pm$ 0.03 & \textbf{2.00 $\pm$ 0.28} & \textbf{1.50 $\pm$ 0.05} & $-$               \\
revised $T_{\rm dis}$           & $<$90     & 1.56 $\pm$ 0.14 & $-$0.40 $\pm$ 0.03 & $>$90     & 1.59 $\pm$ 0.66   & $-$0.19 $\pm$ 0.07 & $-$                      &  $-$                     &  \textbf{90}              \\\hline

N4395**  & $<$20     & noisy TPCF      & noisy TPCF         & $>$20     & 2.60 $\pm$ 0.30   & $-$0.22 $\pm$ 0.02 & $-$                      & $-$                      & \textbf{20}       \\
revised $l_{\rm corr}$        & $<$35     & $<$2.20 $\pm$ 0.41 & $-$0.32 $\pm$ 0.03 & $>$35     & $-$               & $-$0.09 $\pm$ 0.01 & \textbf{$<$2.20 $\pm$ 0.41} & \textbf{$<$1.68 $\pm$ 0.03} & $-$               \\\hline     

N5033*  & $<$20     & 2.32 $\pm$ 0.32 & $-$0.54 $\pm$ 0.07 & $>$20     & 1.47 $\pm$ 0.48   & $-$0.41 $\pm$ 0.16 & \textbf{2.32 $\pm$ 0.32} & \textbf{1.46 $\pm$ 0.07} & $-$       \\
revised $T_{\rm dis}$           & $<$35     & 1.49 $\pm$ 0.27 & $-$0.55 $\pm$ 0.11 & $>$35     & $-$               & $-$0.08 $\pm$ 0.02 & $-$                      & $-$                      &  \textbf{35}              \\\hline

N5194*  & $<$20     & 0.48 $\pm$ 0.09 & $-$0.73 $\pm$ 0.18 & $>$20     & 1.84 $\pm$ 0.21   & $-$0.30 $\pm$ 0.02 & \textbf{0.48 $\pm$ 0.09} & \textbf{1.27 $\pm$ 0.18} &  $-$              \\
revised $T_{\rm dis}$           & $<$150    & 1.78 $\pm$ 0.18            & $-$0.34 $\pm$ 0.02 & $>$150    & 1.78 $\pm$ 0.18   & $-$0.18 $\pm$ 0.05 & $-$                      & $-$                      &  \textbf{150}             \\\hline   

N5457*  & $<$20     & 1.59 $\pm$ 0.15 & $-$0.97 $\pm$ 0.07 & $>$20     & 1.25 $\pm$ 0.08   & $-$0.49 $\pm$ 0.02 & \textbf{1.59 $\pm$ 0.15} & \textbf{1.03 $\pm$ 0.07} & $-$               \\
revised $T_{\rm dis}$          & $<$50     & 1.57 $\pm$ 0.16 & $-$0.75 $\pm$ 0.05 & $>$50     & 0.84 $\pm$ 0.21   & $-$0.24 $\pm$ 0.07 & $-$                      & $-$                      & \textbf{50}       \\\hline

N7793**  & $<$20     & noisy TPCF      & noisy TPCF         & $>$20     & $-$               & $-$0.04 $\pm$ 0.01 &  $-$                     &  $-$                     & \textbf{20}       \\
revised $l_{\rm corr}$        & $<$40     & $<$0.47 $\pm$ 0.06 & $-$0.81 $\pm$ 0.11 & $>$40     & $-$               & $-$0.03 $\pm$ 0.01 & \textbf{$<$0.47 $\pm$ 0.06} & \textbf{$<$1.19 $\pm$ 0.11} & $-$               \\\hline    

Holmberg II & $<$20   & 0.87 $\pm$ 0.10 & $-$0.39 $\pm$ 0.03 & $>$20     & $-$               & $-$0.04 $\pm$ 0.01 & \textbf{0.87 $\pm$ 0.10} & \textbf{1.61 $\pm$ 0.03} & \textbf{20}       \\\hline

WLM*       & $<$20     & noisy TPCF      & noisy TPCF         & $>$20     & $-$               & $-$0.08 $\pm$ 0.02 & $-$                      & $-$                      & \textbf{20}       \\
revised $l_{\rm corr}$        & $<$45     & $<$0.12 $\pm$ 0.01 & $-$0.43 $\pm$ 0.04 & $>$45     & $-$               & $-$0.07 $\pm$ 0.01 & \textbf{$<$0.12 $\pm$ 0.01} & \textbf{$<$1.57 $\pm$ 0.04} & $-$               \\\hline  

\end{tabular}
}%
}
\tablecomments{(1): Galaxy name, (2), (3), and (4): age-range, break scale and the TPCF slope for the \enquote{young} SFCs, (5), (6), and (7): age-range, break scale and the TPCF slope for the \enquote{old} SFCs. * indicates the case II galaxies and the step II TPCF analysis for such galaxies is tabulated as \enquote{revised $T_{\rm dis}$}. ** indicates the case III galaxies and the step II TPCF analysis for such galaxies is tabulated as \enquote{revised $l_{\rm corr}$}. (8), (9), and (10) : finalized hierarchy parameters $l_{\rm corr}$, $D_2$ and $T_{\rm dis}$ values, respectively for the sample galaxies. See Section \ref{sec:obs_TPCF} for more details.}
\label{table3}
\end{table*}

\section{Discussion}
\label{sec:discussion}

\subsection{Interpretation of the correlation length}
\label{subsec:lcorr_interpret}
Mathematically, correlation length marks the transition scale in the broken power law form of the model PW TPCF, which is followed by the young SFCs i.e. $<$20 Myr, $<$A1 Myr and $<$A2 Myr SFCs \citep{2017ApJ...840..113G, 2021MNRAS.507.5542M, shashank2025tracing}. This transition demarcates a regime of a steep negative power-law slope on small scales (a signature of a purely hierarchical distribution) and a near-zero slope on large scales (a signature of a nearly random distribution). Physically, on scales smaller than the correlation length, the SFCs are spatially correlated with each other, and the absolute correlation diminishes in a power-law form as a function of spatial scale. On scales larger than the correlation length, the near-zero slope of the TPCF suggests that the SFC distribution is not hierarchical and may potentially be driven by the large-scale structure and dynamics of the galaxy. Thus, the correlation length represents the largest scale of the hierarchical star-forming structures in a galaxy.\ 

The existence of a correlation length in a galaxy implies that SFCs separated by less than the correlation length are all part of a common hierarchy of stellar structures - sustained by supersonic ISM turbulence \citep{2003MNRAS.343..413B, elmegreen2006hierarchical, 2017ApJ...840..113G, 2013MNRAS.436.1245F, Federrath_2018, 2018ApJ...858...31S, 2018PASP..130g2001G, 2021MNRAS.507.5542M, 2022MNRAS.512.1196M}. Such SFCs may share the same triggering events that led to their birth. Moreover, \cite{2017ApJ...842...25G} showed that star clusters located at smaller separations from each other tend to have similar ages and that the age difference between star cluster pairs increases as a power law function of the star cluster pair separation. They used this finding to argue that star formation is hierarchical in both space and time. As our correlation length values range from a few 100 pc up to a few kpc, we suggest that stellar feedback, HII region expansion or supernova explosions on scales up to a few 100 pc, and disk instabilities, spiral density waves, galactic shear and effect of galactic bars on scales between a few 100 pc up to a few kpc may act as the dominant driving mechanisms for the spatio-temporal hierarchy of star formation \citep{2004RvMP...76..125M, 2004ARA&A..42..211E, 2016MNRAS.458.1671K, 2020MNRAS.496.1803N}. These physical processes inject turbulent energy into the star-forming ISM, which aids in the creation of hierarchical, scale-free over-densities. The gravitational collapse of these overdensities, across spatial scales, may lead to the observed hierarchical spatial distribution of star-forming regions within galaxies \citep{2017ApJ...840..113G, 2017ApJ...842...25G, 2021MNRAS.507.5542M}. Therefore, the combined effect of ISM turbulence and gravitational dynamics shapes the star formation hierarchy up to the correlation length scale within a galaxy.\

\begin{figure*}
    \centering
    \includegraphics[width=0.70\textwidth]{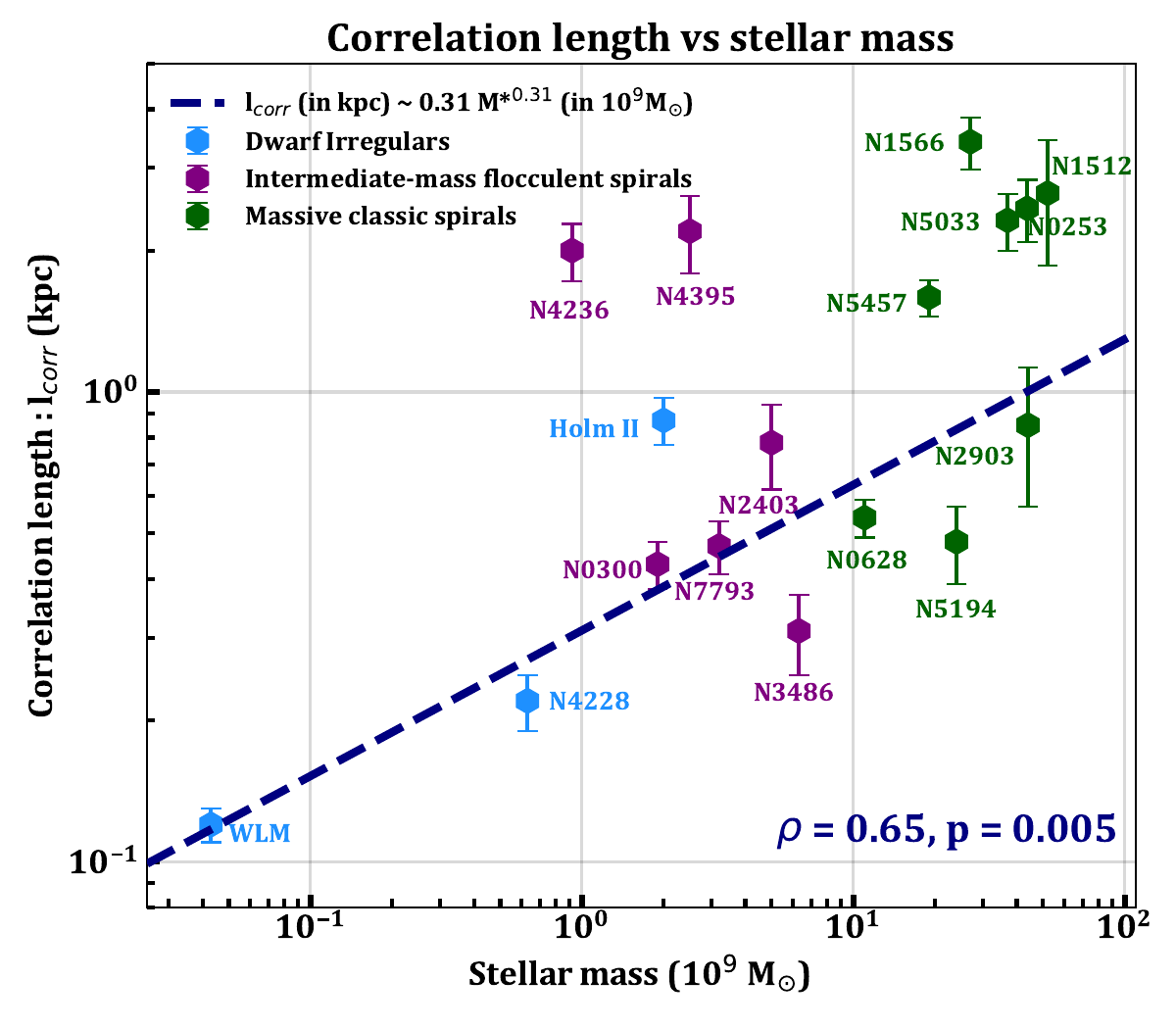}
    \hfill
    \caption{A comparison between the stellar mass and correlation length of the galaxies in log-log space, which reveals a statistically significant, linear dependence.}
    \label{lcorr_M_relation}
\end{figure*}

\cite{2021MNRAS.507.5542M} suggested that galaxies of all morphologies may have their own characteristic correlation length. However, due to partial galaxy coverage, which limited the physical scales they could probe, only lower limits of the correlation length for some of the galaxies (e.g NGC 1566, NGC 5194) could be reported in their paper. Additionally, for the dwarf galaxy NGC 3738, young star clusters did not show any signatures of a hierarchical distribution, which raised the question about the existence of a correlation length in dwarf galaxies. In Paper I, we derived the correlation length of 4 spiral galaxies (including NGC 1566 and NGC 5194) with UVIT data and suggested that all spiral galaxies are supposed to have a correlation length; however, full galaxy coverage is essential for its accurate determination. In this paper, we show that all star-forming galaxies have their own characteristic correlation length (see Table \ref{table3}), irrespective of the galaxy morphology. This makes the correlation length a ubiquitous and fundamental physical scale that characterizes a galaxy's star formation hierarchy. The observed correlation length of our galaxies ranges from a few 100s of pc to $\sim$4 kpc, which is much smaller than the galaxies' sizes. This implies that the star-formation hierarchies within galaxies do not extend to infinitely large spatial scales. 

The physical mechanisms that limit the correlation length of a galaxy require further exploration. In the literature, disk-instabilities and shear have been suggested as candidate mechanisms for setting the sizes of the largest hierarchical structures (and thereby, the correlation lengths) of galaxies \citep{elmegreen2014hierarchical, 2018PASP..130g2001G, 2017ApJ...840..113G, 2021MNRAS.507.5542M, 2023ApJ...944L..18M, lapeer2026feast}. These studies also indicate that large-scale galaxy properties and ISM conditions in galaxies uniquely determine their hierarchy parameters. Motivated by the positive trends of correlation length with stellar mass and Toomre length (a proxy for galactic shear) in \cite{2017ApJ...840..113G, 2021MNRAS.507.5542M} and Paper I, we revisit these trends along with a few additional dependencies for our 17 galaxies in the coming subsections.

\begin{figure*}[t]
    \centering
    \hfill
    \includegraphics[width=0.40\textwidth]{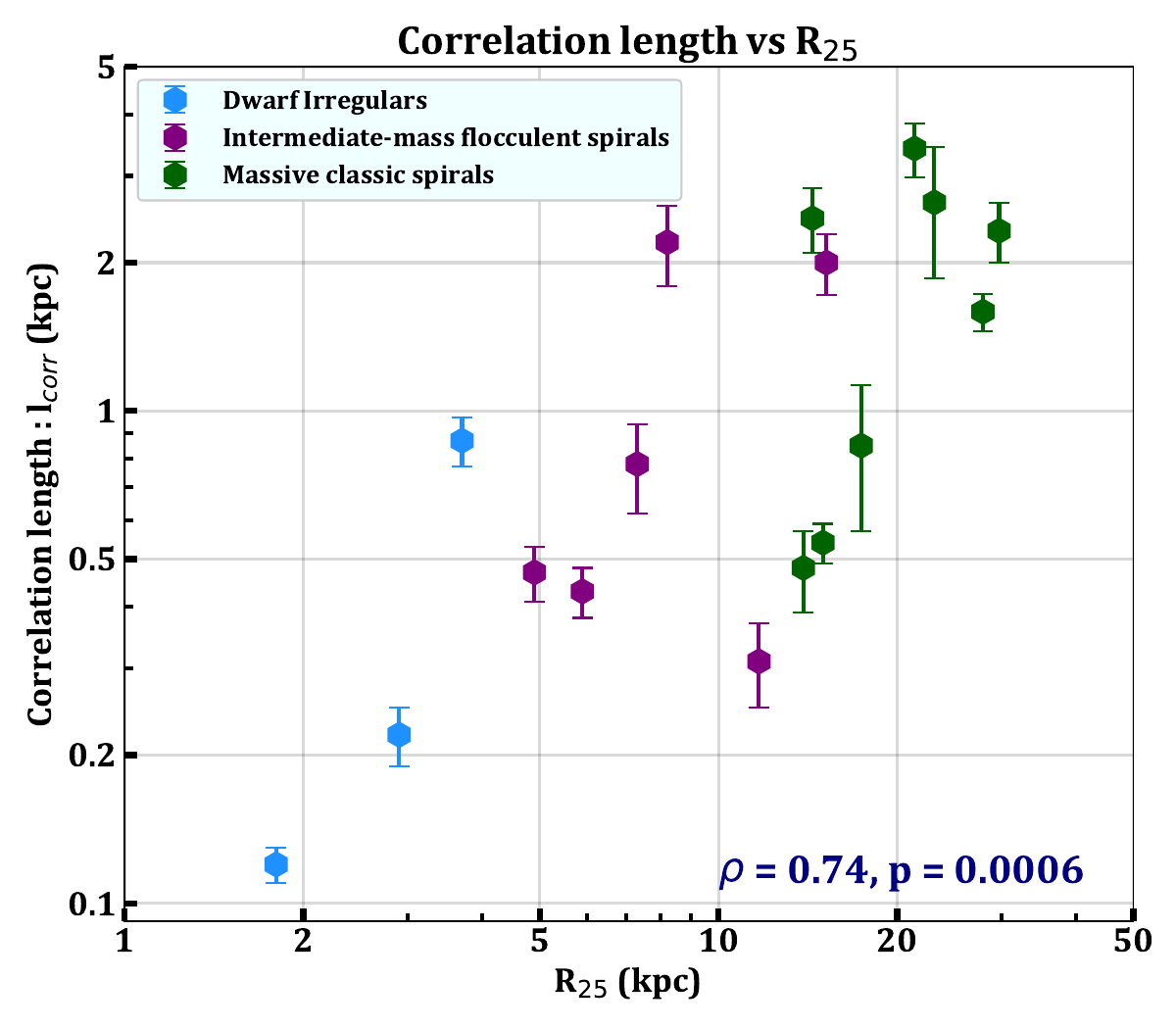}
    \hfill
    \includegraphics[width=0.40\textwidth]{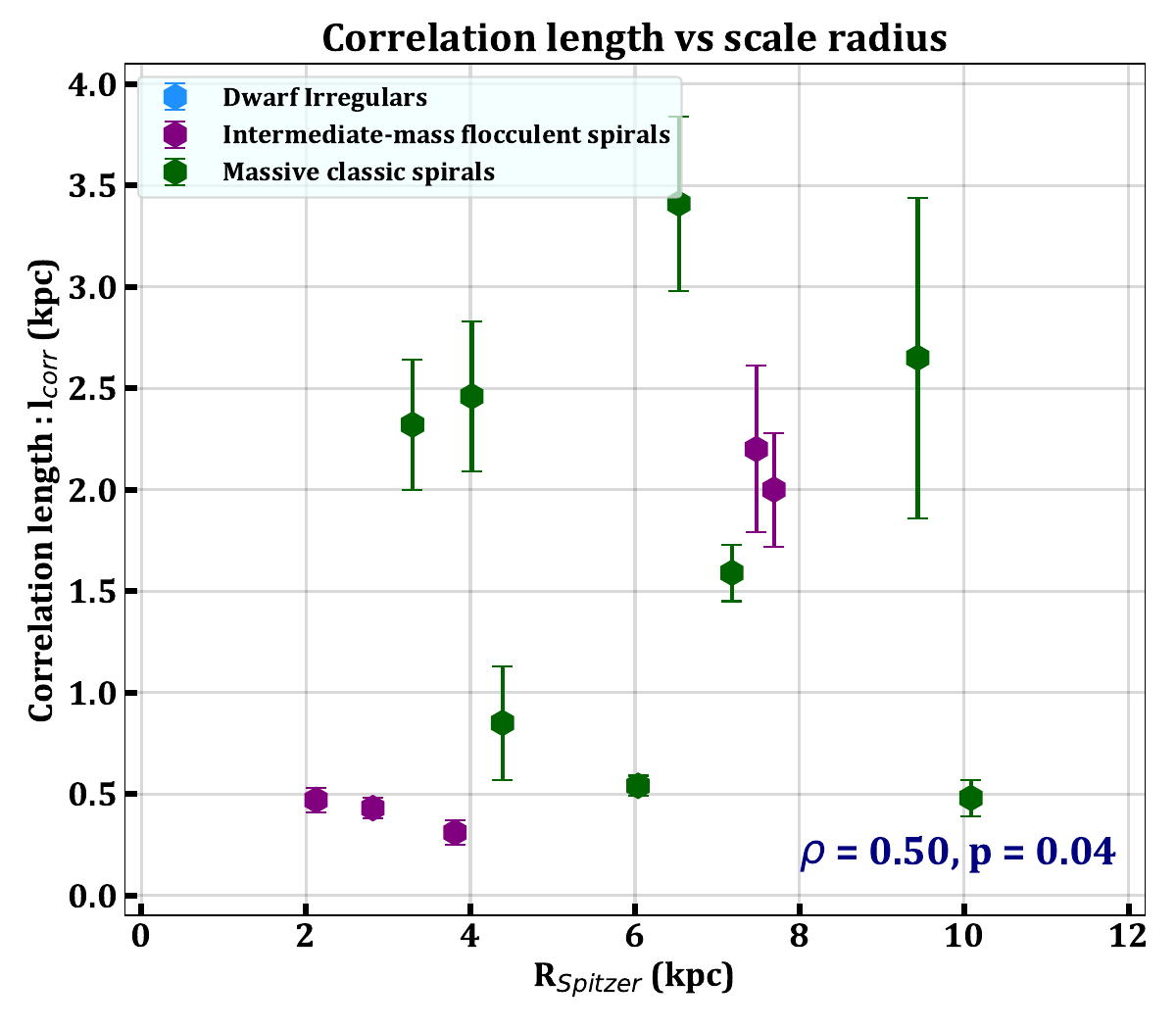}
    \hfill
    \hfill
    \vfill
    \hfill
    \includegraphics[width=0.40\textwidth]{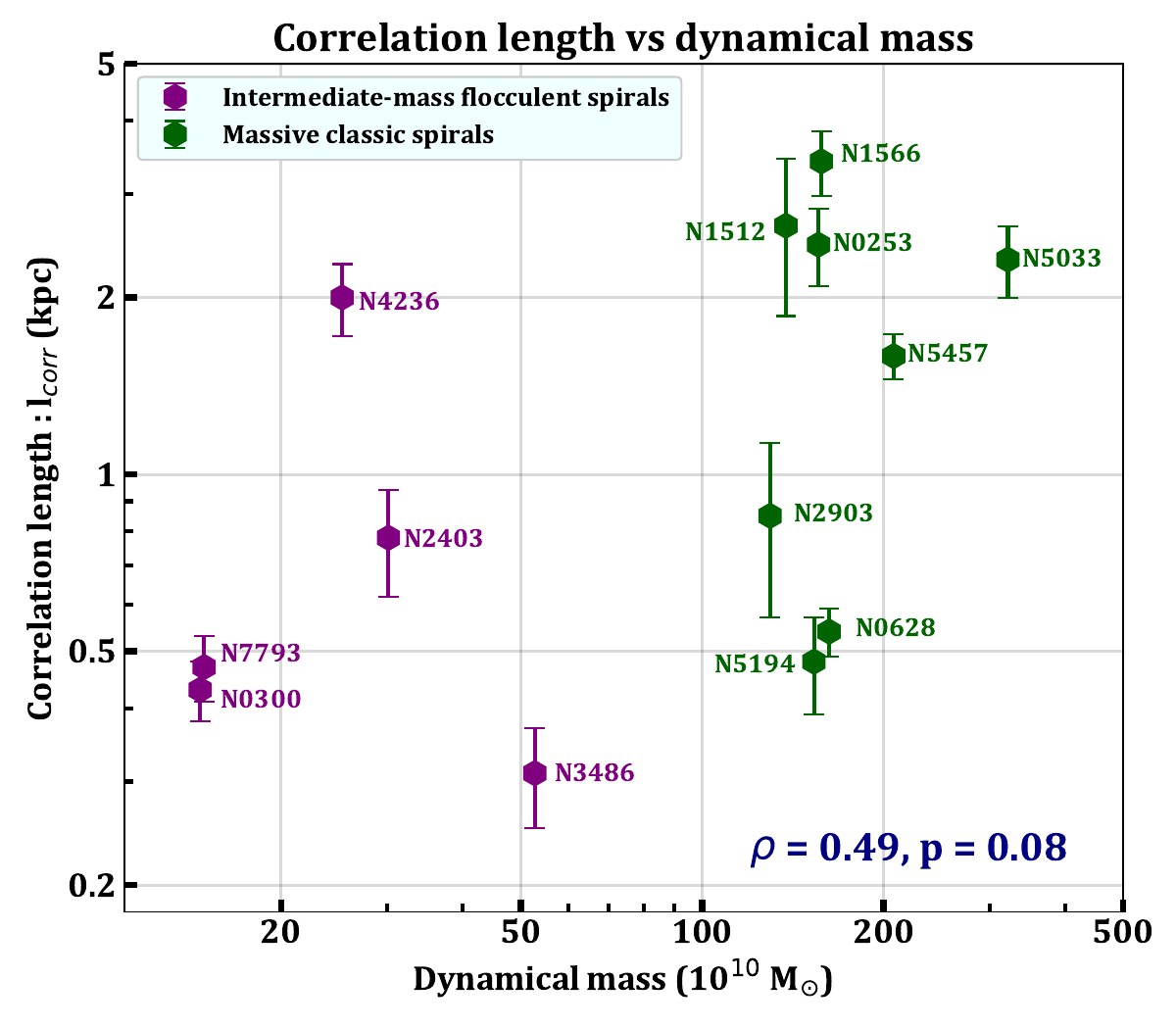}
    \hfill
    \includegraphics[width=0.40\textwidth]{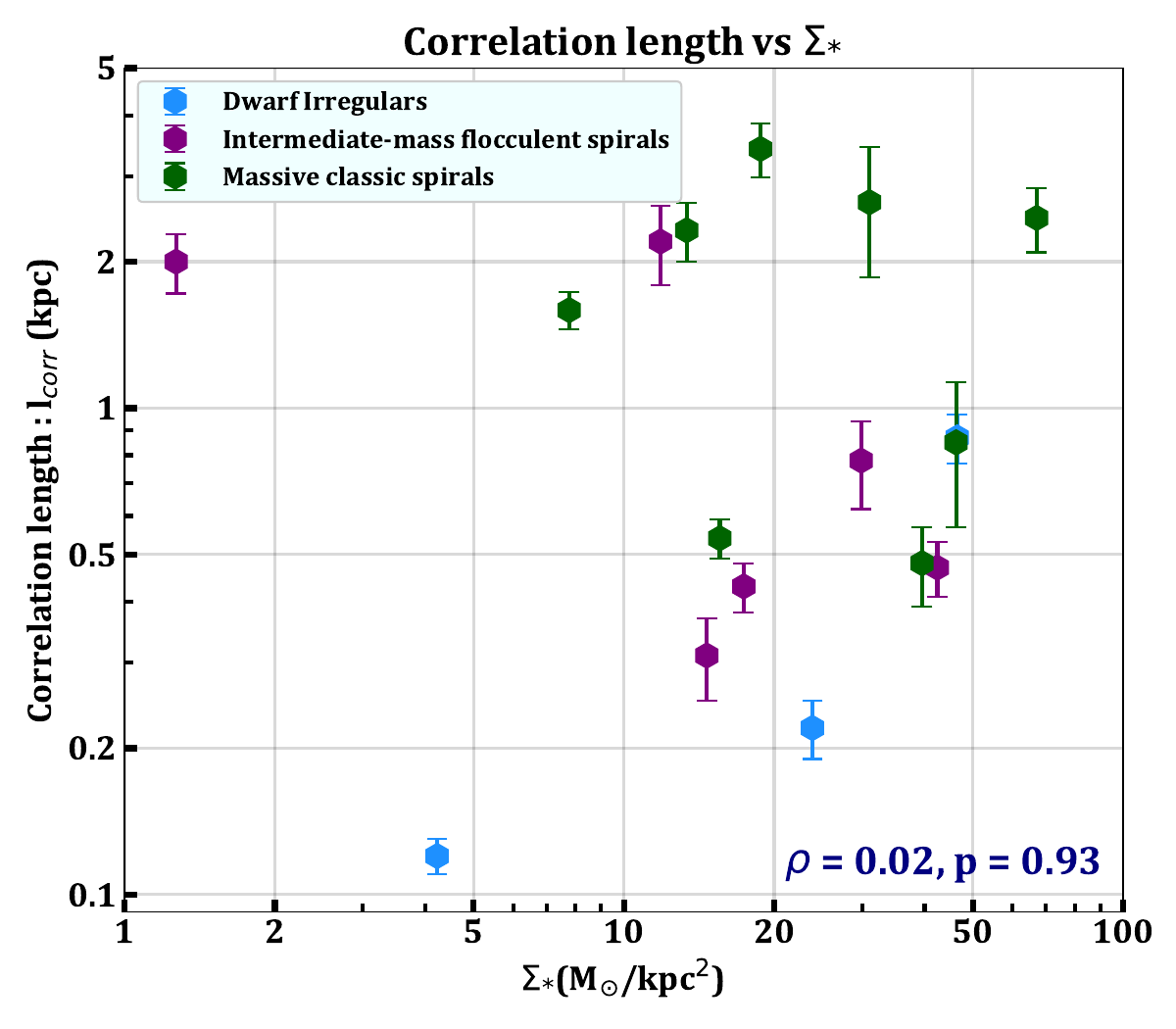}
    \hfill
    \hfill
    \caption{Top row: Results from the trends of the correlation length with the size of the galaxy, $R_{25}$ (left) and R$_{\rm{Spitzer}}$ (right), showing statistically significant positive correlations. Bottom left: Results from the trends of the correlation length with dynamical mass ($M_{\rm{dyn}}$ showing a mild positive correlation.  The dwarf irregular galaxies and NGC 4395 are excluded from this analysis because of the unavailability of V$_{\rm{flat}}$ values in the literature. Bottom right: Results from the trends of the correlation length with stellar mass surface density ($\Sigma_{*}$ showing no correlation. The galaxies are color-coded by morphology. The correlation coefficient ($\rho$) and p-value are provided as navy text.}
    \label{fig:lcorr_more_properties}
\end{figure*}

\subsection{Correlation length as a function of galaxy mass, size and morphology}
\label{sec:lcorr_galaxy_prop}

\begin{table*} 
\caption{Physical quantities associated with the Toomre length estimation and the derived values for 13 spiral galaxies in our sample (for NGC 4395, these quantities could not be found in the available literature).}
\centering
\begin{tabular}{ccccccc}
\hline
Galaxy   & $\Sigma_\mathrm{H_2}$ & $\Sigma_\mathrm{HI}$  & V$_{\rm{flat}}$ & r$_{\rm{median}}$ & H$_2$-based l$_{\rm{Toom}}$ & HI-based l$_{\rm{Toom}}$ \\
         & (M$_{\odot}$/pc$^{2}$) & (M$_{\odot}$/pc$^{2}$) & (km/s)   & (kpc)      & (pc)                  & (pc)               \\
(1) & (2) & (3) & (4) & (5) & (6) & (7) \\\hline
NGC 0253 &  -           & 15.60$^{(c)}$ & 217$^{(u)}$ & 7.50 & - & 1583 \\
NGC 0300 &  -           & 24.87$^{(c)}$ & 104$^{(u)}$ & 3.30 & - & 2127 \\
NGC 0628 & 3.59$^{(o)}$ & 15.60$^{(b)}$ & 217$^{(d)}$ & 4.95 & 159 & 690 \\
NGC 1512 & 1.02$^{(o)}$ & 6.14$^{(c)}$ & 161$^{(v)}$  & 9.00 & 271 & 1630 \\
NGC 1566 & 2.70$^{(p)}$ & 4.00$^{(b)}$ & 179$^{(w)}$  & 7.91 & 448 & 663 \\
NGC 2403 & 0.26$^{(q)}$ & 41.57$^{(d)}$ & 134$^{(d)}$ & 3.45 & 15 & 2341 \\
NGC 2903 & 10.17$^{(r)}$ & 9.26$^{(c)}$ & 180$^{(x)}$ & 5.47 & 798 & 726 \\
NGC 3486 & 0.18$^{(s)}$ & 14.11$^{(s)}$ & 140$^{(s)}$ & 4.21 & 14 & 1084 \\
NGC 4236 & 0.05$^{{(t})}$ & 14.41$^{(t)}$ & 85$^{(h)}$ & 5.72 & 19 & 5543 \\
NGC 5033 & 6.60$^{(s)}$ & 17.05$^{(s)}$ & 217$^{(s)}$ & 15.65 & 2916 & 7533 \\
NGC 5194 & 24.00$^{(p)}$ & 3.80$^{(b)}$ & 219$^{(d)}$ & 5.44 & 1258 & 199 \\
NGC 5457 & 1.66$^{(p)}$ & 7.80$^{(b)}$ & 180$^{(y)}$  & 8.58 & 320 & 1506 \\
NGC 7793 & 2.60$^{(p)}$ & 10.30$^{(b)}$ & 115$^{(d)}$ & 2.60 & 113 & 448 \\
\hline
\end{tabular}
\tablecomments{(1) Galaxy name, (2) H$_{2}$ gas surface density, (3) HI gas surface density, (4) flat rotation velocity, (5) median galactocentric radius of the SFCs within the galaxy, (6) derived H$_{2}$-based Toomre length, (7) derived HI-based Toomre length, for 13 out of 17 of our galaxies. The dwarf irregular galaxies were excluded from this analysis (refer to Section \ref{subsec:lcorr_lToom}. References indicated as bracketed superscripts are as follows, (b) - \cite{2021MNRAS.507.5542M}, (c) - \cite{2021ApJS..257...43L}, (d) - \cite{2008AJ....136.2782L}, (h) - \cite{2007A&A...462..933C}, (o) - \cite{2022MNRAS.516.3006K}, (p) - \cite{shashank2025tracing}, (q) - \cite{1998ApJ...498..541K}, (r) - \cite{2021MNRAS.502.1218R}, (s) - \cite{2016MNRAS.460..689R}, (t) - \cite{2014A&A...563A..31R}, (u) - \cite{2018MNRAS.478.1611K}, (v) - \cite{2019MNRAS.487.2797E}, (w) - \cite{2020ApJ...897..122L}, (x) - \cite{deBlok_2008}, (y) - \cite{Guelin_1970}.}
\label{table4}
\end{table*}

In Figure \ref{lcorr_M_relation}, we show that the correlation length and the stellar mass of galaxies share a statistically significant positive relationship (in a log-log plot) with a Spearman correlation coefficient ($\rho$) of 0.65 and a p-value of 0.005. This strongly indicates that the galaxy's gravitational potential directly impacts the organization of its star formation hierarchy. Physically, the deeper gravitational potential well of any galaxy can facilitate the accumulation, growth, and subsequent collapse of the larger coherent gaseous structures throughout the entire star-forming disk \citep{elmegreen2014hierarchical, 2018PASP..130g2001G, 2021MNRAS.507.5542M}. We also observed that the correlation length values for two of our three dwarf irregular galaxies are less than 300 pc (with the exception of Holmberg II). The correlation length of the majority of our flocculent spiral galaxies (4 out of 6) lies between 300 pc to $\sim$800 pc. Finally, the correlation lengths for the majority of our massive, classic spirals (6 out of 8) lie between $\sim$800 pc to 4 kpc. Taken together, galaxies of different morphologies in our sample roughly occupy mutually exclusive locations in Figure \ref{lcorr_M_relation}. This may indicate that there exist intrinsic differences in the hierarchical star formation properties of galaxies of different morphologies. We believe that these differences can be attributed to the dominant mechanisms driving turbulence in galaxies of different morphologies. This may impact the scales over which these mechanisms affect the structure of the star-forming ISM (see the discussion in Section \ref{subsec:lcorr_implication}).  

In Paper I, we had observed that the correlation length of the flocculent spiral NGC 7793 was $\sim$1/4th that of the 3 massive, classic spiral galaxies - NGC 1566, NGC 5194, and NGC 5457. In this paper, too, we observed that the majority of the flocculent spiral galaxies have significantly smaller correlation length values than the classic spiral galaxies with prominent spiral structure. This may indicate that apart from the mass of the galaxy, the nature of the spiral arm is also an important influencing factor when setting the largest scales of stellar hierarchy in a galaxy. \cite{2003ApJ...590..271E} suggested that self-gravity-driven shear instabilities on a large scale generate the flocculent spiral arms. \cite{2018PASP..130g2001G} posited that disk instabilities in massive spiral galaxies act as swing amplifiers to the spiral structure of the galaxy, which results in a prominent, coherent spiral structure. But, in flocculent spiral galaxies, these instabilities are not strong enough to create a consistent spiral structure, which leads to the flocculent, fragmented arms. As most of the star formation in galaxies is observed within the spiral arms, the nature of spiral arms may justifiably dictate the sizes of the largest, hierarchical star-forming regions. 

We also tested the dependence of the correlation length on galaxy size. We used $R_{25}$ and the near-infrared scale radius (R$_{\rm{Spitzer}}$) as the proxy for galaxy size (Figure \ref{fig:lcorr_more_properties}, top row). The references for $R_{25}$ are provided in Table \ref{table1} and the R$_{\rm{Spitzer}}$ values were taken from the Spitzer Survey of Stellar Structure in Galaxies (S4G; \cite{2015ApJS..219....4S}). We observed statistically significant correlations in both of these relationships: $\rho$ = 0.74, p = 0.0006 for the $R_{25}$ dependence and $\rho$ = 0.50, p = 0.04 for the R$_{\rm{Spitzer}}$ dependence. This indicates that within our sample, larger galaxies tend to have larger correlation lengths. We believe that the correlation length $-$ galaxy size relationship simply reflects the statistically significant dependence of the correlation length on galaxy's stellar mass, given the stellar mass–size scaling of galaxies \citep{2003MNRAS.343..978S, 2014ApJ...788...28V, 2015MNRAS.447.2603L}.

As dynamical mass also traces any galaxy's gravitational potential, we also tested the relationship between correlation length and the dynamical mass (Figure \ref{fig:lcorr_more_properties}, bottom left). The dynamical mass of galaxies was calculated using the formula $M_{\rm{dyn}}$ = V$^2$$_{\rm{flat}}$$R_{25}$/G. The dwarf irregular galaxies and NGC 4395 are excluded from this analysis because of the unavailability of V$_{\rm{flat}}$ values in the literature. In log-log space, we observed a rough linear correlation between the correlation length and the dynamical mass, with $\rho$ = 0.49 and p = 0.08. However, this dependence is not statistically significant. We also observed no correlation between the correlation length and the galaxy averaged stellar mass surface density for our sample of galaxies. As the correlation length exhibits statistically significant relationships with both the stellar mass and the size of galaxies, the absence of a significant correlation between the correlation length and the stellar mass surface density is not surprising.

\begin{figure*}[t]
    \centering
    \hfill
    \includegraphics[width=0.40\textwidth]{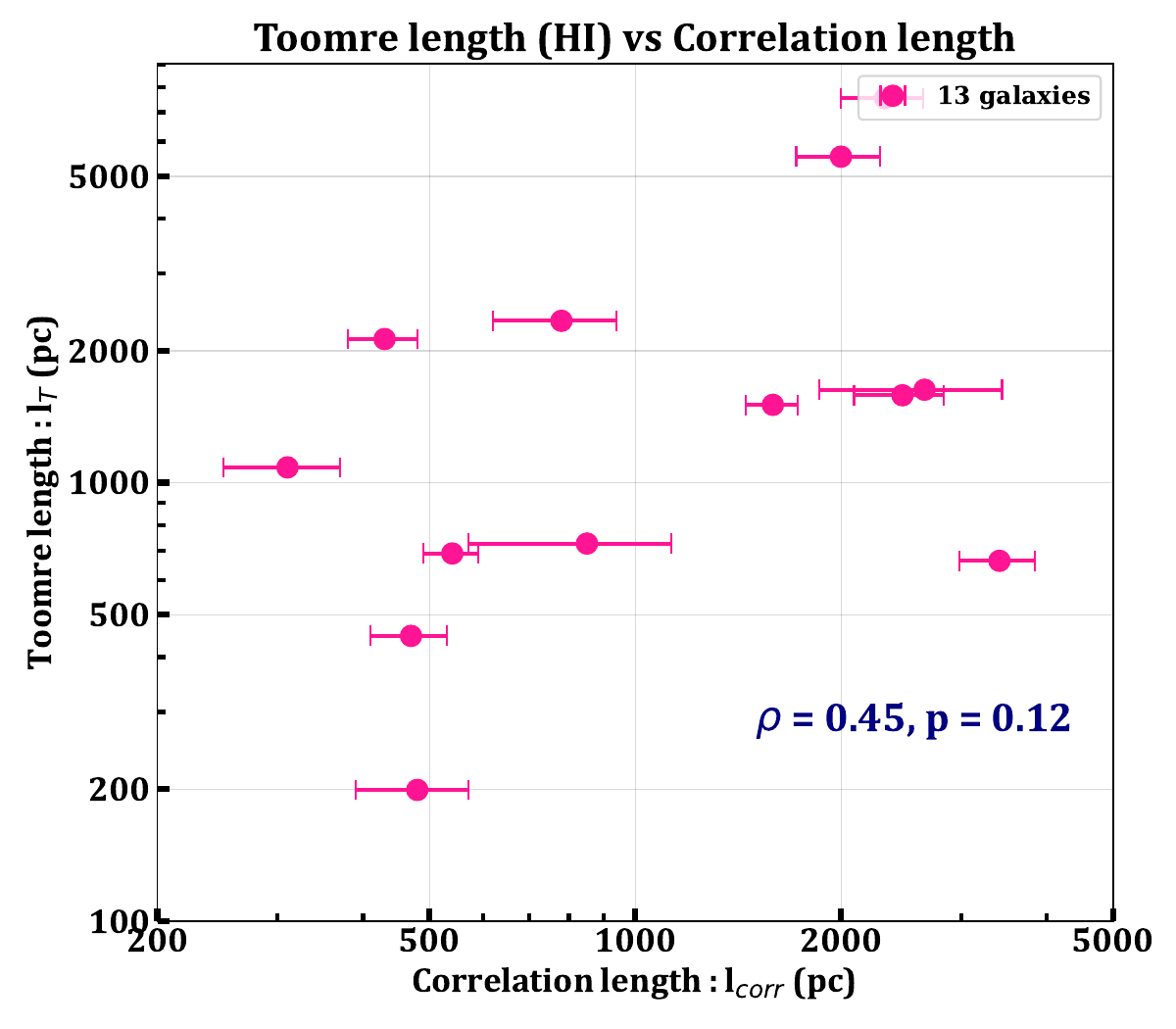}
    \hfill
    \includegraphics[width=0.40\textwidth]{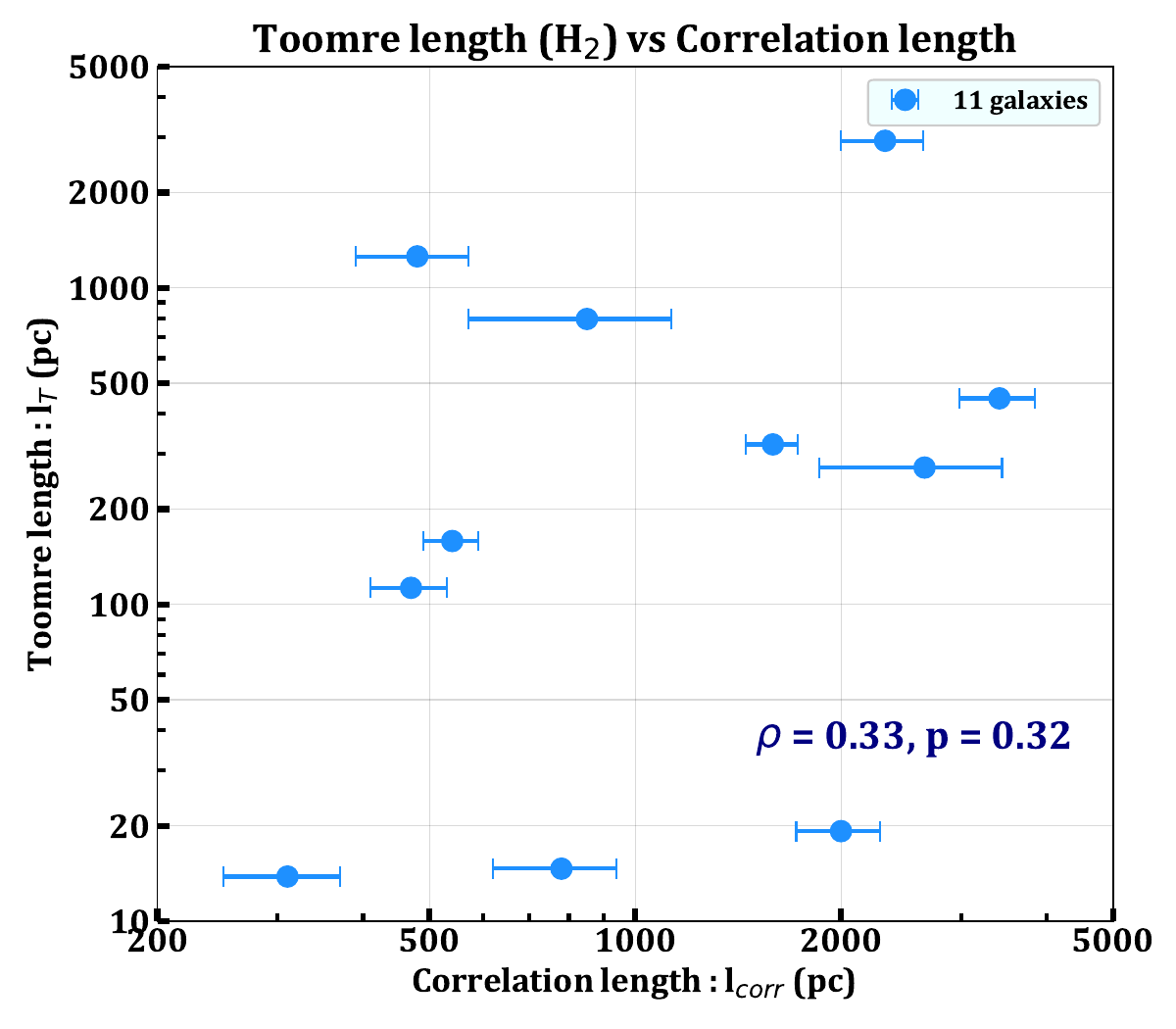}
    \hfill
    \hfill
    \caption{A comparison between the HI (left) and H$_2$ (right) based Toomre length and the correlation length of our galaxies in log-log space. The correlation length exhibits a mild positive correlation with the global HI-based Toomre length (measured for 13 galaxies), but we observe negligible correlation with the global H$_2$-based Toomre length (measured for 11 galaxies). The correlation coefficient ($\rho$) and p-value are provided as navy text.}
    \label{fig:Toom}
\end{figure*}

\subsection{Correlation length v/s Toomre length}
\label{subsec:lcorr_lToom}

In a differentially rotating disk, shear can act as a stabilizing mechanism which can prevent coherent gaseous structures from growing too large \citep{1964ApJ...139.1217T, 1984ApJ...276..114J, 2008ApJ...685L..31E, 2013MNRAS.433.1389R, 2015ApJ...806L..34F, 2013ApJ...779...45M, 2016MNRAS.460.2360R, 2017MNRAS.469..286R}. Galactic shear was also suggested to be one of the mechanisms that can limit the largest scales of star formation hierarchies in recent literature \citep{2017ApJ...842...25G, Grasha_2019, 2021MNRAS.507.5542M}. Numerical calculations by \cite{1964ApJ...139.1217T, 2008ApJ...685L..31E, 2010MNRAS.407.1223R, 2013MNRAS.433.1389R} and \cite{2018A&A...620A..21C} show that the Toomre length of a galaxy can serve as a proxy for the size of the gaseous structures that can survive within a differentially rotating galactic disk. 

The Toomre length is given by,
\begin{equation}
    l_\mathrm{Toom} =  \frac{4 \pi^{2} \mathrm{G} \Sigma_\mathrm{g} \mathrm{r}_{\mathrm{median}}^{2}} {\mathrm{v}_{\mathrm{flat}}^{2}}
\end{equation}
where $\Sigma_{\rm{g}}$ is the gas surface density, r$_{\rm{median}}$ is the median galactocentric radial position of SFCs within the galaxy and V$_{\rm{flat}}$ is the flat rotation speed. Only a single value of V$_{\rm{flat}}$ was adopted per galaxy, from the latest available HI or CO data, assuming that the HI- and H$_2$-based rotation curves agree well with each other. In this paper, we have tested the dependence of correlation length on the HI and H$_2$ based Toomre length for our galaxies, separately. So, we used the HI- and H$_2$- based $\Sigma_{\rm{g}}$ values from the latest available literature wherever measurements were available, and determined the corresponding Toomre length values. $\Sigma_{\rm{g}}$ was either adopted directly from the literature or was measured by dividing the reported HI or H$_2$ gas mass by the galaxy area. The physical quantities used in the Toomre length measurement are tabulated in Table \ref{table4}. As our sample contains 3 dwarf irregular galaxies for which the existence of disk structure is unclear, and therefore the effect of galactic shear is difficult to constrain, we do not include these galaxies in our Toomre length analysis. 

Before discussing the results of the correlation length $-$ Toomre length relationship, we acknowledge two notable caveats associated with this test. Firstly, the spatial distribution of HI and H$_2$ within galaxies may differ spatially. HI extent of spiral galaxies is also known to be significantly larger than the molecular gas extent and $R_{25}$. As we have used the same galaxy area to measure gas surface densities, our HI surface density values and, thereby, the HI-based Toomre length values may represent upper limits. Secondly, we know that shear is maximum at the galaxy centers and it declines at larger radii. $\Sigma_{\rm{g}}$ and galactic rotation speed also vary as a function of galactocentric radius. Overall, the Toomre length may vary significantly as a function of galactocentric radius \citep{2013ApJ...779...45M}. In this paper, however, we are only exploring trends of global hierarchy parameters with large-scale galaxy properties. Therefore, a comparison of correlation length (measured using the SFCs distributed within the entire extent of galaxies) and a globally averaged value of Toomre length is reasonable. The measured Toomre length is therefore seen as the representative average size of shear-limited structures in a galaxy. Local measurements of Toomre length, and thereby investigating the relationship between the properties of star formation hierarchies and galactic shear, are deferred to future papers.

Figure \ref{fig:Toom} presents the results of our comparison between the correlation length and the Toomre length. For the 13 galaxies included in the analysis, we found that the correlation length shares a mild positive correlation with the HI based Toomre length ($\rho$ = 0.45, p = 0.12). In contrast, for the 11 galaxies included in the analysis, we measured a $\rho$ value of 0.33 and a p-value of 0.33 for the correlation length and H$_2$-based Toomre length relationship, indicating no correlation. Previously, \cite{2021MNRAS.507.5542M} had observed a statistically significant positive correlation between the Toomre length and correlation length of 9 galaxies. In Paper I, we also observed a weak positive correlation in this relationship for our 4 galaxies. Though we do not observe, strong and statistically significant dependence of correlation length on the Toomre length, the observed positive trend may indicate that shear acting on the gaseous phase of the ISM may be one of the viable mechanism governing a galaxy's correlation length\footnote{Galactic shear is typically ineffective in dwarf galaxies, so it cannot be a viable mechanism limiting the maximum sizes of correlated stellar structures}. Shear can effectively put an upper limit on the sizes of the largest coherent gaseous structures that can survive within a disk galaxy. These structures can eventually collapse under self-gravity and result in the observed stellar hierarchy.

\begin{figure*}
    \centering
    \hfill
    \includegraphics[width=0.40\textwidth]{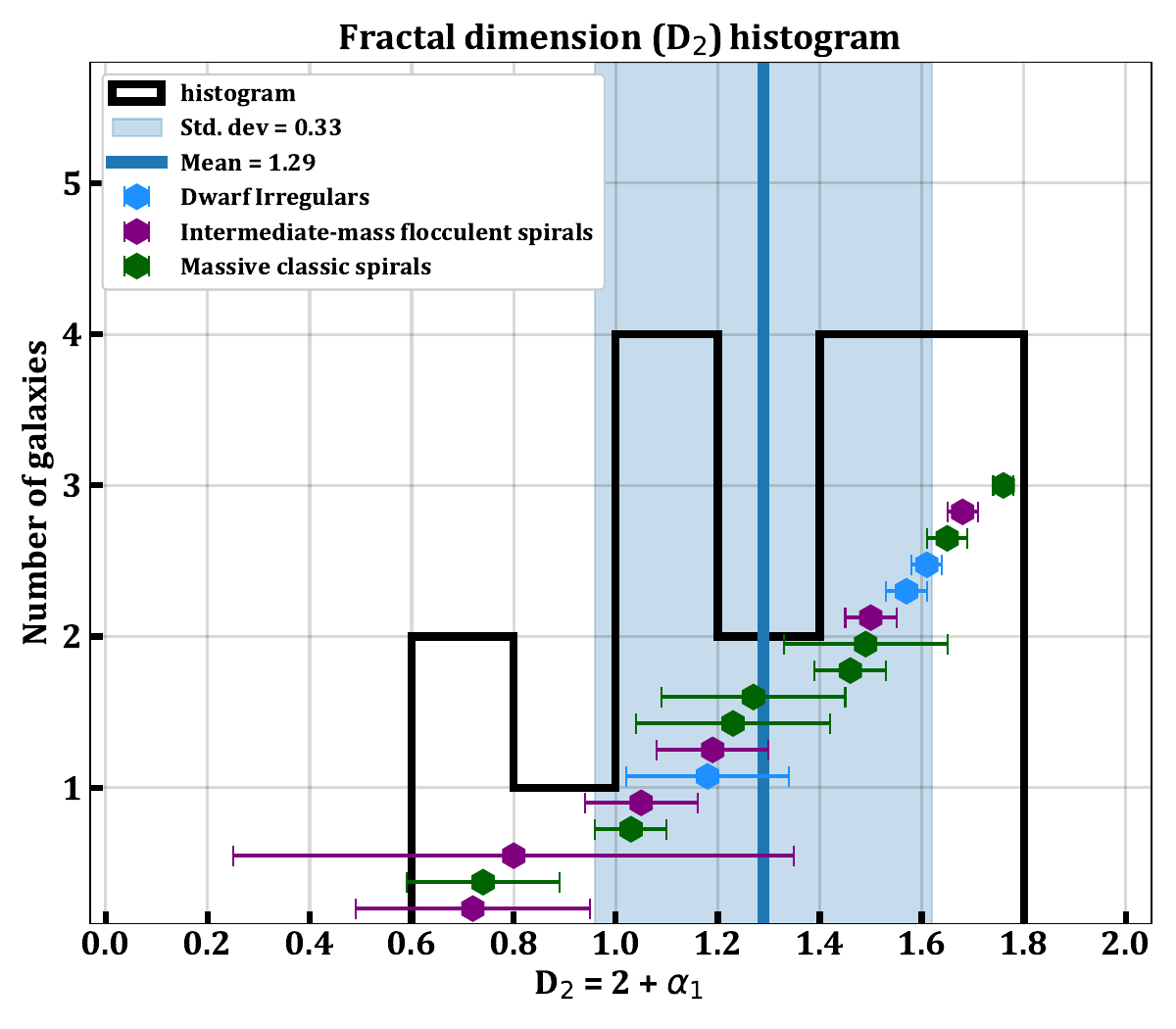}
    \hfill
    \includegraphics[width=0.40\textwidth]{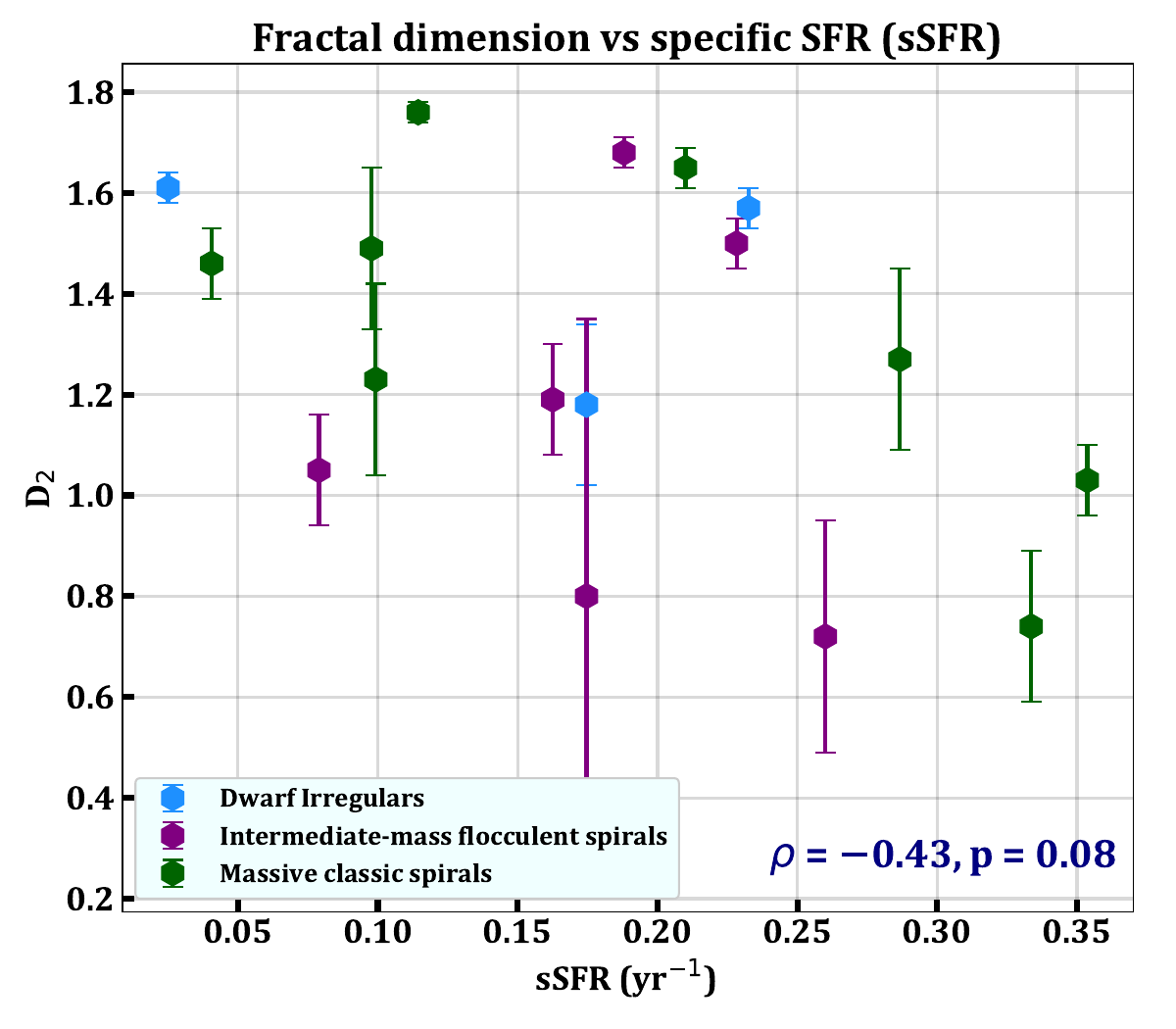}
    \hfill
    \hfill
    \caption{Left: A histogram illustrating the measured fractal dimensions of our galaxies (black lines). The blue vertical line and band represent the mean and standard deviation of the distribution. The colored data points are $D_2$ values for for individual galaxies, color coded in morphologies, and vertically offset from each other on y-axis by 0.25 for better visualization. The scatter in the $D_2$ values for galaxies of the same morphology indicates that fractal dimension and galaxy morphology are not correlated within our galaxy sample. Right : The relationship of $D_2$ with sSFR indicates a weak negative correlation.}
    \label{D2_histogram}
\end{figure*}

\subsection{Non-universality of the fractal dimension}
\label{subsec:D2}

In the literature, several methods have been used to measure the fractal dimension for the distribution of different star formation tracers. Slopes of mass or size distribution function, mass-luminosity function \citep{1996ApJ...471..816E, Shadmehri_2011}, mass-size relation \citep{2010ApJ...723..492R}, perimeter-area relationship \citep{Sun_2017, 2018ApJ...858...31S, 2022MNRAS.512.1196M, 2025ApJ...989..216H} and the TPCF \citep{Sanchez_2008, 2017ApJ...840..113G, 2021MNRAS.507.5542M, 2022MNRAS.516.4612T, lapeer2026feast} are some of the popular ways to estimate the fractal dimension. \cite{1996ApJ...471..816E} derived the three-dimensional fractal dimension for the molecular clouds in different regions of the Milky Way and found that it converges to a constant value of 2.3 $\pm$ 0.3. \cite{2010ApJ...723..492R} used the mass-size relationship for 580 molecular clouds detected in the University of Massachusetts-Stony Brook and Galactic Ring surveys within the Milky Way to derive a fractal dimension value of $\sim$2.36. \cite{Shadmehri_2011} further compiled observations of molecular clouds located within the Local Group and found results consistent with a universal fractal dimension of $\sim$2.3. The universality of the fractal dimension of the ISM has been attributed to supersonic turbulent motions self-consistently shaping its structure \citep{1991AnRFM..23..539S, 1996ApJ...471..816E}.

Within the Milky Way and the Local Group, fractal dimension is measured in three dimensions (D$_3$). However, due to the large distances of external galaxies and the unavailability of a three-dimensional distribution of star formation tracers, the fractal dimension is measured in two dimensions for external galaxies. $D_2$ $\sim$ D$_3$$-$1 is a simple, but perhaps not too realistic conversion that is valid only when the perimeter–area dimension of a projected three-dimensional structure is the same as that of its two-dimensional slice \citep{2004ARA&A..42..211E, 2014MNRAS.439.3775G}. Unlike the fractal dimension measurements in the Milky Way, similar measurements for external galaxies exhibit a deviation from universality. In the LMC, SMC, LMC bar region and the 30 Doradus-N158–N159–N160 star forming complex, using the slope of perimeter-area, mass-size relation, mass distribution function and size distribution function, projected fractal dimension ($D_2$) was measured to be ranging from $\sim$1.30 to $\sim$1.64 \citep{Sun_2017, 2017ApJ...849..149S, 2018ApJ...858...31S, 2022MNRAS.512.1196M, 2025ApJ...989..216H}. \cite{Sanchez_2008} derived the fractal dimension of HII region distribution in 93 nearby galaxies using TPCF and found statistically significant scatter in the fractal dimension measurements. They found that instead of being a universal value, the fractal dimension of galaxies positively correlates with $B$-band magnitude of galaxies. \cite{2021MNRAS.507.5542M} also found using TPCF analysis, a wide range of fractal dimension values between 0.5 and 1.9 for the star cluster distribution in their sample of 12 galaxies. Recently, \cite{lapeer2026feast} studied the distribution of young and emerging star clusters, characterized using HST and JWST observations, in three spiral galaxies - NGC 628, M51 (NGC 5194) and M83, and one dwarf galaxy, NGC 4449. Their fractal dimensions (observed $D_2$ $\sim$1.3), measured using TPCF analysis, were consistent with a universal star formation process and fractal dimension. Though their star cluster catalogs are some of the most complete catalogs to date, in any galaxy, their sample is limited to four galaxies, and their observations do not cover the full star-forming extent of the galaxies. 

Although we too measured a mean $D_2$ value of 1.29 $\pm$ 0.33 for our sample of 17 galaxies, our $D_2$ measurements presented in Table \ref{table3} and the histogram in Figure \ref{D2_histogram}(left) exhibit significant scatter. This indicates a significant deviation from a universal value of the fractal dimension of star formation. \cite{elmegreen2014hierarchical} observed that massive spirals exhibit similar fractal dimension values as the dwarf irregular galaxies. However, visual inspection of the histograms also indicates that all three morphological subclasses in our galaxy sample exhibit similar scatter, with no clear preference for any particular $D_2$ values across subclasses.

While probing the physical mechanisms that govern the galaxy-to-galaxy variation in $D_2$, \cite{2021MNRAS.507.5542M} observed a weak correlation of $D_2$ with stellar mass, star formation rate (SFR), and star formation rate density (SFRD) of galaxies. Motivated by their trends, we revisited these three dependencies for our sample of 17 galaxies (Figure \ref{fig:D2_non_correlations}). Moreover, we tested the relationship of $D_2$ with stellar mass surface density (Figure \ref{fig:D2_non_correlations}) and specific SFR (sSFR) (Figure \ref{D2_histogram}(right)). We used the $D_2$ values tabulated in Table \ref{table3} and the SFR values from literature (see Table \ref{table1}). SFRD and sSFR are calculated by dividing the SFR by the galaxy area and stellar mass, respectively. We found no significant correlation of $D_2$ with stellar mass, SFR, SFRD and stellar mass surface density. These non-correlations are presented in Appendix \ref{d2_non_correlations}. However, $D_2$ exhibits a weak negative correlation with sSFR, with a Spearman correlation coefficient value of $\rho$ = $-$0.43 and a p-value of 0.08 (Figure \ref{D2_histogram}(right)). Overall, our results are consistent with a non-universal fractal dimension of star formation. However, the governing mechanisms that determine the observed fractal dimension in galaxies remain unclear and require further investigation with a much larger galaxy sample. 


\subsection{Dispersal timescales of stellar hierarchies}
\label{subsec:Tdis}

Multiple physical processes taking place in the ISM can contribute to the dispersal of stellar hierarchies with increasing age such as 1) internal galactic dynamics causing the star-forming regions to naturally drift away from each other over time; 2) motion of SFCs in an intrinsically turbulent ISM (shaped by stellar feedback, supernovae explosions, density waves and shocks) with complex velocity structures can take young SFCs away from each other; 3) overlap of successive generations of star formation in the same spatial location gives an appearance of a dispersing hierarchy \citep{2017ApJ...840..113G, 2018ApJ...853...88E, 2021MNRAS.507.5542M, Meena_2025, lapeer2026feast}. The combined effect of all these factors can disperse the initial hierarchical distribution of the SFCs over a few tens of Myrs. We measured a hierarchy dispersal timescale of $\sim$20 Myrs for 12 out of 17 of our galaxies. For two galaxies, NGC 5033 and NGC 5457, the hierarchy dispersal timescale lies between 30 and 50 Myrs. These timescales agree well with the timescales derived in \cite{2017ApJ...840..113G, 2021MNRAS.507.5542M} and Paper I. For the remaining three galaxies, NGC 0628, NGC 4228 and NGC 4236, we derived hierarchy dispersal timescales of 100 Myrs or longer, which agree with the hierarchy dispersal timescales derived for the Large Magellanic Cloud (LMC) and its bar region in \cite{2009MNRAS.392..868B} and \cite{2017ApJ...849..149S}. \cite{2025ApJ...989..216H} also observed hierarchical structuring of UV-selected stellar structures, on a $\sim$150 Myr timescales in the SMC. \cite{2008MNRAS.391L..93G} and \cite{2009MNRAS.392..868B} suggested that random motions in the galactic potential in dwarf galaxies like the LMC and Small Magellanic Cloud (SMC) can transform the hierarchical structures to uniform distributions on approximately one crossing timescale (of order $\sim$100 Myr). NGC 4228 and NGC 4236 both have stellar masses smaller than 10$^{9}$M$_{\odot}$, which is comparable to typical dwarf galaxies. Regardless, galaxy-to-galaxy variations in the hierarchy dispersal timescales are expected and are likely driven by the difference in the galactic dynamics (e.g. rotation speed, shear and spiral structure) and ambient ISM conditions (e.g. ambient stellar and gas density, mid-plane pressure, tidal forces and stellar feedback) within galaxies \citep{Grasha_2018, Grasha_2019}. Additionally, \cite{2017ApJ...849..149S} propose that the hierarchy dispersal timescales can vary within different regions of the same galaxies, owing to the variation in the ambient environmental conditions.

We note that the combined effect of the choice of metallicity, dust attenuation law, and UVIT's filter-dependent extinction coefficients used in Paper II during the age estimation process can make some of our age estimates upper limits (refer to Section 3.4 of Paper II for more details). We aim to provide more reliable constraints for the hierarchy dispersal timescales for our galaxies with spectral energy distribution (SED) fitting based age-estimates. In the future, a comparison of hierarchy dispersal timescales with large-scale galaxy properties and ambient ISM conditions can help us further understand the physical mechanisms that influence the dispersal of stellar hierarchies.\

\subsection{Diversity in the hierarchy parameters of galaxies}
\label{subsec:non_universality}

For our morphologically diverse sample of 17 galaxies, we observed a broad range in the three hierarchy parameters. The correlation length values range from $\sim$100 pc for the dwarf galaxy WLM to $\sim$3.41 kpc for the massive spiral galaxy NGC 1566. Fractal dimension values range from $\sim$0.72 for NGC 2403 to $\sim$1.76 for NGC 0253. We observed hierarchy dispersal timescales of 20 Myrs for the majority of our galaxies, but a few galaxies exhibit hierarchy dispersal scales of $\gtrsim$100 Myr. These results represent a remarkable diversity in the derived hierarchy parameters of galaxies. 

This diversity in hierarchy parameters, particularly in the fractal dimension measurement, contrasts with the studies of the hierarchical structuring of molecular gas in the Milky Way and the Local Group \citep{1981MNRAS.194..809L, 1996ApJ...471..816E, Shadmehri_2011}. These studies suggested that the hierarchical properties of molecular gas, and thereby star formation should be universal. \cite{2021MNRAS.502.1218R} also showed that in a sample of 19 PHANGS-ALMA galaxies, giant molecular clouds follow the form of Larsen's law, which is valid in the Milky Way ($\sigma_{\rm{GMC}}$ $\propto$ size$_{\rm{GMC}} ^{0.5}$). In a paradigm where the hierarchical star formation process and its governing mechanisms are universal, the resulting fractal dimension of star formation is expected to be similar across different galaxies. However, our results indicate that despite the hierarchical properties of molecular gas being universal (as indicated by past studies), the resulting stellar matter distribution does not exhibit a similarly fractal dimension. Instead, the star formation process is strongly impacted by the unique ISM conditions of the host galaxy, and the imprint of these ISM conditions can be observed through its different hierarchy parameters \citep{Sanchez_2008, elmegreen2014hierarchical, 2017ApJ...840..113G, 2021MNRAS.507.5542M}.

\subsection{Implications from the existence of a largest scale for a galaxy's star formation hierarchy}
\label{subsec:lcorr_implication}

\subsubsection{Observational results}
Breaks similar to those seen in our young ($<$20 Myr, $<$A1 Myr and $<$A2 Myr) SFC TPCF have also been observed in the TPCF of various star formation tracers in different galaxies \citep{Sanchez_2005, 2008ApJ...681.1248O, 2017ApJ...840..113G, 2017ApJ...849..149S, 2021MNRAS.507.5542M, lapeer2026feast}. The leading interpretation of TPCF breaks and thereby the correlation length suggests that it represents a transition from turbulence-driven hierarchy on scales smaller than the correlation length to a galactic dynamics and large-scale structure driven distribution of star formation on scales greater than the correlation length. \cite{2022MNRAS.512.1196M} observed a break at $\sim$700 pc in the size distribution function of young stellar structure and suggested that supersonic turbulence dominates on scales smaller than the break, while global galactic structure dominates on larger scales. \cite{2026ApJ..1003...50C} used an ordinal patterns framework to investigate the morphological complexity of galaxies and observed a characteristic scale of $\sim$200 pc in NGC 0628. They proposed that this scale marks the transition from a regime of small-scale structures shaped by star formation and feedback to a regime of larger-scale structures governed by galactic dynamics.

This break scale has sometimes been associated with the disk scale height of the galaxy. In this scenario, the break represents the transition in ISM turbulence from being three-dimensional on scales smaller than the break scale and two-dimensional on scales larger than the break scale \citep{2009MNRAS.392..868B}. However, we know from the observations of edge-on disk galaxies that the scale heights of the stellar disks of massive spiral galaxies are only up to a few hundred pc. \cite{1997A&A...327..966D} and \cite{1998MNRAS.299..595D} performed systematic, statistical studies of disc scale heights in highly inclined galaxies using near-infrared data. They found that though the vertical light profile of the galaxies is independent of galaxy type (S0 to Sd) at all radii, the radial to vertical scale height ratio of spiral galaxies varies with the galaxy type. \cite{2008MNRAS.384L..34D, 2009MNRAS.397L..60D} showed from HI observations that disk scale heights of spiral galaxies can be $\lesssim$500-800 pc. \cite{2025ApJ...986...13E} showed that there exist breaks in the H-$\alpha$, Pa-$\alpha$ and mid-infrared power spectrum of NGC 5194 at 120$-$170 pc, indicating its disk thickness. However, our correlation length measurements of a few kpcs in massive, classic spiral galaxies effectively make it appear unlikely that the TPCF break corresponds to the disk scale height. 

\subsubsection{Possible physical interpretations}

The existence of star formation hierarchies has been extensively linked to supersonic ISM turbulence by several studies in the past \citep{1981MNRAS.194..809L, 1996ApJ...471..816E, elmegreen2006hierarchical, 2009ApJ...692..364F}. \cite{2009MNRAS.397L..60D} and \cite{2020MNRAS.496.1803N} attribute the presence of scale-invariant gaseous structures in galaxies to supersonic turbulent motions in the ISM. In the turbulence-driven hierarchy framework, star formation hierarchies should be observable up to scales comparable to the largest turbulence injection scale. We also know that different physical processes in galaxies inject turbulence in the ISM on vastly different scales \citep{2004ARA&A..42..211E, 2004RvMP...76..125M, 2009ApJ...692..364F}. Turbulence also cascades down from larger scales to smaller scales, so the kpc scale, gravity-based sources of turbulence can also produce turbulence on hundreds and tens of parsec scales, via turbulent cascade. 

\begin{figure*}
    \centering
    \includegraphics[width=0.80\textwidth]{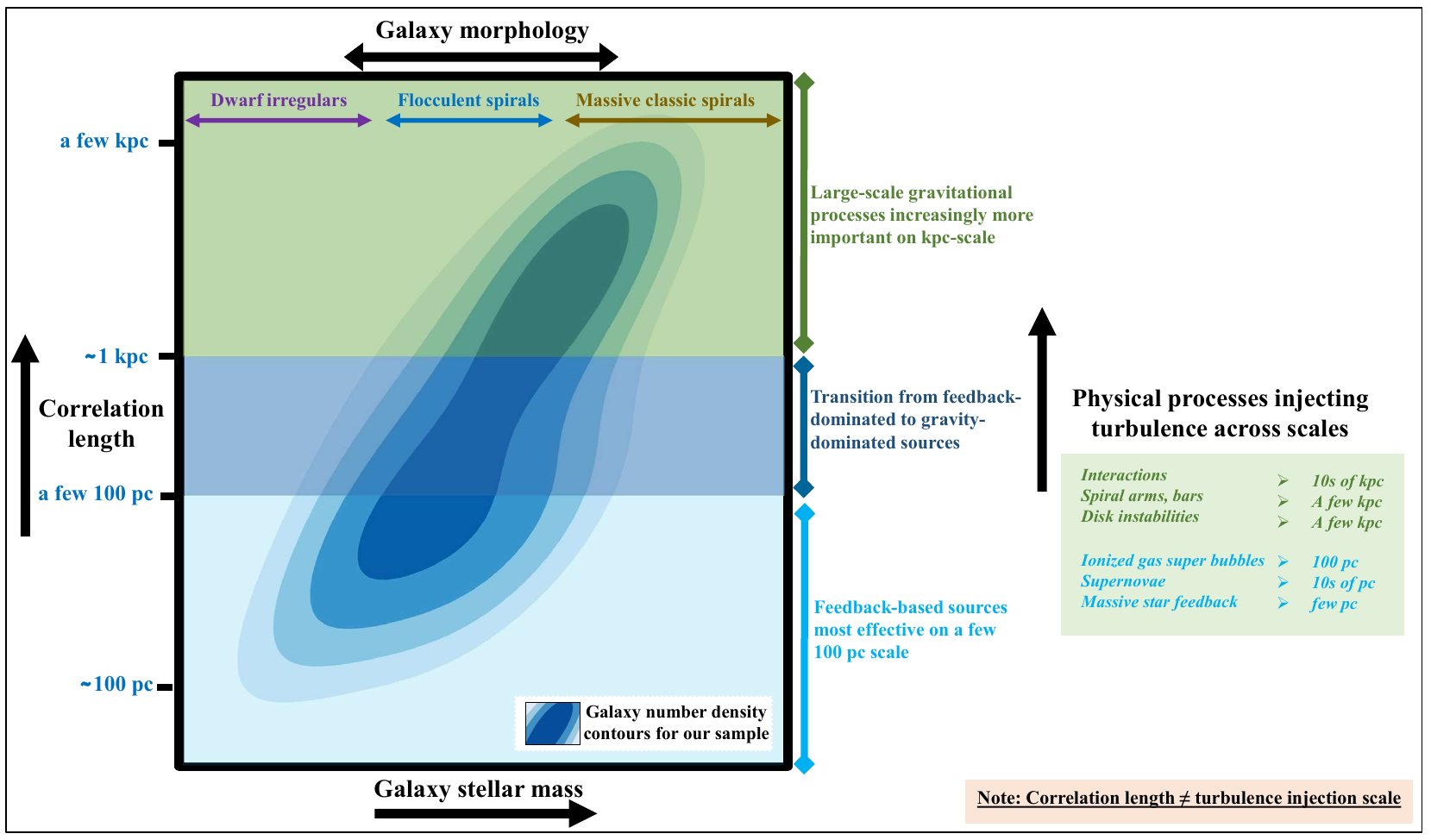}
    \caption{An illustrative schematic connecting the observed correlation length$-$stellar mass relation with the physical mechanisms driving turbulence and their characteristic spatial scales. The blue contour plot is created using the distribution of our 17 galaxies in the correlation length$-$stellar mass plot shown in the Figure \ref{lcorr_M_relation}. We caution that the turbulence injection scale does not directly correspond to the correlation length scale. This is because the turbulence-driven, hierarchically structured gas undergoes multiple physical processes before being converted into the observed SFC distribution which is spatially correlated up to the correlation length scale.}
    \label{fig:HSF_illustration}
\end{figure*}

For the 2/3 dwarf irregular galaxies\footnote[5]{why the correlation length of Holmberg II is so high as compared to the other two dwarf irregular galaxies needs a separate investigation, which can be attempted in the future} whose correlation lengths are less than 300 pc, we suggest that feedback-driven turbulence may be the dominant mechanism that governs their star formation hierarchy. Kpc-scale correlation length values for the massive, classic spiral galaxies imply that stellar feedback alone cannot be responsible for injecting turbulence in the ISM and thereby governing their star formation hierarchy. This is because feedback energy gets blown out of the galaxy's disk on scales beyond the disk scale height i.e. a few 100 pcs \citep{2023ApJ...944L..18M}. By isotropy, it is reasonable to assume that the feedback energy can only propagate within the galactic disks up to similar length scales. The insufficiency of feedback energy as the dominant source of turbulence in massive galaxies was also discussed in \cite{2016MNRAS.458.1671K}, who suggested that the turbulence in massive galaxies can be accounted for by disk instabilities. They posited that it is more likely that the gravity is the source of the large-scale ISM turbulence and wins over the stellar feedback (see also \cite{2021PASP..133j2001B}). \cite{2010MNRAS.409.1088B} and \cite{2018A&A...620A..21C} suggested that turbulence at the large scales is insensitive to the stellar feedback and is mostly driven by gravitational instabilities. They suggested that scale-invariant motions at large scales arise from the spiral waves, and the turbulent energy gets cascaded down to smaller scales. This points towards a scenario where gravity-driven sources of turbulence, such as disk instabilities, spiral structure, galactic shear, and bars, would need to be invoked to explain correlated star formation on kpc scales. Finally, the 300 pc to $\sim$1 kpc correlation length values for most of our intermediate mass, flocculent spiral galaxies can be seen as a transition step in galaxy morphology where gravity-based sources such as disk instabilities, spiral structures and shear take over from stellar feedback in terms of generating turbulence in the ISM and setting their star formation hierarchy. This interpretation is also illustrated in Figure \ref{fig:HSF_illustration}.

It is worth noting that many studies also propose that gravity-driven sources of turbulence can explain the star formation-related structure formation on all scales. \cite{2025A&A...695A.155A} showed that gravity alone can explain the observed filamentary gaseous structures within rotating galaxy disks, without the need for feedback and stellar origins of ISM turbulence. These gaseous structures can, in principle, act as the seeds of the observed stellar hierarchy in galaxies. \cite{2016AJ....152..134M} showed that for The HI Nearby Galaxy Survey (THINGS) galaxies \citep{Walter2008THINGS}, large-scale turbulence is more likely to be driven by disk gravitational instabilities, density waves, or bar-streaming motions and is less likely to be a result of stellar feedback. The HI-based study by \cite{2020MNRAS.496.1803N} found that gravitational instability can drive turbulence at a few kpc scales. Sources such as shear and magneto-rotational instabilities produce solenoidal turbulence at kpc scales whereas, ionizing feedback, expanding HII regions, supernovae, spiral compression drive compressive turbulence on a few tens of parsecs up to a few kpc scales \citep{2004RvMP...76..125M, 2004ARA&A..42..211E, 2013MNRAS.436.1245F, 2016ApJ...832..143F, 2016ApJ...825...30P, 2016ApJ...822...11P, 2016MNRAS.458.1671K, 2018NatAs...2..896O, 2020MNRAS.493.4643M, 2021MNRAS.500.1721M, 2023A&A...672A.193F, 2026MNRAS.547ag359M}. 

A direct relationship between the observed correlation lengths of galaxies and the turbulence injection mechanisms acting on scales ranging from tens of pc up to several kpc still needs to be established. The galaxy averaged stellar, or gas velocity dispersion, can in principle serve as a proxy for ISM turbulence in galaxies. We aim to explore this connection in our future papers, with a larger galaxy sample. More studies are also needed that investigate how gravity-driven dynamics and disk instabilities can explain the observed kpc-scale stellar hierarchies.


\section{Summary and future plans}
\label{sec:summary}
In this paper, we have investigated the hierarchical distribution of UV-bright star-forming clumps (SFCs) in 17 nearby galaxies of varied morphologies and mass range, all located within 20 Mpc. Our sample of 17 galaxies is the largest to date, employed for the investigation of the hierarchical organization of stellar matter in galaxies. It includes 8 massive, classic spirals, 6 intermediate-mass flocculent spirals and 3 dwarf irregular galaxies. This paper extends our previous investigation of extragalactic star formation hierarchies using the UltraViolet Imaging Telescope (UVIT) observations of four galaxies (\citealt{shashank2025tracing}, Paper I) to a larger sample of 17 galaxies. Our aim was to better understand the physical processes governing hierarchical star formation across a wide range of spatial scales, in galaxies spanning diverse morphologies and ISM conditions.

To examine and parametrize the star formation hierarchies, we utilized a catalog of $\sim$25000 SFCs characterized in our 17 galaxies by \cite{2026arXiv260612254S} (Paper II) and used the two-point correlation function (TPCF) for our statistical analysis. We divided the SFCs identified in our galaxies into \enquote{young} and \enquote{old} sets using a two-step approach (see Section \ref{sec:obs_TPCF} and Figure \ref{fig:schematic}). A homogeneous age cut of 20 Myr was adopted in step I across all galaxies. Further, a variable age cut of A1 (or A2) Myrs was adopted in Step II for 10 galaxies, with both A1 and A2 being greater than 20 Myr. Combining this SFC classification with the TPCF analysis allowed us to 1) constrain the hierarchical distribution of star formation in our galaxies, which is primarily governed by supersonic turbulence and gravitational dynamics in the star-forming ISM; and 2) examine the age-evolution of star formation hierarchies as they are subjected to galactic dynamics over tens of Myrs. We characterized the star formation hierarchies in our galaxies using three hierarchy parameters - the largest scale of the hierarchy (or correlation length), projected two-dimensional fractal dimension and the hierarchy dispersal timescale (see Table \ref{table3}). We observed remarkable diversity in these hierarchy parameters so, to understand the physical mechanisms governing the star formation hierarchies, we compared our hierarchy parameters with the host galaxy properties (see Section \ref{sec:discussion}). 

The distinctive feature of our study is facilitated by UVIT's 28\arcmin~field of view (FoV), which provided the complete coverage of the full star-forming extent of our galaxies. This is crucial for statistical studies of star formation hierarchies such as ours, as it allows us to constrain the global hierarchy parameters of galaxies and compare them against large-scale galaxy properties. Through this analysis, we attempted to connect the multi-scale process of star formation with the environmental factors and the physical mechanisms that govern it. The main conclusions of our work are following.\

\begin{itemize}
    \item Supersonic ISM turbulence sustains star formation hierarchies in galaxies, which do not extend to the full galaxy size. Instead, these hierarchies exhibit a galaxy-specific maximum scale named correlation length ($l_{\rm corr}$), which is ubiquitous across different galaxy morphologies.  
        
    \item  The hierarchical distribution of SFCs can be observed between a few tens of pc up to the $l_{\rm corr}$ scale, which ranges between $\sim$100 pc and $\sim$4 kpc in our galaxy sample.
    
    \item $l_{\rm corr}$ exhibits a statistically significant, positive correlation with the galaxy's stellar mass and a weak positive correlation with the dynamical mass. This suggests that the galaxy's gravitational potential strongly dictates the sizes of the largest, hierarchically structured star-forming complexes. 
    
    \item Dwarf irregulars, intermediate-mass flocculent spirals and massive, classic spirals lie on the same $l_{\rm corr}$ - stellar mass trend-line but occupy mutually exclusive regions in this relationship. 
    
    \item Most flocculent spirals and dwarf irregulars have sub-kpc $l_{\rm corr}$ values, whereas the classic spirals exhibit $>$1 kpc $l_{\rm corr}$ values. This might suggest that along with the galaxy's mass, its nature of spiral arms may play a key role in determining the largest sizes of hierarchical stellar structures in galaxies.
    
    \item We observed a weak positive correlation between the $l_{\rm corr}$ and HI-based Toomre length of our galaxies, but no correlation with H$_2$-based Toomre length (see Figure \ref{fig:Toom}). This possibly suggests that galactic shear may limit the sizes of the largest coherent gaseous structures, thereby restricting $l_{\rm corr}$ values to scales much smaller than the galaxy's size. However, this interpretation may not apply to dwarf irregulars. 

    \item Assuming star formation hierarchies are set by turbulent motions in the ISM, contextualizing the $l_{\rm corr}$ values against turbulence injection scales suggests that stellar feedback in dwarf irregulars and gravity-driven processes such as disk instabilities, spiral structure, and galactic shear in flocculent and classic spirals play the dominant role in setting their star formation hierarchies. 
    
    \item Our broad range of observed hierarchy parameters ($l_{\rm corr}$ $\in$ 0.12$-$3.4 kpc, $D_2$ $\in$ 0.71$-$1.73, $T_{\rm dis}$ $\in$ 20$-$150 Myr) point towards a non-universality in the hierarchical nature of star formation. In this paradigm, large-scale galaxy properties and the unique ISM conditions determine the properties of star formation hierarchies.  
\end{itemize}

With this paper, we have investigated hierarchical star formation in 17 star-forming galaxies which had available, archival UVIT FUV and NUV observations, with sufficient exposure times. Recently, \cite{2026ApJ..1002..220A} combined UVIT FUV observation with the Dark Energy Camera Legacy Survey (DECaLS) $g$-band data to investigate star formation hierarchy in NGC 5457 and NGC 1313. They were able to constrain the global, full-galaxy coverage hierarchy parameters of NGC 1313 for the first time, which demonstrated the utility of combining UVIT FUV and optical observations to study star formation hierarchies in nearby galaxies. Along these lines, we are working on a sample of $\sim$25 more galaxies which have UVIT FUV-only observations available. In order to determine the SFC ages, we will be using archival optical data and multi-band SED-fitting with CIGALE \citep{2019A&A...622A.103B}. This larger galaxy sample will enable the exploration of hierarchical star formation in more diverse galaxy environments. It is also expected to clarify the trends observed in this paper and potentially reveal new dependencies of hierarchy parameters on different host galaxy properties. Finally, studies of hierarchical star formation in extreme star formation environments such as mergers, tidal tails and galaxy outskirts are rare in the literature. Recently, Jayanth et al. (in prep.) studied the hierarchical distribution of SFCs in the morphologically disturbed and highly extended NGC 5291 system. Their study brings insights into how hierarchical star formation proceeds and evolves outside the normal disks of galaxies. We aim to further extend our exploration of the hierarchical star formation process onto the interesting environments offered by interacting galaxies.

\section*{acknowledgement}
GS and SS thank Dr. Alessandro Boselli and Dr. Carlo Schimd for their constructive suggestions during virtual meetings. GS and SS also acknowledge the PIs of the UVIT data used in this paper for observing these beautiful galaxies. SS acknowledges support from the Science and Engineering Research Board of India through the POWER research grant (SPG/2021/002672) and from the Alexander von Humboldt Foundation. S.H.M. acknowledges the support of NASA grant No. 80NSSC20K0500, NSF grant AST-2009679, and the Simons Foundation. C.M. acknowledges support from the National Science and Technology Council, Taiwan (grant NSTC 112-2112-M-001-027-MY3) and the Academia Sinica Investigator award (grant AS-IA-112-M04). This publication uses data from the UVIT, which is one of the key instruments on-board the AstroSat mission of the Indian Space Research Organisation (ISRO). The UVIT data is archived at the Indian Space Science Data Centre (ISSDC). We acknowledge the use of Python (\citealt{python09}), ASTROML (\citealt{2012cidu.conf...47V}), scikit-learn (\citealt{JMLR:v12:pedregosa11a}), Matplotlib (\citealt{Hunter07}), NumPy (\citealt{NumPy20}), SciPy (\citealt{SciPy20}), AstroPy (\citealt{astropy_2018}), Astrodendro (\href{http://www.dendrograms.org/}{http://www.dendrograms.org/)}, photutils (\citealt{larry_bradley_2024_10967176}) and CCDLAB (\citealt{2021JApA...42...30P}).


\appendix
\renewcommand{\thefigure}{\Alph{section}}

\section{Robustness of measured $l_{\rm corr}$ against variation in bin choices and fitting range}
\label{binning_effect}

\begin{figure*}[t]
    \centering
    \hfill
    \includegraphics[width=0.40\textwidth]{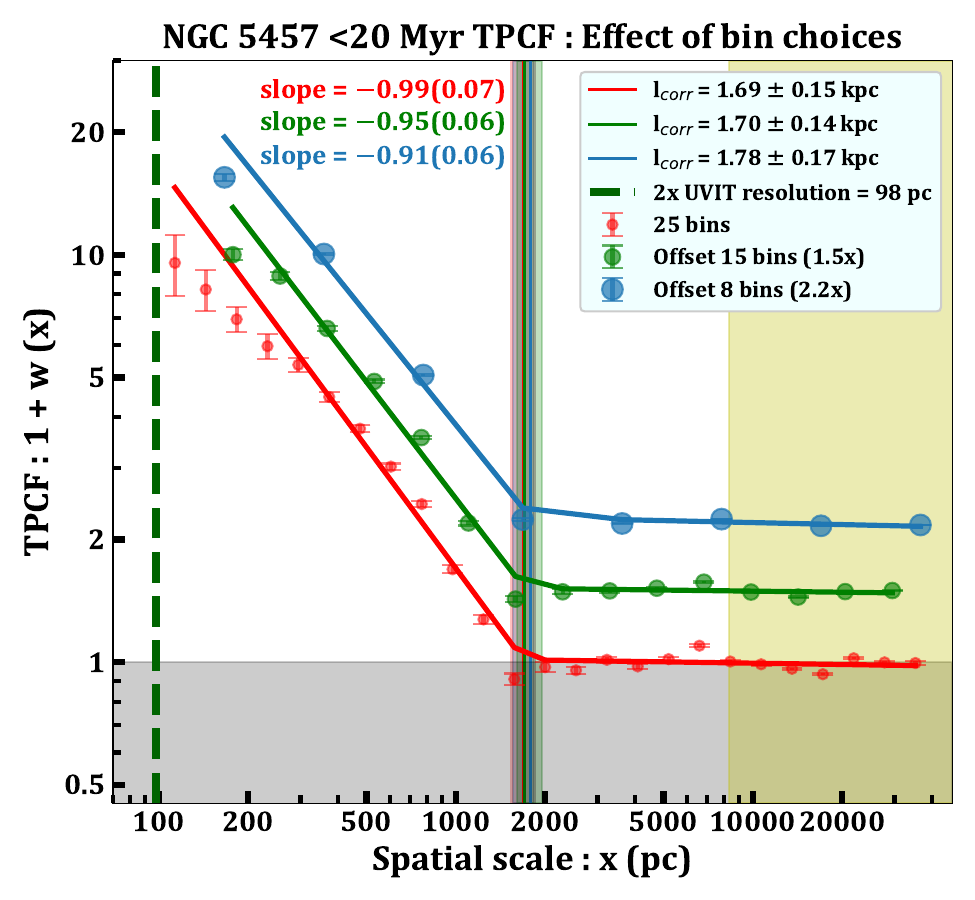}
    \hfill
    \includegraphics[width=0.40\textwidth]{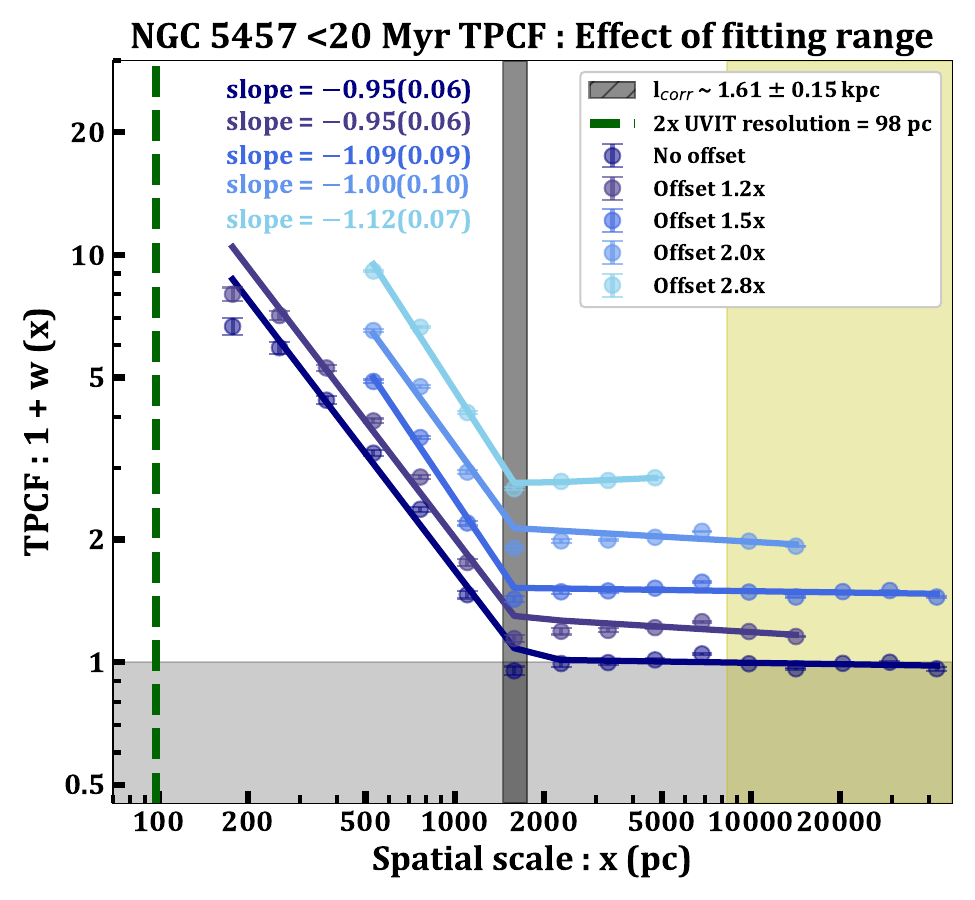}
    \hfill
    \hfill
    \caption{Tests demonstrating the robustness of the measured $l_{\rm corr}$ against a variation in the number of bins used to measured TPCF (left) and the range of TPCF values which are fit by the PW model (right). $<$20 Myr SFCs of NGC 5457 are chosen for this test. In the right-side plot, TPCF plots with different choices of fitting interval are plotted vertically offset from each other for clarity. See Appendix \ref{binning_effect} for details.}
    \label{fig:bin_n_fit}
\end{figure*}

To ensure that out $l_{\rm corr}$ measurement is robust against the number of bins over which TPCF is measured for a given spatial range, we conducted the following test using NGC 5457. For NGC 5457's $<$20 Myr SFCs, we performed the TPCF analysis with 25, 15 and 8 bins between $\sim$100 pc and $\sim$40000 pc. We measured the $l_{\rm corr}$ for each of the three runs and found that the derived $l_{\rm corr}$ and the TPCF slope (and thereby $D_2$) are consistent with each other. The observed TPCF plots are shown in Figure \ref{fig:bin_n_fit} (left), with different runs placed vertically offset from each other for clarity.

Moreover, to test whether $l_{\rm corr}$ measurement is robust against the choice of fitting interval, we artificially removed 3 to 6 points from either side of the TPCF break and fitted the PW model to the $<$20 Myr TPCF of NGC 5457. The results of this test are provided in Figure \ref{fig:bin_n_fit} (right) where we observe that the TPCF break for the different choices of fitting interval (plotted vertically offset from each other for clarity) lie in a slim band of $\sim$1.61 $\pm$ 0.15 kpc. This implies that reducing the number of observed TPCF points used to fit the PW model by three to six on either side of the expected break scale does not significantly change the derived $l_{\rm corr}$ value.


\section{Effect of the galaxy distances and spatial resolution on the measured correlation length values}
\label{appdx:dist_resolution_effect}

\begin{figure*}[!b]
    \centering
    \hfill
    \includegraphics[width=0.335\textwidth]{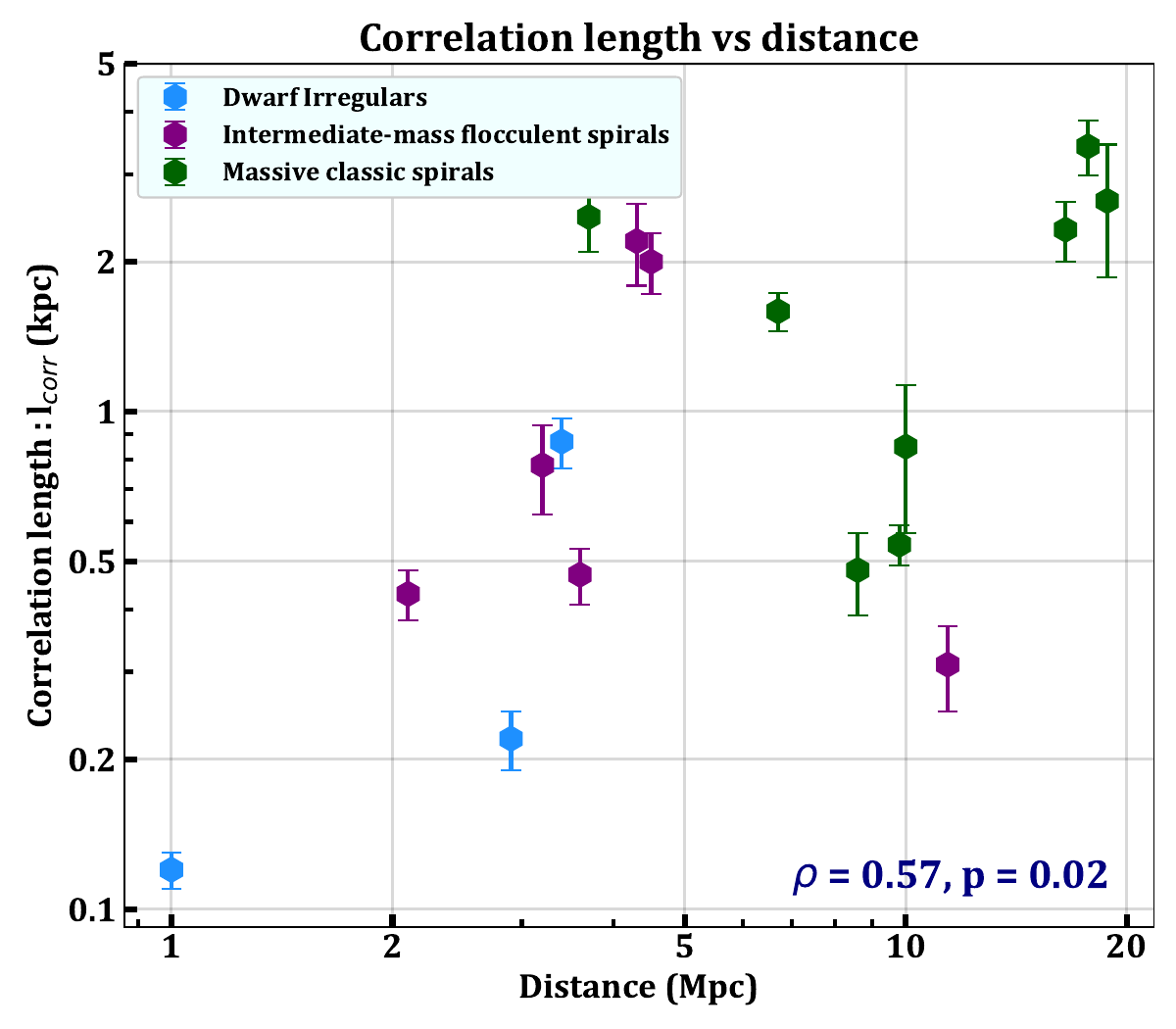}
    \hfill
    \includegraphics[width=0.315\textwidth]{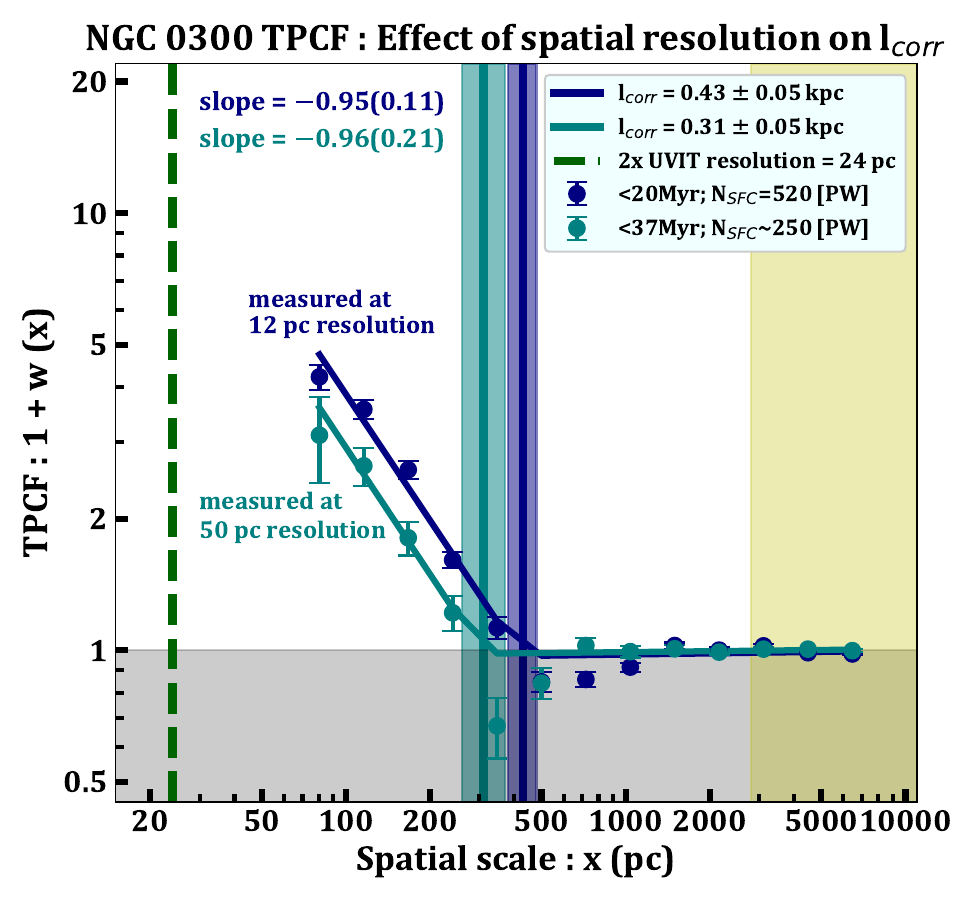}
    \hfill
    \includegraphics[width=0.315\textwidth]{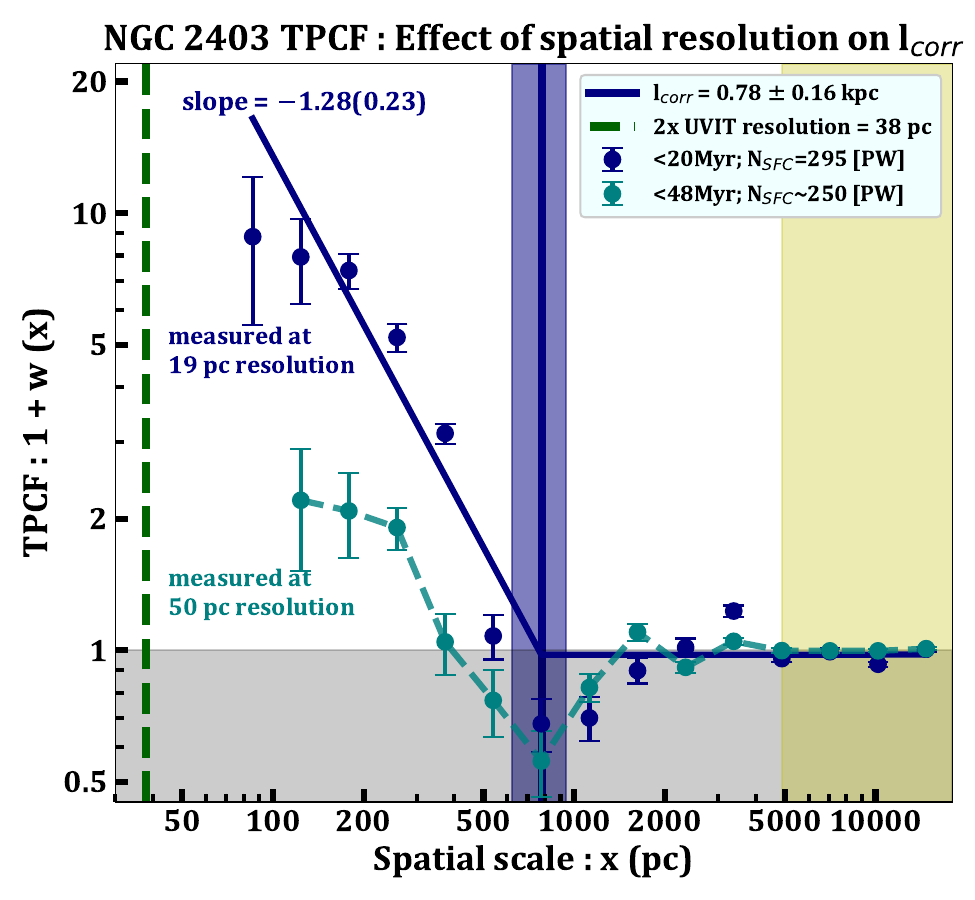}
    \hfill
    \label{fig:dist_and_resolution}
    \caption{Left : The relationship of $l_{\rm corr}$ with the galaxy distance indicates a strong positive correlation, which may be artificially driven by our sample-selection procedure. Middle and Right: Tests demonstrating the robustness of the measured $l_{\rm corr}$ against a variation in spatial resolution for NGC 0300 (middle) and NGC 2403 (right). The TPCF measured at the native resolution is given in navy and at 50 pc resolution is given in teal. See Appendix \ref{appdx:dist_resolution_effect} for details.}
\end{figure*}

Although, we observed that different morphologies occupy distinct regions in $l_{\rm corr}$$-$stellar mass relationship, this result may be affected by the distances of our galaxies. To test this effect, we plotted $l_{\rm corr}$ against galaxy distances and observed a spearman correlation coefficient $\rho$ = 0.57 and a p-value of 0.02 (Figure \ref{fig:dist_and_resolution}(left)). Though this indicates a statistically significant correlation, we believe this is primarily driven by the galaxy sample selection. The criteria of having sufficient ($\sim$250) young SFCs for a robust TPCF measurement limits our smaller size galaxies which naturally contain fewer number of UVIT-detectable SFCs to certain distances, i.e. dwarf irregular galaxies to within $\sim$4 Mpc and flocculent spirals to within $\sim$12 Mpc. The classic, massive spirals will often have sufficient UVIT-detectable SFCs owing to their large absolute sizes, but their large angular extent and UVIT’s 28\arcmin~diameter means that we can only have these sample galaxies at distance between 6 to 20 Mpc. These considerations influenced the selection of our galaxy sample. We aim to address the distance bias in future studies using a larger sample of $\sim$40$-$50 galaxies, with sufficient numbers of galaxies within each morphological subclass, across the 0–20 Mpc distance range. This will allow us to better assess the impact of galaxy distance on the derived hierarchy parameters.

In order to test the effect of spatial resolution on the derived $l_{\rm corr}$, we degraded the UVIT images of NGC 0300 and NGC 2403 from spatial resolutions of 12 and 19 pc, respectively, to a common resolution of 50 pc. This also tests the robustness of $l_{\rm corr}$ measurement against a 2.5 to 4 times increase in galaxy distance. These galaxies were chosen for this analysis because of their high exposure times, proximity and large number of SFCs detected at native resolution which should ensure a sufficient number of SFCs being detected at 50 pc resolution. We aimed to perform TPCF analysis on the \enquote{young} SFCs identified in these poorer resolution image and compare the derived $l_{\rm corr}$ at native and 50 pc spatial resolutions. In the 50 pc resolution images, we characterized the SFCs following a procedure similar to that described in Section \ref{sec:data}. We demonstrated in paper II that SFCs identified at poorer resolutions exhibit systematically older ages. Consequently, we did not have sufficient young SFCs to perform the TPCF analysis with a 20 Myr age cut - similar to the Step I$-$Case III galaxies. So, we performed the TPCF analysis on the youngest $\sim$250 SFCs characterized at 50 pc resolution, following the step II for case III galaxies. This corresponded to a 37 Myr age cut in NGC 0300 and 48 Myr age cut in NGC 2403. In Figure \ref{fig:dist_and_resolution} (middle), we show that the measured $l_{\rm corr}$ at 50 pc resolution and native 12 pc resolution for NGC 0300 are quite similar to each other. In Figure \ref{fig:dist_and_resolution} (right), we show that the young SFC TPCF measured at 50 pc resolution is quite noisy which cannot be adequately fitted by PW model. However, visual inspection of the plot hints that the TPCF break, possibly indicated by the TPCF dip, appears comparable to the l${\rm{corr}}$ measured at 19 pc resolution. This analysis suggests that for galaxies in which measurements are possible, the derived $l_{\rm corr}$ values are robust against a 2.5 to 4 factor variation in spatial resolution and distance.


\section{Correlations of $D_2$ with $M_\star$, SFR, $\Sigma$$_{SFR}$, $\Sigma$$_*$}\ 
\label{d2_non_correlations}

\begin{figure*}[t]
    \centering
    \hfill
    \includegraphics[width=0.40\textwidth]{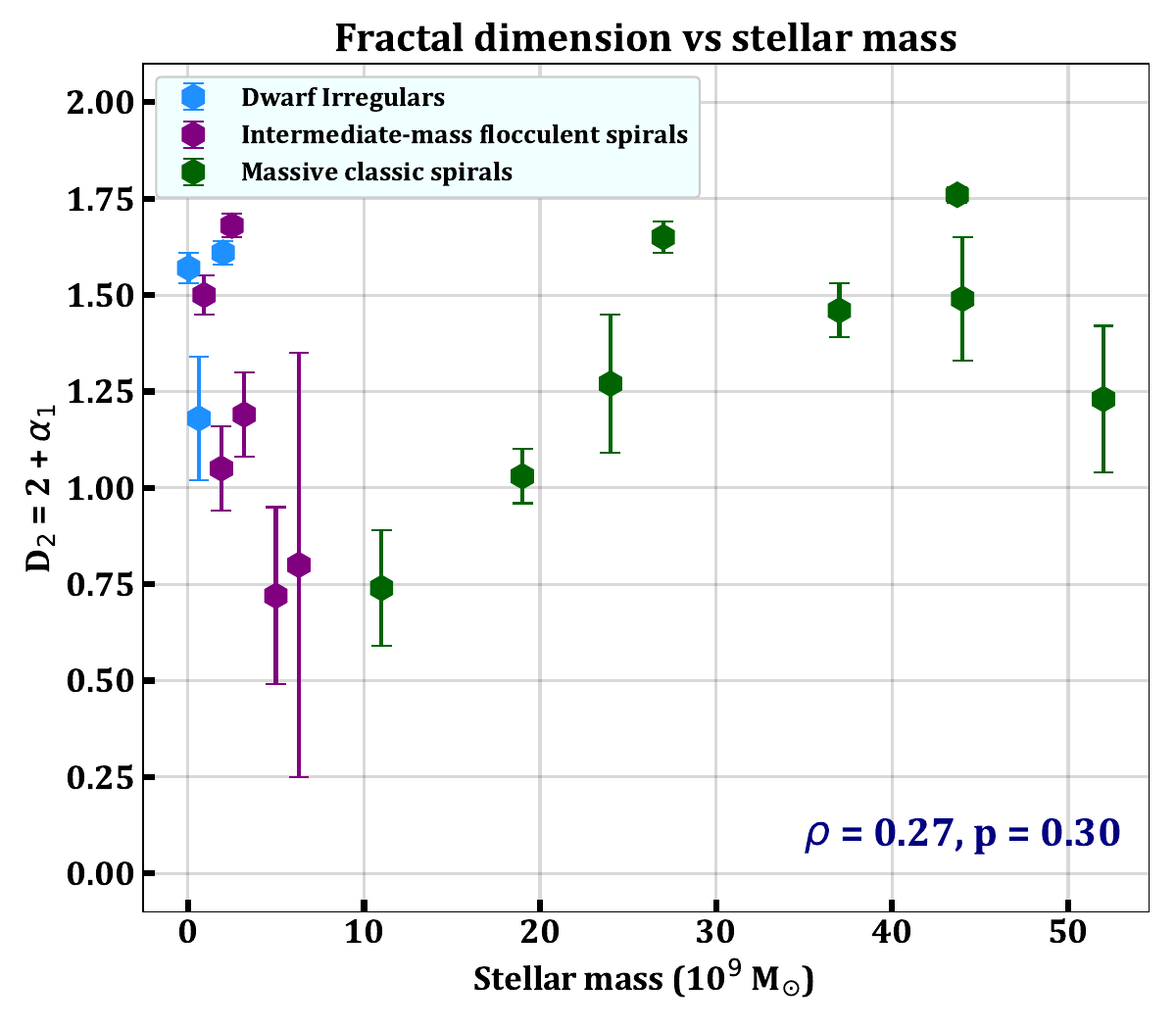}
    \hfill
    \includegraphics[width=0.40\textwidth]{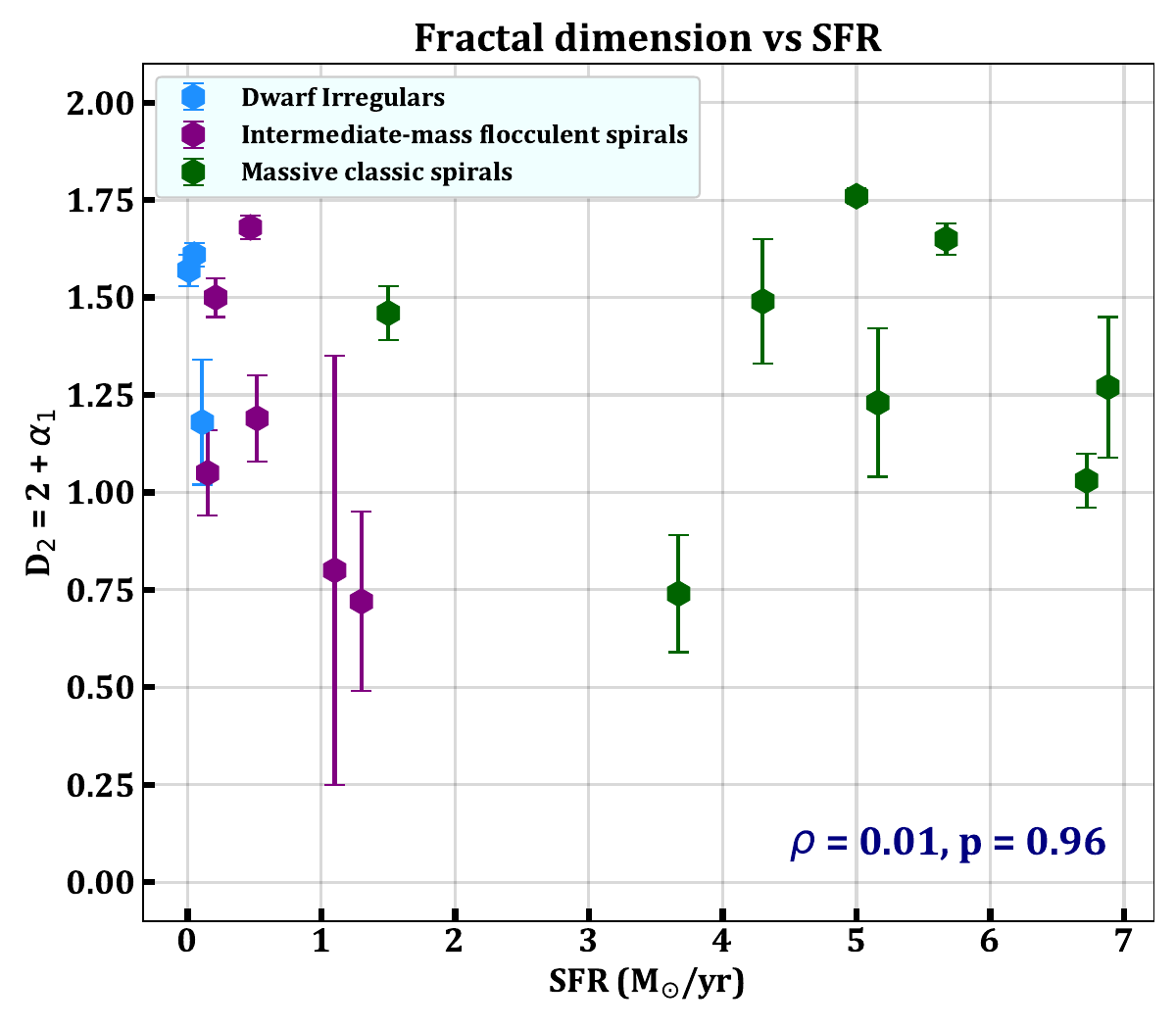}
    \hfill
    \hfill
    \vfill
    \hfill
    \includegraphics[width=0.40\textwidth]{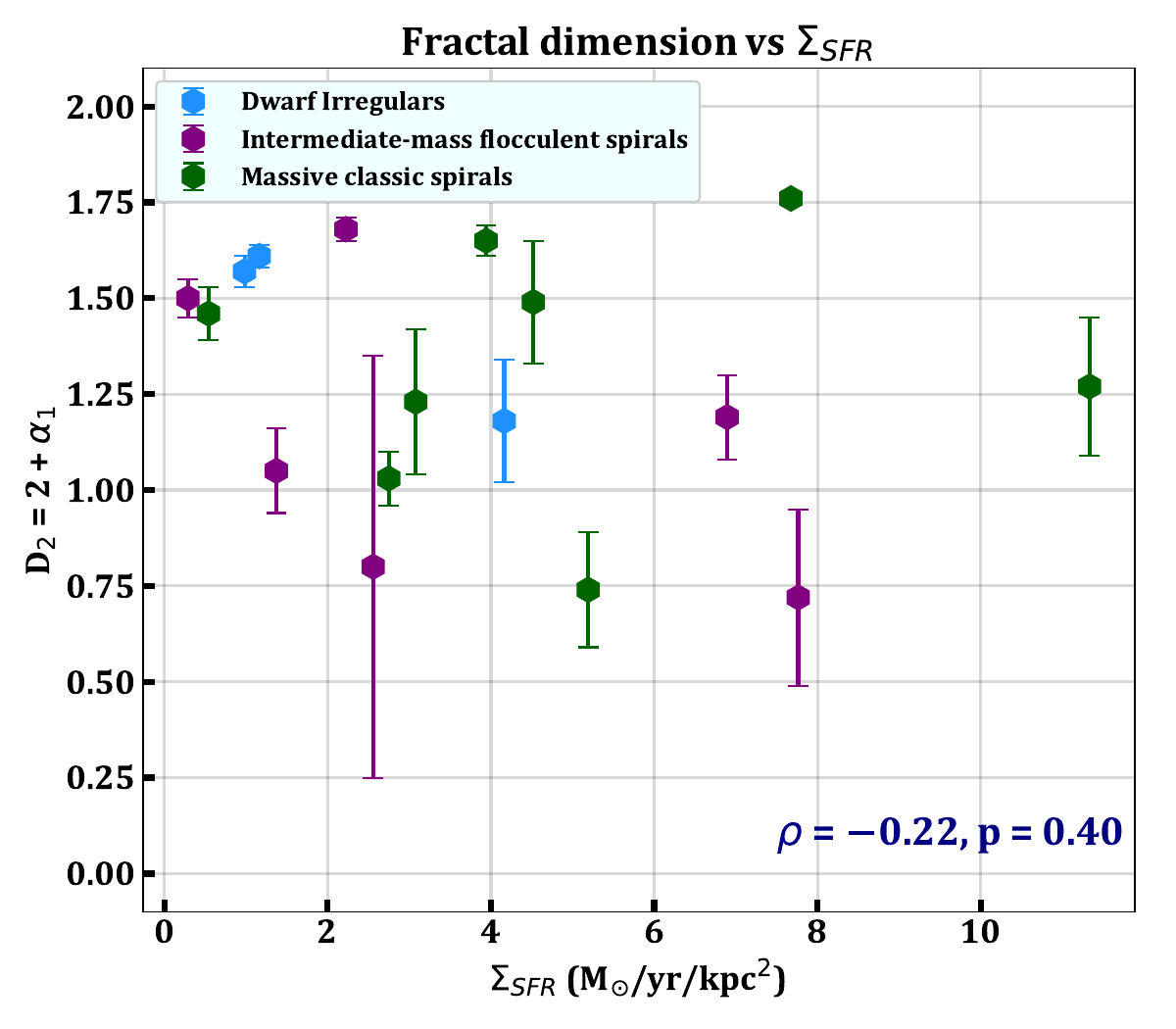}
    \hfill
    \includegraphics[width=0.40\textwidth]{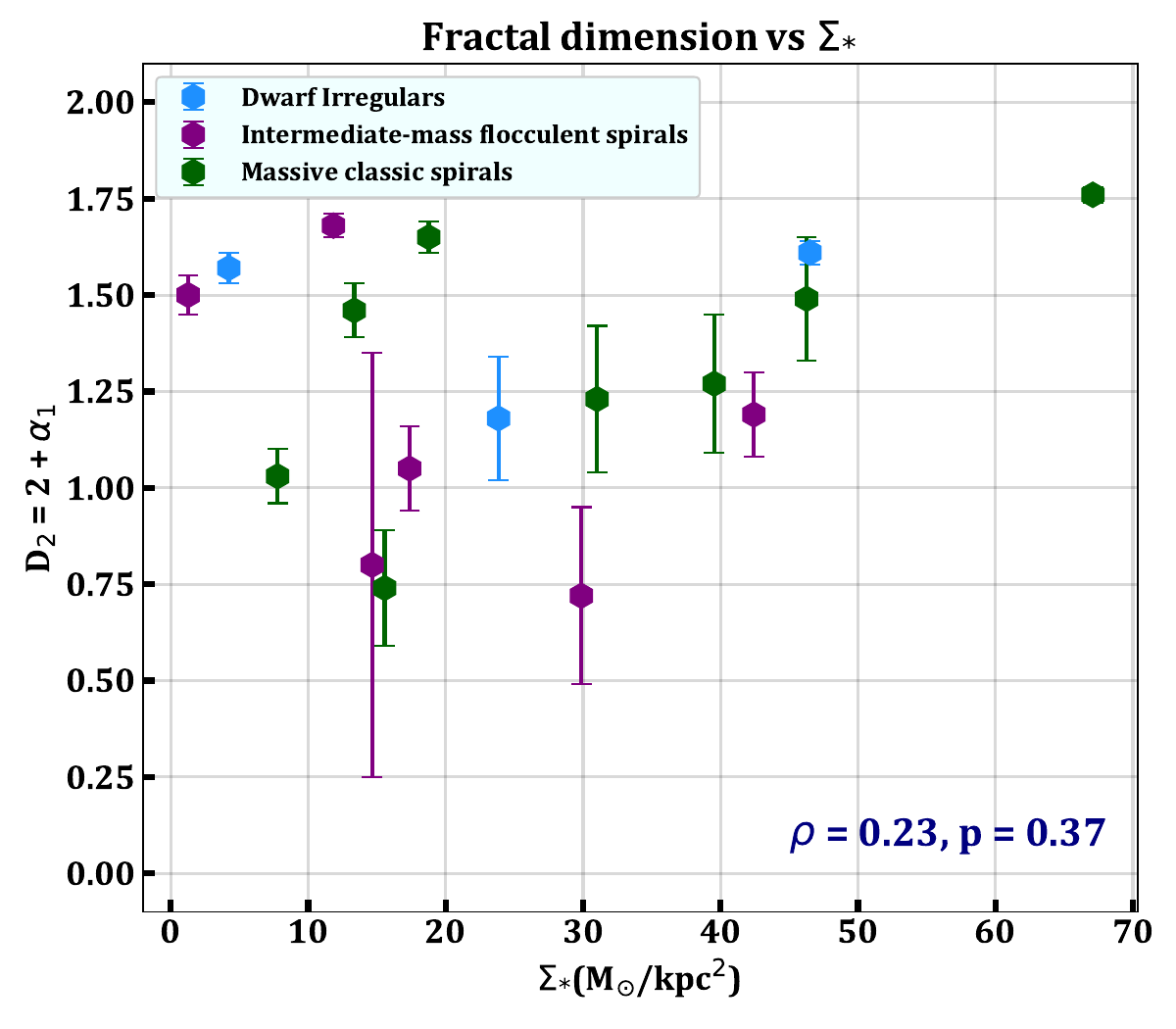}
    \hfill
    \hfill
    \caption{Relationships of$D_2$ with $M_\star$, SFR, $\Sigma$$_{SFR}$, $\Sigma$$_*$, respectively (left to right, top to bottom), showing no significant correlation.}
    \label{fig:D2_non_correlations}
\end{figure*}


We present the non-correlations of $D_2$ with stellar mass, SFR, SFRD ($\Sigma$$_{\rm{SFR}}$) and stellar mass surface density ($\Sigma$$_*$) in this section (see Figure \ref{fig:D2_non_correlations}). 

\bibliography{references3}

@BOOK{1980lssu.book.....P,
       author = {{Peebles}, P.~J.~E.},
        title = "{The large-scale structure of the universe}",
         year = 1980,
       adsurl = {https://ui.adsabs.harvard.edu/abs/1980lssu.book.....P}
}

@ARTICLE{2021MNRAS.507.5542M,
       author = {{Menon}, Shyam H. and {Grasha}, Kathryn and {Elmegreen}, Bruce G. and {Federrath}, Christoph and {Krumholz}, Mark R. and {Calzetti}, Daniela and {S{\'a}nchez}, N{\'e}stor and {Linden}, Sean T. and {Adamo}, Angela and {Messa}, Matteo and {Cook}, David O. and {Dale}, Daniel A. and {Grebel}, Eva K. and {Fumagalli}, Michele and {Sabbi}, Elena and {Johnson}, Kelsey E. and {Smith}, Linda J. and {Kennicutt}, Robert C.},
        title = "{The dependence of the hierarchical distribution of star clusters on galactic environment}",
      journal = {\mnras},
         year = 2021,
        month = nov,
       volume = {507},
       number = {4},
        pages = {5542-5566},
          doi = {10.1093/mnras/stab2413},
archivePrefix = {arXiv},
       eprint = {2108.04387},
 primaryClass = {astro-ph.GA},
       adsurl = {https://ui.adsabs.harvard.edu/abs/2021MNRAS.507.5542M}
}

@ARTICLE{2017ApJ...840..113G,
       author = {{Grasha}, K. and {Calzetti}, D. and {Adamo}, A. and {Kim}, H. and {Elmegreen}, B.~G. and {Gouliermis}, D.~A. and {Dale}, D.~A. and {Fumagalli}, M. and {Grebel}, E.~K. and {Johnson}, K.~E. and {Kahre}, L. and {Kennicutt}, R.~C. and {Messa}, M. and {Pellerin}, A. and {Ryon}, J.~E. and {Smith}, L.~J. and {Shabani}, F. and {Thilker}, D. and {Ubeda}, L.},
        title = "{The Hierarchical Distribution of the Young Stellar Clusters in Six Local Star-forming Galaxies}",
      journal = {\apj},
         year = 2017,
        month = may,
       volume = {840},
       number = {2},
          eid = {113},
        pages = {113},
          doi = {10.3847/1538-4357/aa6f15},
archivePrefix = {arXiv},
       eprint = {1704.06321},
 primaryClass = {astro-ph.GA},
       adsurl = {https://ui.adsabs.harvard.edu/abs/2017ApJ...840..113G}
}

@ARTICLE{2017ApJ...842...25G,
       author = {{Grasha}, K. and {Elmegreen}, B.~G. and {Calzetti}, D. and {Adamo}, A. and {Aloisi}, A. and {Bright}, S.~N. and {Cook}, D.~O. and {Dale}, D.~A. and {Fumagalli}, M. and {Gallagher}, III, J.~S. and {Gouliermis}, D.~A. and {Grebel}, E.~K. and {Kahre}, L. and {Kim}, H. and {Krumholz}, M.~R. and {Lee}, J.~C. and {Messa}, M. and {Ryon}, J.~E. and {Ubeda}, L.},
        title = "{Hierarchical Star Formation in Turbulent Media: Evidence from Young Star Clusters}",
      journal = {\apj},
         year = 2017,
        month = jun,
       volume = {842},
       number = {1},
          eid = {25},
        pages = {25},
          doi = {10.3847/1538-4357/aa740b},
archivePrefix = {arXiv},
       eprint = {1705.06281},
 primaryClass = {astro-ph.GA},
       adsurl = {https://ui.adsabs.harvard.edu/abs/2017ApJ...842...25G}
}

@ARTICLE{1998MNRAS.299..588E,
       author = {{Efremov}, Yuri N. and {Elmegreen}, Bruce G.},
        title = "{Hierarchical star formation from the time-space distribution of star clusters in the Large Magellanic Cloud}",
      journal = {\mnras},
         year = 1998,
        month = sep,
       volume = {299},
       number = {2},
        pages = {588-594},
          doi = {10.1046/j.1365-8711.1998.01819.x},
archivePrefix = {arXiv},
       eprint = {astro-ph/9805259},
 primaryClass = {astro-ph},
       adsurl = {https://ui.adsabs.harvard.edu/abs/1998MNRAS.299..588E}
}

@ARTICLE{1981MNRAS.194..809L,
       author = {{Larson}, R.~B.},
        title = "{Turbulence and star formation in molecular clouds.}",
      journal = {\mnras},
         year = 1981,
        month = mar,
       volume = {194},
        pages = {809-826},
          doi = {10.1093/mnras/194.4.809},
       adsurl = {https://ui.adsabs.harvard.edu/abs/1981MNRAS.194..809L}
}

@ARTICLE{1964ApJ...139.1217T,
       author = {{Toomre}, A.},
        title = "{On the gravitational stability of a disk of stars.}",
      journal = {\apj},
         year = 1964,
        month = may,
       volume = {139},
        pages = {1217-1238},
          doi = {10.1086/147861},
       adsurl = {https://ui.adsabs.harvard.edu/abs/1964ApJ...139.1217T}
}

@ARTICLE{2008ApJ...685L..31E,
       author = {{Escala}, Andr{\'e}s and {Larson}, Richard B.},
        title = "{Stability of Galactic Gas Disks and the Formation of Massive Clusters}",
      journal = {\apjl},
         year = 2008,
        month = sep,
       volume = {685},
       number = {1},
        pages = {L31},
          doi = {10.1086/592271},
archivePrefix = {arXiv},
       eprint = {0806.0853},
 primaryClass = {astro-ph},
       adsurl = {https://ui.adsabs.harvard.edu/abs/2008ApJ...685L..31E}
}

@ARTICLE{2004ARA&A..42..211E,
       author = {{Elmegreen}, Bruce G. and {Scalo}, John},
        title = "{Interstellar Turbulence I: Observations and Processes}",
      journal = {\araa},
         year = 2004,
        month = sep,
       volume = {42},
       number = {1},
        pages = {211-273},
          doi = {10.1146/annurev.astro.41.011802.094859},
archivePrefix = {arXiv},
       eprint = {astro-ph/0404451},
 primaryClass = {astro-ph},
       adsurl = {https://ui.adsabs.harvard.edu/abs/2004ARA&A..42..211E}
}

@ARTICLE{2004RvMP...76..125M,
       author = {{Mac Low}, Mordecai-Mark and {Klessen}, Ralf S.},
        title = "{Control of star formation by supersonic turbulence}",
      journal = {Reviews of Modern Physics},
         year = 2004,
        month = jan,
       volume = {76},
       number = {1},
        pages = {125-194},
          doi = {10.1103/RevModPhys.76.125},
archivePrefix = {arXiv},
       eprint = {astro-ph/0301093},
 primaryClass = {astro-ph},
       adsurl = {https://ui.adsabs.harvard.edu/abs/2004RvMP...76..125M}
}

@ARTICLE{2013MNRAS.436.1245F,
       author = {{Federrath}, Christoph},
        title = "{On the universality of supersonic turbulence}",
      journal = {\mnras},
         year = 2013,
        month = dec,
       volume = {436},
       number = {2},
        pages = {1245-1257},
          doi = {10.1093/mnras/stt1644},
archivePrefix = {arXiv},
       eprint = {1306.3989},
 primaryClass = {astro-ph.SR},
       adsurl = {https://ui.adsabs.harvard.edu/abs/2013MNRAS.436.1245F}
}

@ARTICLE{2010ApJ...720..541S,
       author = {{S{\'a}nchez}, N{\'e}stor and {A{\~n}ez}, Neyda and {Alfaro}, Emilio J. and {Crone Odekon}, Mary},
        title = "{The Fractal Dimension of Star-forming Regions at Different Spatial Scales in M33}",
      journal = {\apj},
         year = 2010,
        month = sep,
       volume = {720},
       number = {1},
        pages = {541-547},
          doi = {10.1088/0004-637X/720/1/541},
archivePrefix = {arXiv},
       eprint = {1007.3621},
 primaryClass = {astro-ph.CO},
       adsurl = {https://ui.adsabs.harvard.edu/abs/2010ApJ...720..541S}
}

@ARTICLE{Sun_2017,
       author = {{Sun}, Ning-Chen and {de Grijs}, Richard and {Subramanian}, Smitha and {Cioni}, Maria-Rosa L. and {Rubele}, Stefano and {Bekki}, Kenji and {Ivanov}, Valentin D. and {Piatti}, Andr{\'e}s E. and {Ripepi}, Vincenzo},
        title = "{The VMC Survey. XXII. Hierarchical Star Formation in the 30 Doradus-N158-N159-N160 Star-forming Complex}",
      journal = {\apj},
         year = 2017,
        month = feb,
       volume = {835},
       number = {2},
          eid = {171},
        pages = {171},
          doi = {10.3847/1538-4357/835/2/171},
archivePrefix = {arXiv},
       eprint = {1611.06508},
 primaryClass = {astro-ph.GA},
       adsurl = {https://ui.adsabs.harvard.edu/abs/2017ApJ...835..171S}
}

@ARTICLE{Rodriguez_2020,
       author = {{Rodr{\'\i}guez}, M.~J. and {Baume}, G. and {Feinstein}, C.},
        title = "{Hierarchical star formation in nearby galaxies}",
      journal = {\aap},
         year = 2020,
        month = dec,
       volume = {644},
          eid = {A101},
        pages = {A101},
          doi = {10.1051/0004-6361/202038970},
archivePrefix = {arXiv},
       eprint = {2010.14419},
 primaryClass = {astro-ph.GA},
       adsurl = {https://ui.adsabs.harvard.edu/abs/2020A&A...644A.101R}
}

@ARTICLE{Sanchez_2008,
       author = {{S{\'a}nchez}, N{\'e}stor and {Alfaro}, Emilio J.},
        title = "{The Fractal Distribution of H II Regions in Disk Galaxies}",
      journal = {\apjs},
         year = 2008,
        month = sep,
       volume = {178},
       number = {1},
        pages = {1-19},
          doi = {10.1086/589653},
archivePrefix = {arXiv},
       eprint = {0804.4554},
 primaryClass = {astro-ph},
       adsurl = {https://ui.adsabs.harvard.edu/abs/2008ApJS..178....1S}
}

@ARTICLE{Grasha_2018,
       author = {{Grasha}, K. and {Calzetti}, D. and {Bittle}, L. and {Johnson}, K.~E. and {Donovan Meyer}, J. and {Kennicutt}, R.~C. and {Elmegreen}, B.~G. and {Adamo}, A. and {Krumholz}, M.~R. and {Fumagalli}, M. and {Grebel}, E.~K. and {Gouliermis}, D.~A. and {Cook}, D.~O. and {Gallagher}, J.~S. and {Aloisi}, A. and {Dale}, D.~A. and {Linden}, S. and {Sacchi}, E. and {Thilker}, D.~A. and {Walterbos}, R.~A.~M. and {Messa}, M. and {Wofford}, A. and {Smith}, L.~J.},
        title = "{Connecting young star clusters to CO molecular gas in NGC 7793 with ALMA-LEGUS}",
      journal = {\mnras},
         year = 2018,
        month = nov,
       volume = {481},
       number = {1},
        pages = {1016-1027},
          doi = {10.1093/mnras/sty2154},
archivePrefix = {arXiv},
       eprint = {1808.02496},
 primaryClass = {astro-ph.GA},
       adsurl = {https://ui.adsabs.harvard.edu/abs/2018MNRAS.481.1016G}
}

@ARTICLE{Zhang_2001,
       author = {{Zhang}, Qing and {Fall}, S. Michael and {Whitmore}, Bradley C.},
        title = "{A Multiwavelength Study of the Young Star Clusters and Interstellar Medium in the Antennae Galaxies}",
      journal = {\apj},
         year = 2001,
        month = nov,
       volume = {561},
       number = {2},
        pages = {727-750},
          doi = {10.1086/322278},
archivePrefix = {arXiv},
       eprint = {astro-ph/0105174},
 primaryClass = {astro-ph},
       adsurl = {https://ui.adsabs.harvard.edu/abs/2001ApJ...561..727Z}
}

@ARTICLE{Chevance_2020,
       author = {{Chevance}, M{\'e}lanie and {Kruijssen}, J.~M. Diederik and {Hygate}, Alexander P.~S. and {Schruba}, Andreas and {Longmore}, Steven N. and {Groves}, Brent and {Henshaw}, Jonathan D. and {Herrera}, Cinthya N. and {Hughes}, Annie and {Jeffreson}, Sarah M.~R. and {Lang}, Philipp and {Leroy}, Adam K. and {Meidt}, Sharon E. and {Pety}, J{\'e}r{\^o}me and {Razza}, Alessandro and {Rosolowsky}, Erik and {Schinnerer}, Eva and {Bigiel}, Frank and {Blanc}, Guillermo A. and {Emsellem}, Eric and {Faesi}, Christopher M. and {Glover}, Simon C.~O. and {Haydon}, Daniel T. and {Ho}, I. -Ting and {Kreckel}, Kathryn and {Lee}, Janice C. and {Liu}, Daizhong and {Querejeta}, Miguel and {Saito}, Toshiki and {Sun}, Jiayi and {Usero}, Antonio and {Utomo}, Dyas},
        title = "{The lifecycle of molecular clouds in nearby star-forming disc galaxies}",
      journal = {\mnras},
         year = 2020,
        month = apr,
       volume = {493},
       number = {2},
        pages = {2872-2909},
          doi = {10.1093/mnras/stz3525},
archivePrefix = {arXiv},
       eprint = {1911.03479},
 primaryClass = {astro-ph.GA},
       adsurl = {https://ui.adsabs.harvard.edu/abs/2020MNRAS.493.2872C}
}

@ARTICLE{Shadmehri_2011,
       author = {{Shadmehri}, Mohsen and {Elmegreen}, Bruce G.},
        title = "{Mass functions in fractal clouds: the role of cloud structure in the stellar initial mass function}",
      journal = {\mnras},
         year = 2011,
        month = jan,
       volume = {410},
       number = {2},
        pages = {788-804},
          doi = {10.1111/j.1365-2966.2010.17481.x},
archivePrefix = {arXiv},
       eprint = {1008.1218},
 primaryClass = {astro-ph.GA},
       adsurl = {https://ui.adsabs.harvard.edu/abs/2011MNRAS.410..788S}
}

@ARTICLE{Sanchez_2005,
       author = {{S{\'a}nchez}, N{\'e}stor and {Alfaro}, Emilio J. and {P{\'e}rez}, Enrique},
        title = "{The Fractal Dimension of Projected Clouds}",
      journal = {\apj},
         year = 2005,
        month = jun,
       volume = {625},
       number = {2},
        pages = {849-856},
          doi = {10.1086/429553},
archivePrefix = {arXiv},
       eprint = {astro-ph/0501573},
 primaryClass = {astro-ph},
       adsurl = {https://ui.adsabs.harvard.edu/abs/2005ApJ...625..849S}
}

@INPROCEEDINGS{2012SPIE.8443E..1NK,
       author = {{Kumar}, Amit and {Ghosh}, S.~K. and {Hutchings}, J. and {Kamath}, P.~U. and {Kathiravan}, S. and {Mahesh}, P.~K. and {Murthy}, J. and {Nagbhushana}, S. and {Pati}, A.~K. and {Rao}, M.~N. and {Rao}, N.~K. and {Sriram}, S. and {Tandon}, S.~N.},
        title = "{Ultra Violet Imaging Telescope (UVIT) on ASTROSAT}",
    booktitle = {Space Telescopes and Instrumentation 2012: Ultraviolet to Gamma Ray},
         year = 2012,
       editor = {{Takahashi}, Tadayuki and {Murray}, Stephen S. and {den Herder}, Jan-Willem A.},
       series = {Society of Photo-Optical Instrumentation Engineers (SPIE) Conference Series},
       volume = {8443},
        month = sep,
          eid = {84431N},
        pages = {84431N},
          doi = {10.1117/12.924507},
archivePrefix = {arXiv},
       eprint = {1208.4670},
 primaryClass = {astro-ph.IM},
       adsurl = {https://ui.adsabs.harvard.edu/abs/2012SPIE.8443E..1NK}
}

@ARTICLE{NumPy20,
  author  = {Harris, Charles R. and Millman, K. Jarrod and van der Walt, Stéfan J and Gommers, Ralf and Virtanen, Pauli and Cournapeau, David and Wieser, Eric and Taylor, Julian and Berg, Sebastian and Smith, Nathaniel J. and Kern, Robert and Picus, Matti and Hoyer, Stephan and van Kerkwijk, Marten H. and Brett, Matthew and Haldane, Allan and Fernández del Río, Jaime and Wiebe, Mark and Peterson, Pearu and Gérard-Marchant, Pierre and Sheppard, Kevin and Reddy, Tyler and Weckesser, Warren and Abbasi, Hameer and Gohlke, Christoph and Oliphant, Travis E.},
  title   = {Array programming with {NumPy}},
  journal = {Nature},
  year    = {2020},
  volume  = {585},
  pages   = {357–362},
  doi     = {10.1038/s41586-020-2649-2}
}

@ARTICLE{1998ApJ...498..541K,
       author = {{Kennicutt}, Robert C., Jr.},
        title = "{The Global Schmidt Law in Star-forming Galaxies}",
      journal = {\apj},
         year = 1998,
        month = may,
       volume = {498},
       number = {2},
        pages = {541-552},
          doi = {10.1086/305588},
archivePrefix = {arXiv},
       eprint = {astro-ph/9712213},
 primaryClass = {astro-ph},
       adsurl = {https://ui.adsabs.harvard.edu/abs/1998ApJ...498..541K}
}

@INPROCEEDINGS{2012cidu.conf...47V,
       author = {{VanderPlas}, J. and {Connolly}, A.~J. and {Ivezic}, Z. and {Gray}, A.},
        title = "{Introduction to astroML: Machine learning for astrophysics}",
    booktitle = {Proceedings of Conference on Intelligent Data Understanding (CIDU},
         year = 2012,
        month = oct,
        pages = {47-54},
          doi = {10.1109/CIDU.2012.6382200},
archivePrefix = {arXiv},
       eprint = {1411.5039},
 primaryClass = {astro-ph.IM},
       adsurl = {https://ui.adsabs.harvard.edu/abs/2012cidu.conf...47V}
}

@software{larry_bradley_2024_10967176,
  author       = {Larry Bradley and
                  Brigitta Sip{\H o}cz and
                  Thomas Robitaille and
                  Erik Tollerud and
                  Z\`e Vin{\'{\i}}cius and
                  Christoph Deil and
                  Kyle Barbary and
                  Tom J Wilson and
                  Ivo Busko and
                  Axel Donath and
                  Hans Moritz G{\"u}nther and
                  Mihai Cara and
                  P. L. Lim and
                  Sebastian Me{\ss}linger and
                  Zach Burnett and
                  Simon Conseil and
                  Michael Droettboom and
                  Azalee Bostroem and
                  E. M. Bray and
                  Lars Andersen Bratholm and
                  William Jamieson and
                  Adam Ginsburg and
                  Geert Barentsen and
                  Matt Craig and
                  Sergio Pascual and
                  Shivangee Rathi and
                  Marshall Perrin and
                  Brett M. Morris and
                  Gabriel Perren},
  title        = {astropy/photutils: 1.12.0},
  month        = apr,
  year         = 2024,
  publisher    = {Zenodo},
  version      = {1.12.0},
  doi          = {10.5281/zenodo.10967176},
  url          = {https://doi.org/10.5281/zenodo.10967176}
}

@ARTICLE{deBlok_2008,
       author = {{de Blok}, W.~J.~G. and {Walter}, F. and {Brinks}, E. and {Trachternach}, C. and {Oh}, S. -H. and {Kennicutt}, R.~C., Jr.},
        title = "{High-Resolution Rotation Curves and Galaxy Mass Models from THINGS}",
      journal = {\aj},
         year = 2008,
        month = dec,
       volume = {136},
       number = {6},
        pages = {2648-2719},
          doi = {10.1088/0004-6256/136/6/2648},
archivePrefix = {arXiv},
       eprint = {0810.2100},
 primaryClass = {astro-ph},
       adsurl = {https://ui.adsabs.harvard.edu/abs/2008AJ....136.2648D}
}

@ARTICLE{2009MNRAS.397L..60D,
       author = {{Dutta}, Prasun and {Begum}, Ayesha and {Bharadwaj}, Somnath and {Chengalur}, Jayaram N.},
        title = "{The scaleheight of NGC 1058 measured from its HI power spectrum}",
      journal = {\mnras},
         year = 2009,
        month = jul,
       volume = {397},
       number = {1},
        pages = {L60-L63},
          doi = {10.1111/j.1745-3933.2009.00684.x},
archivePrefix = {arXiv},
       eprint = {0905.1450},
 primaryClass = {astro-ph.GA},
       adsurl = {https://ui.adsabs.harvard.edu/abs/2009MNRAS.397L..60D}
}

@ARTICLE{2023ApJ...944L..18M,
       author = {{Meidt}, Sharon E. and {Rosolowsky}, Erik and {Sun}, Jiayi and {Koch}, Eric W. and {Klessen}, Ralf S. and {Leroy}, Adam K. and {Schinnerer}, Eva and {Barnes}, Ashley. T. and {Glover}, Simon C.~O. and {Lee}, Janice C. and {van der Wel}, Arjen and {Watkins}, Elizabeth J. and {Williams}, Thomas G. and {Bigiel}, F. and {Boquien}, M{\'e}d{\'e}ric and {Blanc}, Guillermo A. and {Cao}, Yixian and {Chevance}, M{\'e}lanie and {Dale}, Daniel A. and {Egorov}, Oleg V. and {Emsellem}, Eric and {Grasha}, Kathryn and {Henshaw}, Jonathan D. and {Kruijssen}, J.~M. Diederik and {Larson}, Kirsten L. and {Liu}, Daizhong and {Murphy}, Eric J. and {Pety}, J{\'e}r{\^o}me and {Querejeta}, Miguel and {Saito}, Toshiki and {Sandstrom}, Karin M. and {Smith}, Rowan J. and {Sormani}, Mattia C. and {Thilker}, David A.},
        title = "{PHANGS-JWST First Results: Interstellar Medium Structure on the Turbulent Jeans Scale in Four Disk Galaxies Observed by JWST and the Atacama Large Millimeter/submillimeter Array}",
      journal = {\apjl},
         year = 2023,
        month = feb,
       volume = {944},
       number = {2},
          eid = {L18},
        pages = {L18},
          doi = {10.3847/2041-8213/acaaa8},
archivePrefix = {arXiv},
       eprint = {2212.06434},
 primaryClass = {astro-ph.GA},
       adsurl = {https://ui.adsabs.harvard.edu/abs/2023ApJ...944L..18M}
}

@ARTICLE{1991AnRFM..23..539S,
       author = {{Sreenivasan}, K.~R.},
        title = "{Fractals and multifractals in fluid turbulence}",
      journal = {Annual Review of Fluid Mechanics},
         year = 1991,
        month = jan,
       volume = {23},
        pages = {539-600},
          doi = {10.1146/annurev.fl.23.010191.002543},
       adsurl = {https://ui.adsabs.harvard.edu/abs/1991AnRFM..23..539S}
}

@ARTICLE{2003ApJ...590..271E,
       author = {{Elmegreen}, Bruce G. and {Elmegreen}, Debra Meloy and {Leitner}, Samuel N.},
        title = "{A Turbulent Origin for Flocculent Spiral Structure in Galaxies}",
      journal = {\apj},
         year = 2003,
        month = jun,
       volume = {590},
       number = {1},
        pages = {271-283},
          doi = {10.1086/374860},
archivePrefix = {arXiv},
       eprint = {astro-ph/0305049},
 primaryClass = {astro-ph},
       adsurl = {https://ui.adsabs.harvard.edu/abs/2003ApJ...590..271E}
}

@ARTICLE{2008MNRAS.384L..34D,
       author = {{Dutta}, Prasun and {Begum}, Ayesha and {Bharadwaj}, Somnath and {Chengalur}, Jayaram N.},
        title = "{HI power spectrum of the spiral galaxy NGC628}",
      journal = {\mnras},
         year = 2008,
        month = feb,
       volume = {384},
       number = {1},
        pages = {L34-L37},
          doi = {10.1111/j.1745-3933.2007.00417.x},
archivePrefix = {arXiv},
       eprint = {0711.1234},
 primaryClass = {astro-ph},
       adsurl = {https://ui.adsabs.harvard.edu/abs/2008MNRAS.384L..34D}
}

@ARTICLE{2020MNRAS.493.4643M,
       author = {{Menon}, Shyam H. and {Federrath}, Christoph and {Kuiper}, Rolf},
        title = "{On the turbulence driving mode of expanding H II regions}",
      journal = {\mnras},
         year = 2020,
        month = apr,
       volume = {493},
       number = {4},
        pages = {4643-4656},
          doi = {10.1093/mnras/staa580},
archivePrefix = {arXiv},
       eprint = {2002.08707},
 primaryClass = {astro-ph.GA},
       adsurl = {https://ui.adsabs.harvard.edu/abs/2020MNRAS.493.4643M}
}

@ARTICLE{Guelin_1970,
       author = {{Gu{\'e}lin}, M. and {Weliachew}, L.},
        title = "{A neutral hydrogen study of the spiral galaxy NGC 5457.}",
      journal = {\aap},
         year = 1970,
        month = jul,
       volume = {7},
        pages = {141-149},
       adsurl = {https://ui.adsabs.harvard.edu/abs/1970A&A.....7..141G}
}

@ARTICLE{2019MNRAS.487.2797E,
       author = {{Elagali}, A. and {Staveley-Smith}, L. and {Rhee}, J. and {Wong}, O.~I. and {Bosma}, A. and {Westmeier}, T. and {Koribalski}, B.~S. and {Heald}, G. and {For}, B. -Q. and {Kleiner}, D. and {Lee-Waddell}, K. and {Madrid}, J.~P. and {Popping}, A. and {Reynolds}, T.~N. and {Meyer}, M.~J. and {Allison}, J.~R. and {Lagos}, C.~D.~P. and {Voronkov}, M.~A. and {Serra}, P. and {Shao}, L. and {Wang}, J. and {Anderson}, C.~S. and {Bunton}, J.~D. and {Bekiaris}, G. and {Walsh}, W.~M. and {Kilborn}, V.~A. and {Kamphuis}, P. and {Oh}, S. -H.},
        title = "{WALLABY early science - III. An H I study of the spiral galaxy NGC 1566}",
      journal = {\mnras},
         year = 2019,
        month = aug,
       volume = {487},
       number = {2},
        pages = {2797-2817},
          doi = {10.1093/mnras/stz1448},
archivePrefix = {arXiv},
       eprint = {1905.09491},
 primaryClass = {astro-ph.GA},
       adsurl = {https://ui.adsabs.harvard.edu/abs/2019MNRAS.487.2797E}
}

@ARTICLE{2000ApJ...530..277E,
       author = {{Elmegreen}, Bruce G.},
        title = "{Star Formation in a Crossing Time}",
      journal = {\apj},
         year = 2000,
        month = feb,
       volume = {530},
       number = {1},
        pages = {277-281},
          doi = {10.1086/308361},
archivePrefix = {arXiv},
       eprint = {astro-ph/9911172},
 primaryClass = {astro-ph},
       adsurl = {https://ui.adsabs.harvard.edu/abs/2000ApJ...530..277E}
}

@ARTICLE{2019ARA&A..57..227K,
       author = {{Krumholz}, Mark R. and {McKee}, Christopher F. and {Bland-Hawthorn}, Joss},
        title = "{Star Clusters Across Cosmic Time}",
      journal = {\araa},
         year = 2019,
        month = aug,
       volume = {57},
        pages = {227-303},
          doi = {10.1146/annurev-astro-091918-104430},
archivePrefix = {arXiv},
       eprint = {1812.01615},
 primaryClass = {astro-ph.GA},
       adsurl = {https://ui.adsabs.harvard.edu/abs/2019ARA&A..57..227K}
}

@article{shashank2025tracing,
  title={Tracing hierarchical star formation out to kiloparsec scales in nearby spiral galaxies with UVIT},
  author={Shashank, Gairola and Subramanian, Smitha and Muraleedharan, Sreedevi and Menon, Shyam H and Mondal, Chayan and Krishna, Sriram and Das, Mousumi and Subramaniam, Annapurni},
  journal={Astronomy \& Astrophysics},
  volume={693},
  pages={A188},
  year={2025},
  publisher={EDP Sciences}
}

@article{elmegreen2014hierarchical,
  title={Hierarchical Star Formation in Nearby LEGUS Galaxies},
  author={Elmegreen, Debra Meloy and Elmegreen, Bruce G and Adamo, Angela and Aloisi, Alessandra and Andrews, Jennifer and Annibali, Francesca and Bright, Stacey N and Calzetti, Daniela and Cignoni, Michele and Evans, Aaron S and others},
  journal={The Astrophysical Journal Letters},
  volume={787},
  number={1},
  pages={L15},
  year={2014},
  publisher={IOP Publishing}
}

@article{rosolowsky2008structural, 
  title={Structural analysis of molecular clouds: dendrograms},
  author={Rosolowsky, EW and Pineda, JE and Kauffmann, J and Goodman, AA},
  journal={The Astrophysical Journal},
  volume={679},
  number={2},
  pages={1338},
  year={2008},
  publisher={IOP Publishing}
}

@article{leitherer2014effects,
  title={The effects of stellar rotation. II. A comprehensive set of Starburst99 models},
  author={Leitherer, Claus and Ekstr{\"o}m, Sylvia and Meynet, Georges and Schaerer, Daniel and Agienko, Katerina B and Levesque, Emily M},
  journal={The Astrophysical Journal Supplement Series},
  volume={212},
  number={1},
  pages={14},
  year={2014},
  publisher={IOP Publishing}
}

@article{leitherer1999starburst99,
  title={Starburst99: synthesis models for galaxies with active star formation},
  author={Leitherer, Claus and Schaerer, Daniel and Goldader, Jeffrey D and Delgado, Rosa M Gonz{\'a}lez and Robert, Carmelle and Kune, Denis Foo and De Mello, Du{\'\i}lia F and Devost, Daniel and Heckman, Timothy M},
  journal={The Astrophysical Journal Supplement Series},
  volume={123},
  number={1},
  pages={3},
  year={1999},
  publisher={IOP Publishing}
}

@article{elmegreen2006hierarchical,
  title={Hierarchical star formation in the spiral galaxy NGC 628},
  author={Elmegreen, Bruce G and Elmegreen, Debra Meloy and Chandar, Rupali and Whitmore, Brad and Regan, Michael},
  journal={The Astrophysical Journal},
  volume={644},
  number={2},
  pages={879},
  year={2006},
  publisher={IOP Publishing}
}

@ARTICLE{2019MNRAS.484.4897C,
       author = {{Cook}, D.~O. and {Lee}, J.~C. and {Adamo}, A. and {Kim}, H. and {Chandar}, R. and {Whitmore}, B.~C. and {Mok}, A. and {Ryon}, J.~E. and {Dale}, D.~A. and {Calzetti}, D. and {Andrews}, J.~E. and {Aloisi}, A. and {Ashworth}, G. and {Bright}, S.~N. and {Brown}, T.~M. and {Christian}, C. and {Cignoni}, M. and {Clayton}, G.~C. and {da Silva}, R. and {de Mink}, S.~E. and {Dobbs}, C.~L. and {Elmegreen}, B.~G. and {Elmegreen}, D.~M. and {Evans}, A.~S. and {Fumagalli}, M. and {Gallagher}, J.~S. and {Gouliermis}, D.~A. and {Grasha}, K. and {Grebel}, E.~K. and {Herrero}, A. and {Hunter}, D.~A. and {Jensen}, E.~I. and {Johnson}, K.~E. and {Kahre}, L. and {Kennicutt}, R.~C. and {Krumholz}, M.~R. and {Lee}, N.~J. and {Lennon}, D. and {Linden}, S. and {Martin}, C. and {Messa}, M. and {Nair}, P. and {Nota}, A. and {{\"O}stlin}, G. and {Parziale}, R.~C. and {Pellerin}, A. and {Regan}, M.~W. and {Sabbi}, E. and {Sacchi}, E. and {Schaerer}, D. and {Schiminovich}, D. and {Shabani}, F. and {Slane}, F.~A. and {Small}, J. and {Smith}, C.~L. and {Smith}, L.~J. and {Taibi}, S. and {Thilker}, D.~A. and {de la Torre}, I.~C. and {Tosi}, M. and {Turner}, J.~A. and {Ubeda}, L. and {Van Dyk}, S.~D. and {Walterbos}, R. AM and {Wofford}, A.},
        title = "{Star cluster catalogues for the LEGUS dwarf galaxies}",
      journal = {\mnras},
         year = 2019,
        month = apr,
       volume = {484},
       number = {4},
        pages = {4897-4919},
          doi = {10.1093/mnras/stz331},
archivePrefix = {arXiv},
       eprint = {1902.00082},
 primaryClass = {astro-ph.GA},
       adsurl = {https://ui.adsabs.harvard.edu/abs/2019MNRAS.484.4897C}
}

@ARTICLE{2017ApJ...841..131A,
       author = {{Adamo}, A. and {Ryon}, J.~E. and {Messa}, M. and {Kim}, H. and {Grasha}, K. and {Cook}, D.~O. and {Calzetti}, D. and {Lee}, J.~C. and {Whitmore}, B.~C. and {Elmegreen}, B.~G. and {Ubeda}, L. and {Smith}, L.~J. and {Bright}, S.~N. and {Runnholm}, A. and {Andrews}, J.~E. and {Fumagalli}, M. and {Gouliermis}, D.~A. and {Kahre}, L. and {Nair}, P. and {Thilker}, D. and {Walterbos}, R. and {Wofford}, A. and {Aloisi}, A. and {Ashworth}, G. and {Brown}, T.~M. and {Chandar}, R. and {Christian}, C. and {Cignoni}, M. and {Clayton}, G.~C. and {Dale}, D.~A. and {de Mink}, S.~E. and {Dobbs}, C. and {Elmegreen}, D.~M. and {Evans}, A.~S. and {Gallagher}, III, J.~S. and {Grebel}, E.~K. and {Herrero}, A. and {Hunter}, D.~A. and {Johnson}, K.~E. and {Kennicutt}, R.~C. and {Krumholz}, M.~R. and {Lennon}, D. and {Levay}, K. and {Martin}, C. and {Nota}, A. and {{\"O}stlin}, G. and {Pellerin}, A. and {Prieto}, J. and {Regan}, M.~W. and {Sabbi}, E. and {Sacchi}, E. and {Schaerer}, D. and {Schiminovich}, D. and {Shabani}, F. and {Tosi}, M. and {Van Dyk}, S.~D. and {Zackrisson}, E.},
        title = "{Legacy ExtraGalactic UV Survey with The Hubble Space Telescope: Stellar Cluster Catalogs and First Insights Into Cluster Formation and Evolution in NGC 628}",
      journal = {\apj},
         year = 2017,
        month = jun,
       volume = {841},
       number = {2},
          eid = {131},
        pages = {131},
          doi = {10.3847/1538-4357/aa7132},
archivePrefix = {arXiv},
       eprint = {1705.01588},
 primaryClass = {astro-ph.GA},
       adsurl = {https://ui.adsabs.harvard.edu/abs/2017ApJ...841..131A}
}

@ARTICLE{2007ApJ...658.1006M,
       author = {{Mu{\~n}oz-Mateos}, J.~C. and {Gil de Paz}, A. and {Boissier}, S. and {Zamorano}, J. and {Jarrett}, T. and {Gallego}, J. and {Madore}, B.~F.},
        title = "{Specific Star Formation Rate Profiles in Nearby Spiral Galaxies: Quantifying the Inside-Out Formation of Disks}",
      journal = {\apj},
         year = 2007,
        month = apr,
       volume = {658},
       number = {2},
        pages = {1006-1026},
          doi = {10.1086/511812},
archivePrefix = {arXiv},
       eprint = {astro-ph/0612017},
 primaryClass = {astro-ph},
       adsurl = {https://ui.adsabs.harvard.edu/abs/2007ApJ...658.1006M}
}

@article{Mondal_2018,
doi = {10.3847/1538-3881/aad4f6},
url = {https://dx.doi.org/10.3847/1538-3881/aad4f6},
year = {2018},
month = {aug},
publisher = {The American Astronomical Society},
volume = {156},
number = {3},
pages = {109},
author = {Mondal, Chayan and Subramaniam, Annapurni and George, Koshy},
title = {UVIT Imaging of WLM: Demographics of Star-forming Regions in the Nearby Dwarf Irregular Galaxy},
journal = {The Astronomical Journal}
}

@ARTICLE{2016A&A...591A...6B,
       author = {{Boquien}, M. and {Kennicutt}, R. and {Calzetti}, D. and {Dale}, D. and {Galametz}, M. and {Sauvage}, M. and {Croxall}, K. and {Draine}, B. and {Kirkpatrick}, A. and {Kumari}, N. and {Hunt}, L. and {De Looze}, I. and {Pellegrini}, E. and {Rela{\~n}o}, M. and {Smith}, J. -D. and {Tabatabaei}, F.},
        title = "{Towards universal hybrid star formation rate estimators}",
      journal = {\aap},
         year = 2016,
        month = jun,
       volume = {591},
          eid = {A6},
        pages = {A6},
          doi = {10.1051/0004-6361/201527759},
archivePrefix = {arXiv},
       eprint = {1603.09340},
 primaryClass = {astro-ph.GA},
       adsurl = {https://ui.adsabs.harvard.edu/abs/2016A&A...591A...6B}
}

@article{Hassani_2024,
doi = {10.3847/1538-4365/ad152c},
url = {https://dx.doi.org/10.3847/1538-4365/ad152c},
year = {2024},
month = {feb},
publisher = {The American Astronomical Society},
volume = {271},
number = {1},
pages = {2},
author = {Hassani, Hamid and Rosolowsky, Erik and Koch, Eric W. and Postma, Joseph and Nofech, Joseph and Corbould, Harrisen and Thilker, David and Leroy, Adam K. and Schinnerer, Eva and Belfiore, Francesco and Bigiel, Frank and Boquien, Médéric and Chevance, Mélanie and Dale, Daniel A. and Egorov, Oleg V. and Emsellem, Eric and Glover, Simon C. O. and Grasha, Kathryn and Groves, Brent and Henny, Kiana and Kim, Jaeyeon and Klessen, Ralf S. and Kreckel, Kathryn and Kruijssen, J. M. Diederik and Lee, Janice C. and Lopez, Laura A. and Neumann, Justus and Pan, Hsi-An and Sandstrom, Karin M. and Sarbadhicary, Sumit K. and Sun, Jiayi and Williams, Thomas G.},
title = {The PHANGS-AstroSat Atlas of Nearby Star-forming Galaxies},
journal = {The Astrophysical Journal Supplement Series}
}

@ARTICLE{2003ARA&A..41...57L,
       author = {{Lada}, Charles J. and {Lada}, Elizabeth A.},
        title = "{Embedded Clusters in Molecular Clouds}",
      journal = {\araa},
         year = 2003,
        month = jan,
       volume = {41},
        pages = {57-115},
          doi = {10.1146/annurev.astro.41.011802.094844},
archivePrefix = {arXiv},
       eprint = {astro-ph/0301540},
 primaryClass = {astro-ph},
       adsurl = {https://ui.adsabs.harvard.edu/abs/2003ARA&A..41...57L}
}

@ARTICLE{Elmegreen_2006,
       author = {{Elmegreen}, Bruce G. and {Elmegreen}, Debra Meloy and {Chandar}, Rupali and {Whitmore}, Brad and {Regan}, Michael},
        title = "{Hierarchical Star Formation in the Spiral Galaxy NGC 628}",
      journal = {\apj},
         year = 2006,
        month = jun,
       volume = {644},
       number = {2},
        pages = {879-889},
          doi = {10.1086/503797},
archivePrefix = {arXiv},
       eprint = {astro-ph/0605523},
 primaryClass = {astro-ph},
       adsurl = {https://ui.adsabs.harvard.edu/abs/2006ApJ...644..879E}
}

@ARTICLE{2018PASP..130g2001G,
       author = {{Gouliermis}, Dimitrios A.},
        title = "{Unbound Young Stellar Systems: Star Formation on the Loose}",
      journal = {\pasp},
         year = 2018,
        month = jul,
       volume = {130},
       number = {989},
        pages = {072001},
          doi = {10.1088/1538-3873/aac1fd},
archivePrefix = {arXiv},
       eprint = {1806.11541},
 primaryClass = {astro-ph.GA},
       adsurl = {https://ui.adsabs.harvard.edu/abs/2018PASP..130g2001G}
}

@ARTICLE{2015ApJS..219....4S,
       author = {{Salo}, Heikki and {Laurikainen}, Eija and {Laine}, Jarkko and {Comer{\'o}n}, Sebastien and {Gadotti}, Dimitri A. and {Buta}, Ron and {Sheth}, Kartik and {Zaritsky}, Dennis and {Ho}, Luis and {Knapen}, Johan and {Athanassoula}, E. and {Bosma}, Albert and {Laine}, Seppo and {Cisternas}, Mauricio and {Kim}, Taehyun and {Mu{\~n}oz-Mateos}, Juan Carlos and {Regan}, Michael and {Hinz}, Joannah L. and {Gil de Paz}, Armando and {Menendez-Delmestre}, Karin and {Mizusawa}, Trisha and {Erroz-Ferrer}, Santiago and {Meidt}, Sharon E. and {Querejeta}, Miguel},
        title = "{The Spitzer Survey of Stellar Structure in Galaxies (S$^{4}$G): Multi-component Decomposition Strategies and Data Release}",
      journal = {\apjs},
         year = 2015,
        month = jul,
       volume = {219},
       number = {1},
          eid = {4},
        pages = {4},
          doi = {10.1088/0067-0049/219/1/4},
archivePrefix = {arXiv},
       eprint = {1503.06550},
 primaryClass = {astro-ph.GA},
       adsurl = {https://ui.adsabs.harvard.edu/abs/2015ApJS..219....4S}
}

@ARTICLE{2026arXiv260612254S,
       author = {{Shashank}, Gairola and {Subramanian}, Smitha and {Mondal}, Chayan and {Menon}, Shyam H. and {Subramaniam}, Annapurni},
        title = "{Investigating the young stellar populations and hierarchies in nearby galaxies with the UVIT. II. Presenting the properties of \raisebox{-0.5ex}\textasciitilde25,000 UV-detected star-forming clumps}",
      journal = {arXiv e-prints},
         year = 2026,
        month = jun,
          eid = {arXiv:2606.12254},
        pages = {arXiv:2606.12254},
          doi = {10.48550/arXiv.2606.12254},
archivePrefix = {arXiv},
       eprint = {2606.12254},
 primaryClass = {astro-ph.GA},
       adsurl = {https://ui.adsabs.harvard.edu/abs/2026arXiv260612254S}
}

@ARTICLE{2003AJ....125..525J,
       author = {{Jarrett}, T.~H. and {Chester}, T. and {Cutri}, R. and {Schneider}, S.~E. and {Huchra}, J.~P.},
        title = "{The 2MASS Large Galaxy Atlas}",
      journal = {\aj},
         year = 2003,
        month = feb,
       volume = {125},
       number = {2},
        pages = {525-554},
          doi = {10.1086/345794},
       adsurl = {https://ui.adsabs.harvard.edu/abs/2003AJ....125..525J}
}

@ARTICLE{2009ApJ...693.1821D,
       author = {{Dale}, D.~A. and {Smith}, J.~D.~T. and {Schlawin}, E.~A. and {Armus}, L. and {Buckalew}, B.~A. and {Cohen}, S.~A. and {Helou}, G. and {Jarrett}, T.~H. and {Johnson}, L.~C. and {Moustakas}, J. and {Murphy}, E.~J. and {Roussel}, H. and {Sheth}, K. and {Staudaher}, S. and {Bot}, C. and {Calzetti}, D. and {Engelbracht}, C.~W. and {Gordon}, K.~D. and {Hollenbach}, D.~J. and {Kennicutt}, R.~C. and {Malhotra}, S.},
        title = "{The Spitzer Infrared Nearby Galaxies Survey: A High-Resolution Spectroscopy Anthology}",
      journal = {\apj},
         year = 2009,
        month = mar,
       volume = {693},
       number = {2},
        pages = {1821-1834},
          doi = {10.1088/0004-637X/693/2/1821},
archivePrefix = {arXiv},
       eprint = {0811.4190},
 primaryClass = {astro-ph},
       adsurl = {https://ui.adsabs.harvard.edu/abs/2009ApJ...693.1821D}
}

@ARTICLE{2019A&A...622A.103B,
       author = {{Boquien}, M. and {Burgarella}, D. and {Roehlly}, Y. and {Buat}, V. and {Ciesla}, L. and {Corre}, D. and {Inoue}, A.~K. and {Salas}, H.},
        title = "{CIGALE: a python Code Investigating GALaxy Emission}",
      journal = {\aap},
         year = 2019,
        month = feb,
       volume = {622},
          eid = {A103},
        pages = {A103},
          doi = {10.1051/0004-6361/201834156},
archivePrefix = {arXiv},
       eprint = {1811.03094},
 primaryClass = {astro-ph.GA},
       adsurl = {https://ui.adsabs.harvard.edu/abs/2019A&A...622A.103B}
}

@ARTICLE{2026ApJ..1003...50C,
       author = {{Chanu}, Athokpam Langlen and {Amrutha}, S. and {Chingangbam}, Pravabati and {Park}, Changbom},
        title = "{Morphological Complexity of NGC 628{\textemdash}A Multiwavelength Multiscale Analysis Using the Ordinal Pattern Framework}",
      journal = {\apj},
         year = 2026,
        month = may,
       volume = {1003},
       number = {1},
          eid = {50},
        pages = {50},
          doi = {10.3847/1538-4357/ae5c98},
archivePrefix = {arXiv},
       eprint = {2604.08409},
 primaryClass = {astro-ph.GA},
       adsurl = {https://ui.adsabs.harvard.edu/abs/2026ApJ..1003...50C}
}

@ARTICLE{2003MNRAS.343..978S,
       author = {{Shen}, Shiyin and {Mo}, H.~J. and {White}, Simon D.~M. and {Blanton}, Michael R. and {Kauffmann}, Guinevere and {Voges}, Wolfgang and {Brinkmann}, J. and {Csabai}, Istvan},
        title = "{The size distribution of galaxies in the Sloan Digital Sky Survey}",
      journal = {\mnras},
         year = 2003,
        month = aug,
       volume = {343},
       number = {3},
        pages = {978-994},
          doi = {10.1046/j.1365-8711.2003.06740.x},
archivePrefix = {arXiv},
       eprint = {astro-ph/0301527},
 primaryClass = {astro-ph},
       adsurl = {https://ui.adsabs.harvard.edu/abs/2003MNRAS.343..978S}
}

@ARTICLE{2014ApJ...788...28V,
       author = {{van der Wel}, A. and {Franx}, M. and {van Dokkum}, P.~G. and {Skelton}, R.~E. and {Momcheva}, I.~G. and {Whitaker}, K.~E. and {Brammer}, G.~B. and {Bell}, E.~F. and {Rix}, H.-W. and {Wuyts}, S. and {Ferguson}, H.~C. and {Holden}, B.~P. and {Barro}, G. and {Koekemoer}, A.~M. and {Chang}, Yu-Yen and {McGrath}, E.~J. and {H{\"a}ussler}, B. and {Dekel}, A. and {Behroozi}, P. and {Fumagalli}, M. and {Leja}, J. and {Lundgren}, B.~F. and {Maseda}, M.~V. and {Nelson}, E.~J. and {Wake}, D.~A. and {Patel}, S.~G. and {Labb{\'e}}, I. and {Faber}, S.~M. and {Grogin}, N.~A. and {Kocevski}, D.~D.},
        title = "{3D-HST+CANDELS: The Evolution of the Galaxy Size-Mass Distribution since z = 3}",
      journal = {\apj},
         year = 2014,
        month = jun,
       volume = {788},
       number = {1},
          eid = {28},
        pages = {28},
          doi = {10.1088/0004-637X/788/1/28},
archivePrefix = {arXiv},
       eprint = {1404.2844},
 primaryClass = {astro-ph.GA},
       adsurl = {https://ui.adsabs.harvard.edu/abs/2014ApJ...788...28V}
}

@ARTICLE{2015MNRAS.447.2603L,
       author = {{Lange}, Rebecca and {Driver}, Simon P. and {Robotham}, Aaron S.~G. and {Kelvin}, Lee S. and {Graham}, Alister W. and {Alpaslan}, Mehmet and {Andrews}, Stephen K. and {Baldry}, Ivan K. and {Bamford}, Steven and {Bland-Hawthorn}, Joss and {Brough}, Sarah and {Cluver}, Michelle E. and {Conselice}, Christopher J. and {Davies}, Luke J.~M. and {Haeussler}, Boris and {Konstantopoulos}, Iraklis S. and {Loveday}, Jon and {Moffett}, Amanda J. and {Norberg}, Peder and {Phillipps}, Steven and {Taylor}, Edward N. and {L{\'o}pez-S{\'a}nchez}, {\'A}ngel R. and {Wilkins}, Stephen M.},
        title = "{Galaxy And Mass Assembly (GAMA): mass-size relations of z < 0.1 galaxies subdivided by S{\'e}rsic index, colour and morphology}",
      journal = {\mnras},
         year = 2015,
        month = mar,
       volume = {447},
       number = {3},
        pages = {2603-2630},
          doi = {10.1093/mnras/stu2467},
archivePrefix = {arXiv},
       eprint = {1411.6355},
 primaryClass = {astro-ph.GA},
       adsurl = {https://ui.adsabs.harvard.edu/abs/2015MNRAS.447.2603L}
}

@ARTICLE{2006ApJ...652.1339M,
       author = {{McClure-Griffiths}, N.~M. and {Dickey}, J.~M. and {Gaensler}, B.~M. and {Green}, A.~J. and {Haverkorn}, Marijke},
        title = "{Magnetically Dominated Strands of Cold Hydrogen in the Riegel-Crutcher Cloud}",
      journal = {\apj},
         year = 2006,
        month = dec,
       volume = {652},
       number = {2},
        pages = {1339-1347},
          doi = {10.1086/508706},
archivePrefix = {arXiv},
       eprint = {astro-ph/0608585},
 primaryClass = {astro-ph},
       adsurl = {https://ui.adsabs.harvard.edu/abs/2006ApJ...652.1339M}
}

@ARTICLE{2009MNRAS.398..887D,
       author = {{Dutta}, Prasun and {Begum}, Ayesha and {Bharadwaj}, Somnath and {Chengalur}, Jayaram N.},
        title = "{A study of interstellar medium of dwarf galaxies using HI power spectrum analysis}",
      journal = {\mnras},
         year = 2009,
        month = sep,
       volume = {398},
       number = {2},
        pages = {887-897},
          doi = {10.1111/j.1365-2966.2009.15105.x},
archivePrefix = {arXiv},
       eprint = {0905.1756},
 primaryClass = {astro-ph.GA},
       adsurl = {https://ui.adsabs.harvard.edu/abs/2009MNRAS.398..887D}
}

@ARTICLE{1987ApJ...312L..45B,
       author = {{Bally}, John and {Langer}, William D. and {Stark}, Antony A. and {Wilson}, Robert W.},
        title = "{Filamentary Structure in the Orion Molecular Cloud}",
      journal = {\apjl},
         year = 1987,
        month = jan,
       volume = {312},
        pages = {L45},
          doi = {10.1086/184817},
       adsurl = {https://ui.adsabs.harvard.edu/abs/1987ApJ...312L..45B}
}

@ARTICLE{2007ARA&A..45..565M,
       author = {{McKee}, Christopher F. and {Ostriker}, Eve C.},
        title = "{Theory of Star Formation}",
      journal = {\araa},
         year = 2007,
        month = sep,
       volume = {45},
       number = {1},
        pages = {565-687},
          doi = {10.1146/annurev.astro.45.051806.110602},
archivePrefix = {arXiv},
       eprint = {0707.3514},
 primaryClass = {astro-ph},
       adsurl = {https://ui.adsabs.harvard.edu/abs/2007ARA&A..45..565M}
}

@ARTICLE{2013NewA...19...89D,
       author = {{Dutta}, Prasun and {Begum}, Ayesha and {Bharadwaj}, Somnath and {Chengalur}, Jayaram N.},
        title = "{Probing interstellar turbulence in spiral galaxies using H I power spectrum analysis}",
      journal = {\na},
         year = 2013,
        month = feb,
       volume = {19},
        pages = {89-98},
          doi = {10.1016/j.newast.2012.08.008},
archivePrefix = {arXiv},
       eprint = {1208.5386},
 primaryClass = {astro-ph.GA},
       adsurl = {https://ui.adsabs.harvard.edu/abs/2013NewA...19...89D}
}

@Article{Hunter07,
  Author    = {Hunter, J. D.},
  Title     = {Matplotlib: A 2D graphics environment},
  Journal   = {Computing in Science \& Engineering},
  Volume    = {9},
  Number    = {3},
  Pages     = {90--95},
  publisher = {IEEE COMPUTER SOC},
  doi       = {10.1109/MCSE.2007.55},
  year      = 2007
}

@book{python09,
 author = {Van Rossum, Guido and Drake, Fred L.},
 title = {Python 3 Reference Manual},
 year = {2009},
 isbn = {1441412697},
 publisher = {CreateSpace},
 address = {Scotts Valley, CA}
}

@ARTICLE{astropy_2018,
       author = {{Astropy Collaboration} and {Price-Whelan}, A.~M. and
         {Sip{\H{o}}cz}, B.~M. and {G{\"u}nther}, H.~M. and {Lim}, P.~L. and
         {Crawford}, S.~M. and {Conseil}, S. and {Shupe}, D.~L. and
         {Craig}, M.~W. and {Dencheva}, N. and {Ginsburg}, A. and {Vand
        erPlas}, J.~T. and {Bradley}, L.~D. and {P{\'e}rez-Su{\'a}rez}, D. and
         {de Val-Borro}, M. and {Aldcroft}, T.~L. and {Cruz}, K.~L. and
         {Robitaille}, T.~P. and {Tollerud}, E.~J. and {Ardelean}, C. and
         {Babej}, T. and {Bach}, Y.~P. and {Bachetti}, M. and {Bakanov}, A.~V. and
         {Bamford}, S.~P. and {Barentsen}, G. and {Barmby}, P. and
         {Baumbach}, A. and {Berry}, K.~L. and {Biscani}, F. and {Boquien}, M. and
         {Bostroem}, K.~A. and {Bouma}, L.~G. and {Brammer}, G.~B. and
         {Bray}, E.~M. and {Breytenbach}, H. and {Buddelmeijer}, H. and
         {Burke}, D.~J. and {Calderone}, G. and {Cano Rodr{\'\i}guez}, J.~L. and
         {Cara}, M. and {Cardoso}, J.~V.~M. and {Cheedella}, S. and {Copin}, Y. and
         {Corrales}, L. and {Crichton}, D. and {D'Avella}, D. and {Deil}, C. and
         {Depagne}, {\'E}. and {Dietrich}, J.~P. and {Donath}, A. and
         {Droettboom}, M. and {Earl}, N. and {Erben}, T. and {Fabbro}, S. and
         {Ferreira}, L.~A. and {Finethy}, T. and {Fox}, R.~T. and
         {Garrison}, L.~H. and {Gibbons}, S.~L.~J. and {Goldstein}, D.~A. and
         {Gommers}, R. and {Greco}, J.~P. and {Greenfield}, P. and
         {Groener}, A.~M. and {Grollier}, F. and {Hagen}, A. and {Hirst}, P. and
         {Homeier}, D. and {Horton}, A.~J. and {Hosseinzadeh}, G. and {Hu}, L. and
         {Hunkeler}, J.~S. and {Ivezi{\'c}}, {\v{Z}}. and {Jain}, A. and
         {Jenness}, T. and {Kanarek}, G. and {Kendrew}, S. and {Kern}, N.~S. and
         {Kerzendorf}, W.~E. and {Khvalko}, A. and {King}, J. and {Kirkby}, D. and
         {Kulkarni}, A.~M. and {Kumar}, A. and {Lee}, A. and {Lenz}, D. and
         {Littlefair}, S.~P. and {Ma}, Z. and {Macleod}, D.~M. and
         {Mastropietro}, M. and {McCully}, C. and {Montagnac}, S. and
         {Morris}, B.~M. and {Mueller}, M. and {Mumford}, S.~J. and {Muna}, D. and
         {Murphy}, N.~A. and {Nelson}, S. and {Nguyen}, G.~H. and
         {Ninan}, J.~P. and {N{\"o}the}, M. and {Ogaz}, S. and {Oh}, S. and
         {Parejko}, J.~K. and {Parley}, N. and {Pascual}, S. and {Patil}, R. and
         {Patil}, A.~A. and {Plunkett}, A.~L. and {Prochaska}, J.~X. and
         {Rastogi}, T. and {Reddy Janga}, V. and {Sabater}, J. and
         {Sakurikar}, P. and {Seifert}, M. and {Sherbert}, L.~E. and
         {Sherwood-Taylor}, H. and {Shih}, A.~Y. and {Sick}, J. and
         {Silbiger}, M.~T. and {Singanamalla}, S. and {Singer}, L.~P. and
         {Sladen}, P.~H. and {Sooley}, K.~A. and {Sornarajah}, S. and
         {Streicher}, O. and {Teuben}, P. and {Thomas}, S.~W. and
         {Tremblay}, G.~R. and {Turner}, J.~E.~H. and {Terr{\'o}n}, V. and
         {van Kerkwijk}, M.~H. and {de la Vega}, A. and {Watkins}, L.~L. and
         {Weaver}, B.~A. and {Whitmore}, J.~B. and {Woillez}, J. and
         {Zabalza}, V. and {Astropy Contributors}},
        title = "{The Astropy Project: Building an Open-science Project and Status of the v2.0 Core Package}",
      journal = {\aj},
         year = 2018,
        month = sep,
       volume = {156},
       number = {3},
          eid = {123},
        pages = {123},
          doi = {10.3847/1538-3881/aabc4f},
archivePrefix = {arXiv},
       eprint = {1801.02634},
 primaryClass = {astro-ph.IM},
       adsurl = {https://ui.adsabs.harvard.edu/abs/2018AJ....156..123A}
}

@article{bianchi2005recent,
  title={Recent star formation in nearby galaxies from Galaxy Evolution Explorer imaging: M101 and M51},
  author={Bianchi, Luciana and Thilker, David A and Burgarella, Denis and Friedman, Peter G and Hoopes, Charles G and Boissier, Samuel and De Paz, Armando Gil and Barlow, Tom A and Byun, Yong-Ik and Donas, Jose and others},
  journal={The Astrophysical Journal},
  volume={619},
  number={1},
  pages={L71},
  year={2005},
  publisher={IOP Publishing}
}

@ARTICLE{2008AJ....136.2782L,
       author = {{Leroy}, Adam K. and {Walter}, Fabian and {Brinks}, Elias and {Bigiel}, Frank and {de Blok}, W.~J.~G. and {Madore}, Barry and {Thornley}, M.~D.},
        title = "{The Star Formation Efficiency in Nearby Galaxies: Measuring Where Gas Forms Stars Effectively}",
      journal = {\aj},
         year = 2008,
        month = dec,
       volume = {136},
       number = {6},
        pages = {2782-2845},
          doi = {10.1088/0004-6256/136/6/2782},
archivePrefix = {arXiv},
       eprint = {0810.2556},
 primaryClass = {astro-ph},
       adsurl = {https://ui.adsabs.harvard.edu/abs/2008AJ....136.2782L}
}

@ARTICLE{2019MNRAS.490..467Z,
       author = {{Zheng}, Yong and {Putman}, Mary E. and {Emerick}, Andrew and {McQuinn}, Kristen B.~W. and {Werk}, Jessica K. and {Lockman}, Felix J. and {Oppenheimer}, Benjamin D. and {Fox}, Andrew J. and {Kirby}, Evan N. and {Burchett}, Joseph N.},
        title = "{Tentative detection of the circumgalactic medium of the isolated low-mass dwarf galaxy WLM}",
      journal = {\mnras},
         year = 2019,
        month = nov,
       volume = {490},
       number = {1},
        pages = {467-477},
          doi = {10.1093/mnras/stz2563},
archivePrefix = {arXiv},
       eprint = {1909.05407},
 primaryClass = {astro-ph.GA},
       adsurl = {https://ui.adsabs.harvard.edu/abs/2019MNRAS.490..467Z}
}

@ARTICLE{2022MNRAS.515.3270S,
       author = {{Smith}, Madison V. and {van Zee}, L. and {Dale}, D.~A. and {Hunter}, L.~C. and {Staudaher}, S. and {Wrock}, T.},
        title = "{A multiwavelength study of star formation in nearby galaxies: evidence for inside-out growth of the stellar disc}",
      journal = {\mnras},
         year = 2022,
        month = sep,
       volume = {515},
       number = {3},
        pages = {3270-3298},
          doi = {10.1093/mnras/stac1974},
       adsurl = {https://ui.adsabs.harvard.edu/abs/2022MNRAS.515.3270S}
}

@ARTICLE{2023ApJ...950...81N,
       author = {{Nandi}, Payel and {Stalin}, C.~S. and {Saikia}, D.~J. and {Muneer}, S. and {Mountrichas}, George and {Wylezalek}, Dominika and {Sagar}, R. and {Kissler-Patig}, Markus},
        title = "{Star Formation in the Dwarf Seyfert Galaxy NGC 4395: Evidence for Both AGN and SN Feedback?}",
      journal = {\apj},
         year = 2023,
        month = jun,
       volume = {950},
       number = {2},
          eid = {81},
        pages = {81},
          doi = {10.3847/1538-4357/accf1e},
archivePrefix = {arXiv},
       eprint = {2304.08986},
 primaryClass = {astro-ph.GA},
       adsurl = {https://ui.adsabs.harvard.edu/abs/2023ApJ...950...81N}
}

@ARTICLE{2015AJ....149....1Z,
       author = {{Zhou}, Zhi-Min and {Cao}, Chen and {Wu}, Hong},
        title = "{Star Formation Properties in Barred Galaxies. III. Statistical Study of Bar-Driven Secular Evolution Using a Sample of Nearby Barred Spirals}",
      journal = {\aj},
         year = 2015,
        month = jan,
       volume = {149},
       number = {1},
          eid = {1},
        pages = {1},
          doi = {10.1088/0004-6256/149/1/1},
archivePrefix = {arXiv},
       eprint = {1409.3045},
 primaryClass = {astro-ph.GA},
       adsurl = {https://ui.adsabs.harvard.edu/abs/2015AJ....149....1Z}
}

@ARTICLE{2021ApJS..257...43L,
       author = {{Leroy}, Adam K. and {Schinnerer}, Eva and {Hughes}, Annie and {Rosolowsky}, Erik and {Pety}, J{\'e}r{\^o}me and {Schruba}, Andreas and {Usero}, Antonio and {Blanc}, Guillermo A. and {Chevance}, M{\'e}lanie and {Emsellem}, Eric and {Faesi}, Christopher M. and {Herrera}, Cinthya N. and {Liu}, Daizhong and {Meidt}, Sharon E. and {Querejeta}, Miguel and {Saito}, Toshiki and {Sandstrom}, Karin M. and {Sun}, Jiayi and {Williams}, Thomas G. and {Anand}, Gagandeep S. and {Barnes}, Ashley T. and {Behrens}, Erica A. and {Belfiore}, Francesco and {Benincasa}, Samantha M. and {Be{\v{s}}li{\'c}}, Ivana and {Bigiel}, Frank and {Bolatto}, Alberto D. and {den Brok}, Jakob S. and {Cao}, Yixian and {Chandar}, Rupali and {Chastenet}, J{\'e}r{\'e}my and {Chiang}, I-Da and {Congiu}, Enrico and {Dale}, Daniel A. and {Deger}, Sinan and {Eibensteiner}, Cosima and {Egorov}, Oleg V. and {Garc{\'\i}a-Rodr{\'\i}guez}, Axel and {Glover}, Simon C.~O. and {Grasha}, Kathryn and {Henshaw}, Jonathan D. and {Ho}, I.-Ting and {Kepley}, Amanda A. and {Kim}, Jaeyeon and {Klessen}, Ralf S. and {Kreckel}, Kathryn and {Koch}, Eric W. and {Kruijssen}, J.~M. Diederik and {Larson}, Kirsten L. and {Lee}, Janice C. and {Lopez}, Laura A. and {Machado}, Josh and {Mayker}, Ness and {McElroy}, Rebecca and {Murphy}, Eric J. and {Ostriker}, Eve C. and {Pan}, Hsi-An and {Pessa}, Ismael and {Puschnig}, Johannes and {Razza}, Alessandro and {S{\'a}nchez-Bl{\'a}zquez}, Patricia and {Santoro}, Francesco and {Sardone}, Amy and {Scheuermann}, Fabian and {Sliwa}, Kazimierz and {Sormani}, Mattia C. and {Stuber}, Sophia K. and {Thilker}, David A. and {Turner}, Jordan A. and {Utomo}, Dyas and {Watkins}, Elizabeth J. and {Whitmore}, Bradley},
        title = "{PHANGS-ALMA: Arcsecond CO(2-1) Imaging of Nearby Star-forming Galaxies}",
      journal = {\apjs},
         year = 2021,
        month = dec,
       volume = {257},
       number = {2},
          eid = {43},
        pages = {43},
          doi = {10.3847/1538-4365/ac17f3},
archivePrefix = {arXiv},
       eprint = {2104.07739},
 primaryClass = {astro-ph.GA},
       adsurl = {https://ui.adsabs.harvard.edu/abs/2021ApJS..257...43L}
}

@ARTICLE{2007A&A...462..933C,
       author = {{Chy{\.z}y}, K.~T. and {Bomans}, D.~J. and {Krause}, M. and {Beck}, R. and {Soida}, M. and {Urbanik}, M.},
        title = "{Magnetic fields and ionized gas in nearby late type galaxies}",
      journal = {\aap},
         year = 2007,
        month = feb,
       volume = {462},
       number = {3},
        pages = {933-941},
          doi = {10.1051/0004-6361:20065932},
archivePrefix = {arXiv},
       eprint = {astro-ph/0611316},
 primaryClass = {astro-ph},
       adsurl = {https://ui.adsabs.harvard.edu/abs/2007A&A...462..933C}
}

@ARTICLE{2019A&A...621A..51H,
       author = {{Hunt}, L.~K. and {De Looze}, I. and {Boquien}, M. and {Nikutta}, R. and {Rossi}, A. and {Bianchi}, S. and {Dale}, D.~A. and {Granato}, G.~L. and {Kennicutt}, R.~C. and {Silva}, L. and {Ciesla}, L. and {Rela{\~n}o}, M. and {Viaene}, S. and {Brandl}, B. and {Calzetti}, D. and {Croxall}, K.~V. and {Draine}, B.~T. and {Galametz}, M. and {Gordon}, K.~D. and {Groves}, B.~A. and {Helou}, G. and {Herrera-Camus}, R. and {Hinz}, J.~L. and {Koda}, J. and {Salim}, S. and {Sandstrom}, K.~M. and {Smith}, J.~D. and {Wilson}, C.~D. and {Zibetti}, S.},
        title = "{Comprehensive comparison of models for spectral energy distributions from 0.1 {\ensuremath{\mu}}m to 1 mm of nearby star-forming galaxies}",
      journal = {\aap},
         year = 2019,
        month = jan,
       volume = {621},
          eid = {A51},
        pages = {A51},
          doi = {10.1051/0004-6361/201834212},
archivePrefix = {arXiv},
       eprint = {1809.04088},
 primaryClass = {astro-ph.GA},
       adsurl = {https://ui.adsabs.harvard.edu/abs/2019A&A...621A..51H}
}

@ARTICLE{1997MNRAS.290...15T,
       author = {{Thean}, A.~H.~C. and {Mundell}, C.~G. and {Pedlar}, A. and {Nicholson}, R.~A.},
        title = "{A neutral hydrogen study of the Seyfert galaxy NGC 5033}",
      journal = {\mnras},
         year = 1997,
        month = sep,
       volume = {290},
       number = {1},
        pages = {15-24},
          doi = {10.1093/mnras/290.1.15},
       adsurl = {https://ui.adsabs.harvard.edu/abs/1997MNRAS.290...15T}
}

@ARTICLE{2019MNRAS.488.3826B,
       author = {{Bresolin}, Fabio},
        title = "{Metallicity gradients in small and nearby spiral galaxies}",
      journal = {\mnras},
         year = 2019,
        month = sep,
       volume = {488},
       number = {3},
        pages = {3826-3843},
          doi = {10.1093/mnras/stz1947},
archivePrefix = {arXiv},
       eprint = {1907.05071},
 primaryClass = {astro-ph.GA},
       adsurl = {https://ui.adsabs.harvard.edu/abs/2019MNRAS.488.3826B}
}

@ARTICLE{1996ApJ...471..816E,
       author = {{Elmegreen}, Bruce G. and {Falgarone}, Edith},
        title = "{A Fractal Origin for the Mass Spectrum of Interstellar Clouds}",
      journal = {\apj},
         year = 1996,
        month = nov,
       volume = {471},
        pages = {816},
          doi = {10.1086/178009},
       adsurl = {https://ui.adsabs.harvard.edu/abs/1996ApJ...471..816E}
}

@ARTICLE{Federrath_2018,
       author = {{Federrath}, Christoph},
        title = "{The turbulent formation of stars}",
      journal = {Physics Today},
         year = 2018,
        month = jun,
       volume = {71},
       number = {6},
        pages = {38-42},
          doi = {10.1063/PT.3.3947},
archivePrefix = {arXiv},
       eprint = {1806.05312},
 primaryClass = {astro-ph.SR},
       adsurl = {https://ui.adsabs.harvard.edu/abs/2018PhT....71f..38F}
}

@ARTICLE{2018ApJ...858...31S,
       author = {{Sun}, Ning-Chen and {de Grijs}, Richard and {Cioni}, Maria-Rosa L. and {Rubele}, Stefano and {Subramanian}, Smitha and {van Loon}, Jacco Th. and {Bekki}, Kenji and {Bell}, Cameron P.~M. and {Ivanov}, Valentin D. and {Marconi}, Marcella and {Muraveva}, Tatiana and {Oliveira}, Joana M. and {Ripepi}, Vincenzo},
        title = "{The VMC Survey. XXIX. Turbulence-controlled Hierarchical Star Formation in the Small Magellanic Cloud}",
      journal = {\apj},
         year = 2018,
        month = may,
       volume = {858},
       number = {1},
          eid = {31},
        pages = {31},
          doi = {10.3847/1538-4357/aabc50},
archivePrefix = {arXiv},
       eprint = {1804.01652},
 primaryClass = {astro-ph.GA},
       adsurl = {https://ui.adsabs.harvard.edu/abs/2018ApJ...858...31S}
}

@ARTICLE{Grasha_2019,
       author = {{Grasha}, K. and {Calzetti}, D. and {Adamo}, A. and {Kennicutt}, R.~C. and {Elmegreen}, B.~G. and {Messa}, M. and {Dale}, D.~A. and {Fedorenko}, K. and {Mahadevan}, S. and {Grebel}, E.~K. and {Fumagalli}, M. and {Kim}, H. and {Dobbs}, C.~L. and {Gouliermis}, D.~A. and {Ashworth}, G. and {Gallagher}, J.~S. and {Smith}, L.~J. and {Tosi}, M. and {Whitmore}, B.~C. and {Schinnerer}, E. and {Colombo}, D. and {Hughes}, A. and {Leroy}, A.~K. and {Meidt}, S.~E.},
        title = "{The spatial relation between young star clusters and molecular clouds in M51 with LEGUS}",
      journal = {\mnras},
         year = 2019,
        month = mar,
       volume = {483},
       number = {4},
        pages = {4707-4723},
          doi = {10.1093/mnras/sty3424},
archivePrefix = {arXiv},
       eprint = {1812.06109},
 primaryClass = {astro-ph.GA},
       adsurl = {https://ui.adsabs.harvard.edu/abs/2019MNRAS.483.4707G}
}

@ARTICLE{2018A&A...620A..21C,
       author = {{Colling}, C{\'e}dric and {Hennebelle}, Patrick and {Geen}, Sam and {Iffrig}, Olivier and {Bournaud}, Fr{\'e}d{\'e}ric},
        title = "{Impact of galactic shear and stellar feedback on star formation}",
      journal = {\aap},
         year = 2018,
        month = dec,
       volume = {620},
          eid = {A21},
        pages = {A21},
          doi = {10.1051/0004-6361/201833161},
archivePrefix = {arXiv},
       eprint = {1809.01037},
 primaryClass = {astro-ph.GA},
       adsurl = {https://ui.adsabs.harvard.edu/abs/2018A&A...620A..21C}
}

@ARTICLE{1984ApJ...276..114J,
       author = {{Jog}, C.~J. and {Solomon}, P.~M.},
        title = "{Two-fluid gravitational instabilities in a galactic disk}",
      journal = {\apj},
         year = 1984,
        month = jan,
       volume = {276},
        pages = {114-126},
          doi = {10.1086/161597},
       adsurl = {https://ui.adsabs.harvard.edu/abs/1984ApJ...276..114J}
}

@ARTICLE{2016MNRAS.458.1671K,
       author = {{Krumholz}, Mark R. and {Burkhart}, Blakesley},
        title = "{Is turbulence in the interstellar medium driven by feedback or gravity? An observational test}",
      journal = {\mnras},
         year = 2016,
        month = may,
       volume = {458},
       number = {2},
        pages = {1671-1677},
          doi = {10.1093/mnras/stw434},
archivePrefix = {arXiv},
       eprint = {1512.03439},
 primaryClass = {astro-ph.GA},
       adsurl = {https://ui.adsabs.harvard.edu/abs/2016MNRAS.458.1671K}
}

@ARTICLE{2025A&A...695A.155A,
       author = {{Arora}, Raghav and {Federrath}, Christoph and {Krumholz}, Mark and {Banerjee}, Robi},
        title = "{Formation of filaments and feathers in disc galaxies: Is self-gravity enough?}",
      journal = {\aap},
         year = 2025,
        month = mar,
       volume = {695},
          eid = {A155},
        pages = {A155},
          doi = {10.1051/0004-6361/202453501},
archivePrefix = {arXiv},
       eprint = {2502.18565},
 primaryClass = {astro-ph.GA},
       adsurl = {https://ui.adsabs.harvard.edu/abs/2025A&A...695A.155A}
}

@ARTICLE{2025ApJ...986...13E,
       author = {{Elmegreen}, Bruce G. and {Calzetti}, Daniela and {Adamo}, Angela and {Sandstrom}, Karin and {Dale}, Daniel and {Bajaj}, Varun and {Boyer}, Martha L. and {Duarte-Cabral}, Ana and {Chown}, Ryan and {Correnti}, Matteo and {Dalcanton}, Julianne J. and {Draine}, Bruce T. and {Gaches}, Brandt and {Gallagher}, John S. and {Grasha}, Kathryn and {Gregg}, Benjamin and {Hunt}, Leslie K. and {Johnson}, Kelsey E. and {Kennicutt}, Robert and {Klessen}, Ralf S. and {Leroy}, Adam K. and {Linden}, Sean and {McLeod}, Anna F. and {Messa}, Matteo and {{\"O}stlin}, G{\"o}ran and {Padave}, Mansi and {Roman-Duval}, Julia and {Smith}, J.~D. and {Walter}, Fabian and {Weinbeck}, Tony D.},
        title = "{An Investigation of Disk Thickness in M51 from H{\ensuremath{\alpha}}, Pa{\ensuremath{\alpha}}, and Mid-infrared Power Spectra}",
      journal = {\apj},
         year = 2025,
        month = jun,
       volume = {986},
       number = {1},
          eid = {13},
        pages = {13},
          doi = {10.3847/1538-4357/adcee6},
archivePrefix = {arXiv},
       eprint = {2504.05430},
 primaryClass = {astro-ph.GA},
       adsurl = {https://ui.adsabs.harvard.edu/abs/2025ApJ...986...13E}
}

@ARTICLE{2003MNRAS.343..413B,
       author = {{Bonnell}, Ian A. and {Bate}, Matthew R. and {Vine}, Stephen G.},
        title = "{The hierarchical formation of a stellar cluster}",
      journal = {\mnras},
         year = 2003,
        month = aug,
       volume = {343},
       number = {2},
        pages = {413-418},
          doi = {10.1046/j.1365-8711.2003.06687.x},
archivePrefix = {arXiv},
       eprint = {astro-ph/0305082},
 primaryClass = {astro-ph},
       adsurl = {https://ui.adsabs.harvard.edu/abs/2003MNRAS.343..413B}
}

@ARTICLE{2020MNRAS.496.1803N,
       author = {{Nandakumar}, Meera and {Dutta}, Prasun},
        title = "{Evidence of large-scale energy cascade in the spiral galaxy NGC 5236}",
      journal = {\mnras},
         year = 2020,
        month = aug,
       volume = {496},
       number = {2},
        pages = {1803-1810},
          doi = {10.1093/mnras/staa1651},
archivePrefix = {arXiv},
       eprint = {2006.04575},
 primaryClass = {astro-ph.GA},
       adsurl = {https://ui.adsabs.harvard.edu/abs/2020MNRAS.496.1803N}
}

@ARTICLE{2013ApJ...779...45M,
       author = {{Meidt}, Sharon E. and {Schinnerer}, Eva and {Garc{\'\i}a-Burillo}, Santiago and {Hughes}, Annie and {Colombo}, Dario and {Pety}, J{\'e}r{\^o}me and {Dobbs}, Clare L. and {Schuster}, Karl F. and {Kramer}, Carsten and {Leroy}, Adam K. and {Dumas}, Galle and {Thompson}, Todd A.},
        title = "{Gas Kinematics on Giant Molecular Cloud Scales in M51 with PAWS: Cloud Stabilization through Dynamical Pressure}",
      journal = {\apj},
         year = 2013,
        month = dec,
       volume = {779},
       number = {1},
          eid = {45},
        pages = {45},
          doi = {10.1088/0004-637X/779/1/45},
archivePrefix = {arXiv},
       eprint = {1304.7910},
 primaryClass = {astro-ph.CO},
       adsurl = {https://ui.adsabs.harvard.edu/abs/2013ApJ...779...45M}
}

@article{Meena_2025,
doi = {10.3847/1538-4357/add475},
url = {https://doi.org/10.3847/1538-4357/add475},
year = {2025},
month = {jun},
publisher = {The American Astronomical Society},
volume = {987},
number = {1},
pages = {33},
author = {Meena, Beena and Sabbi, Elena and Zeidler, Peter and Elmegreen, Bruce G. and Eldridge, Jan J. and Bajaj, Varun and Gennaro, Mario and Pasquali, Anna and Elmegreen, Debra M. and Klessen, Ralf S. and Smith, Linda J. and Bianchi, Luciana and Wofford, Aida and Facchini, Pietro and Gallagher, John S. and Calzetti, Daniela and Grebel, Eva K. and Adamo, Angela and (GULP)},
title = {GULP. II. Hierarchical Distribution and Evolution of Young Stellar Structures in NGC 4449},
journal = {The Astrophysical Journal}
}

@article{lapeer2026feast,
  title={FEAST: Probing Hierarchical Star Formation with the Spatial Distributions of Young Star Clusters},
  author={Lapeer, Drew and Calzetti, Daniela and Grasha, Kathryn and Adamo, Angela and Elmegreen, Bruce G and Bik, Arjan and Bortolini, Giacomo and Buckner, Anne and Cignoni, Michele and Correntim, Matteo and others},
  journal={arXiv preprint arXiv:2601.11434},
  year={2026}
}

@ARTICLE{2025ApJ...989..216H,
       author = {{Hota}, Sipra and {de Grijs}, Richard and {Subramaniam}, Annapurni},
        title = "{UVIT Study of the Magellanic Clouds (U-SMAC). III. Hierarchical Star Formation in the Small Magellanic Cloud Regulated by Turbulence}",
      journal = {\apj},
         year = 2025,
        month = aug,
       volume = {989},
       number = {2},
          eid = {216},
        pages = {216},
          doi = {10.3847/1538-4357/adec84},
archivePrefix = {arXiv},
       eprint = {2506.08951},
 primaryClass = {astro-ph.GA},
       adsurl = {https://ui.adsabs.harvard.edu/abs/2025ApJ...989..216H}
}

@ARTICLE{2009MNRAS.392..868B,
       author = {{Bastian}, Nate and {Gieles}, Mark and {Ercolano}, Barbara and {Gutermuth}, Rob},
        title = "{The spatial evolution of stellar structures in the Large Magellanic Cloud}",
      journal = {\mnras},
         year = 2009,
        month = jan,
       volume = {392},
       number = {2},
        pages = {868-878},
          doi = {10.1111/j.1365-2966.2008.14107.x},
archivePrefix = {arXiv},
       eprint = {0809.1943},
 primaryClass = {astro-ph},
       adsurl = {https://ui.adsabs.harvard.edu/abs/2009MNRAS.392..868B}
}

@ARTICLE{2008MNRAS.391L..93G,
       author = {{Gieles}, M. and {Bastian}, N. and {Ercolano}, B.},
        title = "{Evolution of stellar structure in the Small Magellanic Cloud}",
      journal = {\mnras},
         year = 2008,
        month = nov,
       volume = {391},
       number = {1},
        pages = {L93-L97},
          doi = {10.1111/j.1745-3933.2008.00563.x},
archivePrefix = {arXiv},
       eprint = {0809.2295},
 primaryClass = {astro-ph},
       adsurl = {https://ui.adsabs.harvard.edu/abs/2008MNRAS.391L..93G}
}

@ARTICLE{2022MNRAS.512.1196M,
       author = {{Miller}, Amy E. and {Cioni}, Maria-Rosa L. and {de Grijs}, Richard and {Sun}, Ning-Chen and {Bell}, Cameron P.~M. and {Choudhury}, Samyaday and {Ivanov}, Valentin D. and {Marconi}, Marcella and {Oliveira}, Joana M. and {Petr-Gotzens}, Monika and {Ripepi}, Vincenzo and {van Loon}, Jacco Th},
        title = "{The VMC survey - XLVII. Turbulence-controlled hierarchical star formation in the Large Magellanic Cloud}",
      journal = {\mnras},
         year = 2022,
        month = may,
       volume = {512},
       number = {1},
        pages = {1196-1213},
          doi = {10.1093/mnras/stac508},
archivePrefix = {arXiv},
       eprint = {2202.09267},
 primaryClass = {astro-ph.GA},
       adsurl = {https://ui.adsabs.harvard.edu/abs/2022MNRAS.512.1196M}
}

@ARTICLE{2022MNRAS.516.4612T,
       author = {{Turner}, Jordan A. and {Dale}, Daniel A. and {Lilly}, James and {Boquien}, Mederic and {Deger}, Sinan and {Lee}, Janice C. and {Whitmore}, Bradley C. and {Anand}, Gagandeep S. and {Benincasa}, Samantha M. and {Bigiel}, Frank and {Blanc}, Guillermo A. and {Chevance}, M{\'e}lanie and {Emsellem}, Eric and {Faesi}, Christopher M. and {Glover}, Simon C.~O. and {Grasha}, Kathryn and {Hughes}, Annie and {Klessen}, Ralf S. and {Kreckel}, Kathryn and {Kruijssen}, J.~M. Diederik and {Leroy}, Adam K. and {Pan}, Hsi-An and {Rosolowsky}, Erik and {Schruba}, Andreas and {Williams}, Thomas G.},
        title = "{PHANGS: constraining star formation time-scales using the spatial correlations of star clusters and giant molecular clouds}",
      journal = {\mnras},
         year = 2022,
        month = nov,
       volume = {516},
       number = {3},
        pages = {4612-4626},
          doi = {10.1093/mnras/stac2559},
archivePrefix = {arXiv},
       eprint = {2209.02872},
 primaryClass = {astro-ph.GA},
       adsurl = {https://ui.adsabs.harvard.edu/abs/2022MNRAS.516.4612T}
}

@ARTICLE{2010ApJ...723..492R,
       author = {{Roman-Duval}, Julia and {Jackson}, James M. and {Heyer}, Mark and {Rathborne}, Jill and {Simon}, Robert},
        title = "{Physical Properties and Galactic Distribution of Molecular Clouds Identified in the Galactic Ring Survey}",
      journal = {\apj},
         year = 2010,
        month = nov,
       volume = {723},
       number = {1},
        pages = {492-507},
          doi = {10.1088/0004-637X/723/1/492},
archivePrefix = {arXiv},
       eprint = {1010.2798},
 primaryClass = {astro-ph.GA},
       adsurl = {https://ui.adsabs.harvard.edu/abs/2010ApJ...723..492R}
}

@ARTICLE{2017ApJ...849..149S,
       author = {{Sun}, Ning-Chen and {de Grijs}, Richard and {Subramanian}, Smitha and {Bekki}, Kenji and {Bell}, Cameron P.~M. and {Cioni}, Maria-Rosa L. and {Ivanov}, Valentin D. and {Marconi}, Marcella and {Oliveira}, Joana M. and {Piatti}, Andr{\'e}s E. and {Ripepi}, Vincenzo and {Rubele}, Stefano and {Tatton}, Ben L. and {van Loon}, Jacco Th.},
        title = "{The VMC Survey. XXVII. Young Stellar Structures in the LMC{\textquoteright}s Bar Star-forming Complex}",
      journal = {\apj},
         year = 2017,
        month = nov,
       volume = {849},
       number = {2},
          eid = {149},
        pages = {149},
          doi = {10.3847/1538-4357/aa911e},
archivePrefix = {arXiv},
       eprint = {1710.00984},
 primaryClass = {astro-ph.GA},
       adsurl = {https://ui.adsabs.harvard.edu/abs/2017ApJ...849..149S}
}

@article{Iglesias-Paramo:2004mnq,
    author = "Iglesias-Paramo, Jorge and Boselli, A. and Gavazzi, G. and Zaccardo, A.",
    title = "{Tracing the star formation history of cluster galaxies using the H-alpha / UV flux ratio}",
    eprint = "astro-ph/0403620",
    archivePrefix = "arXiv",
    doi = "10.1051/0004-6361:20034572",
    journal = "Astron. Astrophys.",
    volume = "421",
    pages = "887--897",
    year = "2004"
}

@ARTICLE{2005ApJ...619L..79T,
       author = {{Thilker}, David A. and {Bianchi}, Luciana and {Boissier}, Samuel and {Gil de Paz}, Armando and {Madore}, Barry F. and {Martin}, D. Christopher and {Meurer}, Gerhardt R. and {Neff}, Susan G. and {Rich}, R. Michael and {Schiminovich}, David and {Seibert}, Mark and {Wyder}, Ted K. and {Barlow}, Tom A. and {Byun}, Yong-Ik and {Donas}, Jose and {Forster}, Karl and {Friedman}, Peter G. and {Heckman}, Timothy M. and {Jelinsky}, Patrick N. and {Lee}, Young-Wook and {Malina}, Roger F. and {Milliard}, Bruno and {Morrissey}, Patrick and {Siegmund}, Oswald H.~W. and {Small}, Todd and {Szalay}, Alex S. and {Welsh}, Barry Y.},
        title = "{Recent Star Formation in the Extreme Outer Disk of M83}",
      journal = {\apjl},
         year = 2005,
        month = jan,
       volume = {619},
       number = {1},
        pages = {L79-L82},
          doi = {10.1086/425251},
archivePrefix = {arXiv},
       eprint = {astro-ph/0411306},
 primaryClass = {astro-ph},
       adsurl = {https://ui.adsabs.harvard.edu/abs/2005ApJ...619L..79T}
}

@article{Pasquali_2008,
doi = {10.1086/591658},
url = {https://doi.org/10.1086/591658},
year = {2008},
month = {nov},
publisher = {},
volume = {687},
number = {2},
pages = {1004},
author = {Pasquali, A. and Leroy, A. and Rix, H.-W. and Walter, F. and Herbst, T. and Giallongo, E. and Ragazzoni, R. and Baruffolo, A. and Speziali, R. and Hill, J. and Beccari, G. and Bouché, N. and Buschkamp, P. and Kochanek, C. and Skillman, E. and Bechtold, J.},
title = {The Large Binocular Telescope Panoramic View of the Recent Star Formation Activity in IC 2574},
journal = {The Astrophysical Journal}
}

@ARTICLE{2021JApA...42...30P,
       author = {{Postma}, Joseph E. and {Leahy}, Denis},
        title = "{UVIT data reduction pipeline: A CCDLAB and UVIT tutorial}",
      journal = {Journal of Astrophysics and Astronomy},
         year = 2021,
        month = oct,
       volume = {42},
       number = {2},
          eid = {30},
        pages = {30},
          doi = {10.1007/s12036-020-09689-w},
       adsurl = {https://ui.adsabs.harvard.edu/abs/2021JApA...42...30P}
}

@ARTICLE{2021MNRAS.502.1218R,
       author = {{Rosolowsky}, Erik and {Hughes}, Annie and {Leroy}, Adam K. and {Sun}, Jiayi and {Querejeta}, Miguel and {Schruba}, Andreas and {Usero}, Antonio and {Herrera}, Cinthya N. and {Liu}, Daizhong and {Pety}, J{\'e}r{\^o}me and {Saito}, Toshiki and {Be{\v{s}}li{\'c}}, Ivana and {Bigiel}, Frank and {Blanc}, Guillermo and {Chevance}, M{\'e}lanie and {Dale}, Daniel A. and {Deger}, Sinan and {Faesi}, Christopher M. and {Glover}, Simon C.~O. and {Henshaw}, Jonathan D. and {Klessen}, Ralf S. and {Kruijssen}, J.~M. Diederik and {Larson}, Kirsten and {Lee}, Janice and {Meidt}, Sharon and {Mok}, Angus and {Schinnerer}, Eva and {Thilker}, David A. and {Williams}, Thomas G.},
        title = "{Giant molecular cloud catalogues for PHANGS-ALMA: methods and initial results}",
      journal = {\mnras},
         year = 2021,
        month = mar,
       volume = {502},
       number = {1},
        pages = {1218-1245},
          doi = {10.1093/mnras/stab085},
archivePrefix = {arXiv},
       eprint = {2101.04697},
 primaryClass = {astro-ph.GA},
       adsurl = {https://ui.adsabs.harvard.edu/abs/2021MNRAS.502.1218R}
}

@article{Landy_1993,
author = {Landy, Stephen D. and Szalay, Alexander S.},
doi = {10.1086/172900},
issn = {0004-637X},
journal = {The Astrophysical Journal},
month = {jul},
pages = {64},
publisher = {IOP Publishing},
title = {{Bias and variance of angular correlation functions}},
url = {http://adsabs.harvard.edu/doi/10.1086/172900},
volume = {412},
year = {1993}
}

@ARTICLE{SciPy20,
  author  = {Virtanen, Pauli and Gommers, Ralf and Oliphant, Travis E. and
            Haberland, Matt and Reddy, Tyler and Cournapeau, David and
            Burovski, Evgeni and Peterson, Pearu and Weckesser, Warren and
            Bright, Jonathan and {van der Walt}, St{\'e}fan J. and
            Brett, Matthew and Wilson, Joshua and Millman, K. Jarrod and
            Mayorov, Nikolay and Nelson, Andrew R. J. and Jones, Eric and
            Kern, Robert and Larson, Eric and Carey, C J and
            Polat, {\.I}lhan and Feng, Yu and Moore, Eric W. and
            {VanderPlas}, Jake and Laxalde, Denis and Perktold, Josef and
            Cimrman, Robert and Henriksen, Ian and Quintero, E. A. and
            Harris, Charles R. and Archibald, Anne M. and
            Ribeiro, Ant{\^o}nio H. and Pedregosa, Fabian and
            {van Mulbregt}, Paul and {SciPy 1.0 Contributors}},
  title   = {{{SciPy} 1.0: Fundamental Algorithms for Scientific
            Computing in Python}},
  journal = {Nature Methods},
  year    = {2020},
  volume  = {17},
  pages   = {261--272},
  adsurl  = {https://rdcu.be/b08Wh},
  doi     = {10.1038/s41592-019-0686-2},
}

@ARTICLE{2022MNRAS.516.3006K,
       author = {{Kim}, Jaeyeon and {Chevance}, M{\'e}lanie and {Kruijssen}, J.~M. Diederik and {Leroy}, Adam K. and {Schruba}, Andreas and {Barnes}, Ashley T. and {Bigiel}, Frank and {Blanc}, Guillermo A. and {Cao}, Yixian and {Congiu}, Enrico and {Dale}, Daniel A. and {Faesi}, Christopher M. and {Glover}, Simon C.~O. and {Grasha}, Kathryn and {Groves}, Brent and {Hughes}, Annie and {Klessen}, Ralf S. and {Kreckel}, Kathryn and {McElroy}, Rebecca and {Pan}, Hsi-An and {Pety}, J{\'e}r{\^o}me and {Querejeta}, Miguel and {Razza}, Alessandro and {Rosolowsky}, Erik and {Saito}, Toshiki and {Schinnerer}, Eva and {Sun}, Jiayi and {Tomi{\v{c}}i{\'c}}, Neven and {Usero}, Antonio and {Williams}, Thomas G.},
        title = "{Environmental dependence of the molecular cloud lifecycle in 54 main-sequence galaxies}",
      journal = {\mnras},
         year = 2022,
        month = oct,
       volume = {516},
       number = {2},
        pages = {3006-3028},
          doi = {10.1093/mnras/stac2339},
archivePrefix = {arXiv},
       eprint = {2206.09857},
 primaryClass = {astro-ph.GA},
       adsurl = {https://ui.adsabs.harvard.edu/abs/2022MNRAS.516.3006K}
}

@ARTICLE{2020ApJ...897..122L,
       author = {{Lang}, Philipp and {Meidt}, Sharon E. and {Rosolowsky}, Erik and {Nofech}, Joseph and {Schinnerer}, Eva and {Leroy}, Adam K. and {Emsellem}, Eric and {Pessa}, Ismael and {Glover}, Simon C.~O. and {Groves}, Brent and {Hughes}, Annie and {Kruijssen}, J.~M. Diederik and {Querejeta}, Miguel and {Schruba}, Andreas and {Bigiel}, Frank and {Blanc}, Guillermo A. and {Chevance}, M{\'e}lanie and {Colombo}, Dario and {Faesi}, Christopher and {Henshaw}, Jonathan D. and {Herrera}, Cinthya N. and {Liu}, Daizhong and {Pety}, J{\'e}r{\^o}me and {Puschnig}, Johannes and {Saito}, Toshiki and {Sun}, Jiayi and {Usero}, Antonio},
        title = "{PHANGS CO Kinematics: Disk Orientations and Rotation Curves at 150 pc Resolution}",
      journal = {\apj},
         year = 2020,
        month = jul,
       volume = {897},
       number = {2},
          eid = {122},
        pages = {122},
          doi = {10.3847/1538-4357/ab9953},
archivePrefix = {arXiv},
       eprint = {2005.11709},
 primaryClass = {astro-ph.GA},
       adsurl = {https://ui.adsabs.harvard.edu/abs/2020ApJ...897..122L}
}

@ARTICLE{2016MNRAS.460..689R,
       author = {{Richards}, Emily E. and {van Zee}, L. and {Barnes}, K.~L. and {Staudaher}, S. and {Dale}, D.~A. and {Braun}, T.~T. and {Wavle}, D.~C. and {Dalcanton}, J.~J. and {Bullock}, J.~S. and {Chandar}, R.},
        title = "{Baryonic distributions in galaxy dark matter haloes - I. New observations of neutral and ionized gas kinematics}",
      journal = {\mnras},
         year = 2016,
        month = jul,
       volume = {460},
       number = {1},
        pages = {689-728},
          doi = {10.1093/mnras/stw1016},
archivePrefix = {arXiv},
       eprint = {1605.01638},
 primaryClass = {astro-ph.GA},
       adsurl = {https://ui.adsabs.harvard.edu/abs/2016MNRAS.460..689R}
}

@ARTICLE{2014A&A...563A..31R,
       author = {{R{\'e}my-Ruyer}, A. and {Madden}, S.~C. and {Galliano}, F. and {Galametz}, M. and {Takeuchi}, T.~T. and {Asano}, R.~S. and {Zhukovska}, S. and {Lebouteiller}, V. and {Cormier}, D. and {Jones}, A. and {Bocchio}, M. and {Baes}, M. and {Bendo}, G.~J. and {Boquien}, M. and {Boselli}, A. and {DeLooze}, I. and {Doublier-Pritchard}, V. and {Hughes}, T. and {Karczewski}, O. {\L}. and {Spinoglio}, L.},
        title = "{Gas-to-dust mass ratios in local galaxies over a 2 dex metallicity range}",
      journal = {\aap},
         year = 2014,
        month = mar,
       volume = {563},
          eid = {A31},
        pages = {A31},
          doi = {10.1051/0004-6361/201322803},
archivePrefix = {arXiv},
       eprint = {1312.3442},
 primaryClass = {astro-ph.GA},
       adsurl = {https://ui.adsabs.harvard.edu/abs/2014A&A...563A..31R}
}

@ARTICLE{2016AJ....152..134M,
       author = {{Maier}, Erin and {Chien}, Li-Hsin and {Hunter}, Deidre A.},
        title = "{Turbulence and Star Formation in a Sample of Spiral Galaxies}",
      journal = {\aj},
         year = 2016,
        month = nov,
       volume = {152},
       number = {5},
          eid = {134},
        pages = {134},
          doi = {10.3847/0004-6256/152/5/134},
archivePrefix = {arXiv},
       eprint = {1608.02321},
 primaryClass = {astro-ph.GA},
       adsurl = {https://ui.adsabs.harvard.edu/abs/2016AJ....152..134M}
}

@ARTICLE{2010MNRAS.409.1088B,
       author = {{Bournaud}, Fr{\'e}d{\'e}ric and {Elmegreen}, Bruce G. and {Teyssier}, Romain and {Block}, David L. and {Puerari}, Iv{\^a}nio},
        title = "{ISM properties in hydrodynamic galaxy simulations: turbulence cascades, cloud formation, role of gravity and feedback}",
      journal = {\mnras},
         year = 2010,
        month = dec,
       volume = {409},
       number = {3},
        pages = {1088-1099},
          doi = {10.1111/j.1365-2966.2010.17370.x},
archivePrefix = {arXiv},
       eprint = {1007.2566},
 primaryClass = {astro-ph.CO},
       adsurl = {https://ui.adsabs.harvard.edu/abs/2010MNRAS.409.1088B}
}

@ARTICLE{2018MNRAS.478.1611K,
       author = {{Koribalski}, B{\"a}rbel S. and {Wang}, Jing and {Kamphuis}, P. and {Westmeier}, T. and {Staveley-Smith}, L. and {Oh}, S.-H. and {L{\'o}pez-S{\'a}nchez}, {\'A}. R. and {Wong}, O.~I. and {Ott}, J. and {de Blok}, W.~J.~G. and {Shao}, L.},
        title = "{The Local Volume H I Survey (LVHIS)}",
      journal = {\mnras},
         year = 2018,
        month = aug,
       volume = {478},
       number = {2},
        pages = {1611-1648},
          doi = {10.1093/mnras/sty479},
archivePrefix = {arXiv},
       eprint = {1904.09648},
 primaryClass = {astro-ph.GA},
       adsurl = {https://ui.adsabs.harvard.edu/abs/2018MNRAS.478.1611K}
}

@article{JMLR:v12:pedregosa11a,
  author  = {Fabian Pedregosa and Ga{{\"e}}l Varoquaux and Alexandre Gramfort and Vincent Michel and Bertrand Thirion and Olivier Grisel and Mathieu Blondel and Peter Prettenhofer and Ron Weiss and Vincent Dubourg and Jake Vanderplas and Alexandre Passos and David Cournapeau and Matthieu Brucher and Matthieu Perrot and {{\'E}}douard Duchesnay},
  title   = {Scikit-learn: Machine Learning in Python},
  journal = {Journal of Machine Learning Research},
  year    = {2011},
  volume  = {12},
  number  = {85},
  pages   = {2825--2830},
  url     = {http://jmlr.org/papers/v12/pedregosa11a.html}
}

@ARTICLE{2009ApJ...692..364F,
       author = {{Federrath}, Christoph and {Klessen}, Ralf S. and {Schmidt}, Wolfram},
        title = "{The Fractal Density Structure in Supersonic Isothermal Turbulence: Solenoidal Versus Compressive Energy Injection}",
      journal = {\apj},
         year = 2009,
        month = feb,
       volume = {692},
       number = {1},
        pages = {364-374},
          doi = {10.1088/0004-637X/692/1/364},
archivePrefix = {arXiv},
       eprint = {0710.1359},
 primaryClass = {astro-ph},
       adsurl = {https://ui.adsabs.harvard.edu/abs/2009ApJ...692..364F}
}

@ARTICLE{2021PASP..133j2001B,
       author = {{Burkhart}, Blakesley},
        title = "{Diagnosing Turbulence in the Neutral and Molecular Interstellar Medium of Galaxies}",
      journal = {\pasp},
         year = 2021,
        month = oct,
       volume = {133},
       number = {1028},
          eid = {102001},
        pages = {102001},
          doi = {10.1088/1538-3873/ac25cf},
archivePrefix = {arXiv},
       eprint = {2106.02239},
 primaryClass = {astro-ph.GA},
       adsurl = {https://ui.adsabs.harvard.edu/abs/2021PASP..133j2001B}
}

@ARTICLE{2018ApJ...853...88E,
       author = {{Elmegreen}, Bruce G.},
        title = "{On the Dispersal of Young Stellar Hierarchies}",
      journal = {\apj},
         year = 2018,
        month = jan,
       volume = {853},
       number = {1},
          eid = {88},
        pages = {88},
          doi = {10.3847/1538-4357/aaa252},
archivePrefix = {arXiv},
       eprint = {1712.05967},
 primaryClass = {astro-ph.GA},
       adsurl = {https://ui.adsabs.harvard.edu/abs/2018ApJ...853...88E}
}

@ARTICLE{1997A&A...327..966D,
       author = {{de Grijs}, R. and {Peletier}, R.~F. and {van der Kruit}, P.~C.},
        title = "{The z-structure of disk galaxies towards the galaxy planes}",
      journal = {\aap},
         year = 1997,
        month = nov,
       volume = {327},
        pages = {966-982},
          doi = {10.48550/arXiv.astro-ph/9707074},
archivePrefix = {arXiv},
       eprint = {astro-ph/9707074},
 primaryClass = {astro-ph},
       adsurl = {https://ui.adsabs.harvard.edu/abs/1997A&A...327..966D}
}

@ARTICLE{2014MNRAS.439.3775G,
       author = {{Gouliermis}, Dimitrios A. and {Hony}, Sacha and {Klessen}, Ralf S.},
        title = "{The complex distribution of recently formed stars. Bimodal stellar clustering in the star-forming region NGC 346}",
      journal = {\mnras},
         year = 2014,
        month = apr,
       volume = {439},
       number = {4},
        pages = {3775-3789},
          doi = {10.1093/mnras/stu228},
archivePrefix = {arXiv},
       eprint = {1402.0078},
 primaryClass = {astro-ph.GA},
       adsurl = {https://ui.adsabs.harvard.edu/abs/2014MNRAS.439.3775G}
}

@article{Walter2008THINGS,
  author = {Walter, Fabian and Brinks, Erwin and de Blok, W. J. G.
            and Bigiel, Frank and Kennicutt, Robert C., Jr.
            and Thornley, Michele D. and Leroy, Adam},
  title = {THINGS: The HI Nearby Galaxy Survey},
  journal = {The Astronomical Journal},
  volume = {136},
  number = {6},
  pages = {2563--2647},
  year = {2008},
  doi = {10.1088/0004-6256/136/6/2563}
}

@ARTICLE{1998MNRAS.299..595D,
       author = {{de Grijs}, R.},
        title = "{The global structure of galactic discs}",
      journal = {\mnras},
         year = 1998,
        month = sep,
       volume = {299},
       number = {2},
        pages = {595-610},
          doi = {10.1046/j.1365-8711.1998.01896.x},
archivePrefix = {arXiv},
       eprint = {astro-ph/9804337},
 primaryClass = {astro-ph},
       adsurl = {https://ui.adsabs.harvard.edu/abs/1998MNRAS.299..595D}
}

@ARTICLE{2018NatAs...2..896O,
       author = {{Offner}, Stella S.~R. and {Liu}, Yue},
        title = "{Turbulent action at a distance due to stellar feedback in magnetized clouds}",
      journal = {Nature Astronomy},
         year = 2018,
        month = sep,
       volume = {2},
        pages = {896-900},
          doi = {10.1038/s41550-018-0566-1},
archivePrefix = {arXiv},
       eprint = {1809.03513},
 primaryClass = {astro-ph.SR},
       adsurl = {https://ui.adsabs.harvard.edu/abs/2018NatAs...2..896O}
}

@ARTICLE{2016ApJ...832..143F,
       author = {{Federrath}, C. and {Rathborne}, J.~M. and {Longmore}, S.~N. and {Kruijssen}, J.~M.~D. and {Bally}, J. and {Contreras}, Y. and {Crocker}, R.~M. and {Garay}, G. and {Jackson}, J.~M. and {Testi}, L. and {Walsh}, A.~J.},
        title = "{The Link between Turbulence, Magnetic Fields, Filaments, and Star Formation in the Central Molecular Zone Cloud G0.253+0.016}",
      journal = {\apj},
         year = 2016,
        month = dec,
       volume = {832},
       number = {2},
          eid = {143},
        pages = {143},
          doi = {10.3847/0004-637X/832/2/143},
archivePrefix = {arXiv},
       eprint = {1609.05911},
 primaryClass = {astro-ph.GA},
       adsurl = {https://ui.adsabs.harvard.edu/abs/2016ApJ...832..143F}
}

@ARTICLE{2016ApJ...825...30P,
       author = {{Pan}, Liubin and {Padoan}, Paolo and {Haugb{\o}lle}, Troels and {Nordlund}, {\r{A}}ke},
        title = "{Supernova Driving. II. Compressive Ratio in Molecular-cloud Turbulence}",
      journal = {\apj},
         year = 2016,
        month = jul,
       volume = {825},
       number = {1},
          eid = {30},
        pages = {30},
          doi = {10.3847/0004-637X/825/1/30},
archivePrefix = {arXiv},
       eprint = {1510.04742},
 primaryClass = {astro-ph.GA},
       adsurl = {https://ui.adsabs.harvard.edu/abs/2016ApJ...825...30P}
}

@ARTICLE{2026MNRAS.547ag359M,
       author = {{Miller}, Lewis J. and {Grasha}, Kathryn and {Federrath}, Christoph},
        title = "{The turbulence driving mode in NGC7793 and NGC1313}",
      journal = {\mnras},
         year = 2026,
        month = apr,
       volume = {547},
       number = {2},
          eid = {stag359},
        pages = {stag359},
          doi = {10.1093/mnras/stag359},
archivePrefix = {arXiv},
       eprint = {2602.21405},
 primaryClass = {astro-ph.GA},
       adsurl = {https://ui.adsabs.harvard.edu/abs/2026MNRAS.547ag359M}
}

@ARTICLE{2026arXiv260407450H,
       author = {{He}, Hao and {Leroy}, Adam and {Rosolowsky}, Erik and {Hughes}, Annie and {Sun}, Jiayi and {Machado}, Joshua and {Bigiel}, Frank and {Barnes}, Ashley and {Bazzi}, Zein and {Cao}, Yixian and {Chevance}, Melanie and {Colombo}, Dario and {Glover}, Simon C.~O. and {Henshaw}, Jonathan D. and {Koch}, Eric W. and {Meidt}, Sharon E. and {Pan}, Hsi-An and {Saito}, Toshiki and {Sarbadhicary}, Sumit K. and {Schinnerer}, Eva and {Smith}, Rowan J. and {Usero}, Antonio and {Weinberg}, David H. and {Williams}, Thomas G.},
        title = "{The Structure of Molecular Gas in PHANGS-ALMA Galaxies: Cloud Spacing, Two-Point Correlation and Stacked Intensity Profiles}",
      journal = {arXiv e-prints},
         year = 2026,
        month = apr,
          eid = {arXiv:2604.07450},
        pages = {arXiv:2604.07450},
          doi = {10.48550/arXiv.2604.07450},
archivePrefix = {arXiv},
       eprint = {2604.07450},
 primaryClass = {astro-ph.GA},
       adsurl = {https://ui.adsabs.harvard.edu/abs/2026arXiv260407450H}
}

@ARTICLE{2016ApJ...822...11P,
       author = {{Padoan}, Paolo and {Pan}, Liubin and {Haugb{\o}lle}, Troels and {Nordlund}, {\r{A}}ke},
        title = "{Supernova Driving. I. The Origin of Molecular Cloud Turbulence}",
      journal = {\apj},
         year = 2016,
        month = may,
       volume = {822},
       number = {1},
          eid = {11},
        pages = {11},
          doi = {10.3847/0004-637X/822/1/11},
archivePrefix = {arXiv},
       eprint = {1509.04663},
 primaryClass = {astro-ph.GA},
       adsurl = {https://ui.adsabs.harvard.edu/abs/2016ApJ...822...11P}
}

@ARTICLE{2026ApJ..1002..220A,
       author = {{Ananthu}, Sanal and {Shashank}, Gairola and {Subramanian}, Smitha and {Jayanth}, Rao C. and {Menon}, Shyam H. and {Mondal}, Chayan and {Muraleedharan}, Sreedevi},
        title = "{An FUV-optical Approach for Studying Hierarchical Star Formation in Nearby Galaxies with UVIT}",
      journal = {\apj},
         year = 2026,
        month = may,
       volume = {1002},
       number = {2},
          eid = {220},
        pages = {220},
          doi = {10.3847/1538-4357/ae5bb9},
archivePrefix = {arXiv},
       eprint = {2602.22860},
 primaryClass = {astro-ph.GA},
       adsurl = {https://ui.adsabs.harvard.edu/abs/2026ApJ..1002..220A}
}

@ARTICLE{2008ApJ...681.1248O,
       author = {{Odekon}, Mary Crone},
        title = "{Characteristic Scales in Stellar Clustering: A Transition Near the Disk Scale Height}",
      journal = {\apj},
         year = 2008,
        month = jul,
       volume = {681},
       number = {2},
        pages = {1248-1253},
          doi = {10.1086/589141},
       adsurl = {https://ui.adsabs.harvard.edu/abs/2008ApJ...681.1248O}
}

@ARTICLE{2021MNRAS.500.1721M,
       author = {{Menon}, Shyam H. and {Federrath}, Christoph and {Klaassen}, Pamela and {Kuiper}, Rolf and {Reiter}, Megan},
        title = "{On the compressive nature of turbulence driven by ionizing feedback in the pillars of the Carina Nebula}",
      journal = {\mnras},
         year = 2021,
        month = jan,
       volume = {500},
       number = {2},
        pages = {1721-1740},
          doi = {10.1093/mnras/staa3271},
archivePrefix = {arXiv},
       eprint = {2010.09861},
 primaryClass = {astro-ph.GA},
       adsurl = {https://ui.adsabs.harvard.edu/abs/2021MNRAS.500.1721M}
}

@ARTICLE{2017MNRAS.469..286R,
       author = {{Romeo}, Alessandro B. and {Mogotsi}, Keoikantse Moses},
        title = "{What drives gravitational instability in nearby star-forming spirals? The impact of CO and H I velocity dispersions}",
      journal = {\mnras},
         year = 2017,
        month = jul,
       volume = {469},
       number = {1},
        pages = {286-294},
          doi = {10.1093/mnras/stx844},
archivePrefix = {arXiv},
       eprint = {1701.02138},
 primaryClass = {astro-ph.GA},
       adsurl = {https://ui.adsabs.harvard.edu/abs/2017MNRAS.469..286R}
}

@ARTICLE{2016MNRAS.460.2360R,
       author = {{Romeo}, Alessandro B. and {Fathi}, Kambiz},
        title = "{What powers the starburst activity of NGC 1068? Star-driven gravitational instabilities caught in the act}",
      journal = {\mnras},
         year = 2016,
        month = aug,
       volume = {460},
       number = {3},
        pages = {2360-2367},
          doi = {10.1093/mnras/stw1147},
archivePrefix = {arXiv},
       eprint = {1602.03049},
 primaryClass = {astro-ph.GA},
       adsurl = {https://ui.adsabs.harvard.edu/abs/2016MNRAS.460.2360R}
}

@ARTICLE{2010MNRAS.407.1223R,
       author = {{Romeo}, Alessandro B. and {Burkert}, Andreas and {Agertz}, Oscar},
        title = "{A Toomre-like stability criterion for the clumpy and turbulent interstellar medium}",
      journal = {\mnras},
         year = 2010,
        month = sep,
       volume = {407},
       number = {2},
        pages = {1223-1230},
          doi = {10.1111/j.1365-2966.2010.16975.x},
archivePrefix = {arXiv},
       eprint = {1001.4732},
 primaryClass = {astro-ph.CO},
       adsurl = {https://ui.adsabs.harvard.edu/abs/2010MNRAS.407.1223R}
}

@ARTICLE{2015ApJ...806L..34F,
       author = {{Fathi}, Kambiz and {Izumi}, Takuma and {Romeo}, Alessandro B. and {Mart{\'\i}n}, Sergio and {Imanishi}, Masatoshi and {Hatziminaoglou}, Evanthia and {Aalto}, Susanne and {Espada}, Daniel and {Kohno}, Kotaro and {Krips}, Melanie and {Matsushita}, Satoki and {Meier}, David S. and {Nakai}, Naomasa and {Terashima}, Yuichi},
        title = "{Local Instability Signatures in ALMA Observations of Dense Gas in NGC 7469}",
      journal = {\apjl},
         year = 2015,
        month = jun,
       volume = {806},
       number = {2},
          eid = {L34},
        pages = {L34},
          doi = {10.1088/2041-8205/806/2/L34},
archivePrefix = {arXiv},
       eprint = {1506.01157},
 primaryClass = {astro-ph.GA},
       adsurl = {https://ui.adsabs.harvard.edu/abs/2015ApJ...806L..34F}
}

@ARTICLE{2013MNRAS.433.1389R,
       author = {{Romeo}, Alessandro B. and {Falstad}, Niklas},
        title = "{A simple and accurate approximation for the Q stability parameter in multicomponent and realistically thick discs}",
      journal = {\mnras},
         year = 2013,
        month = aug,
       volume = {433},
       number = {2},
        pages = {1389-1397},
          doi = {10.1093/mnras/stt809},
archivePrefix = {arXiv},
       eprint = {1302.4291},
 primaryClass = {astro-ph.CO},
       adsurl = {https://ui.adsabs.harvard.edu/abs/2013MNRAS.433.1389R}
}

@ARTICLE{2023A&A...672A.193F,
       author = {{Fensch}, J{\'e}r{\'e}my and {Bournaud}, Fr{\'e}d{\'e}ric and {Brucy}, No{\'e} and {Dubois}, Yohan and {Hennebelle}, Patrick and {Rosdahl}, Joakim},
        title = "{Universal gravity-driven isothermal turbulence cascade in disk galaxies}",
      journal = {\aap},
         year = 2023,
        month = apr,
       volume = {672},
          eid = {A193},
        pages = {A193},
          doi = {10.1051/0004-6361/202245491},
archivePrefix = {arXiv},
       eprint = {2301.13221},
 primaryClass = {astro-ph.GA},
       adsurl = {https://ui.adsabs.harvard.edu/abs/2023A&A...672A.193F}
}


\end{document}